\documentclass[11pt, a4paper]{article}
\pdfoutput=1

\usepackage{jcappub}
\usepackage{mypackages}
\usepackage{hyperref} %Automatically links \label and \ref commands; Always load last
\hypersetup{
    colorlinks=true,       % false: boxed links; true: colored links
    linkcolor=red,          % color of internal links
    citecolor=blue,        % color of links to bibliography
    filecolor=magenta,      % color of file links
    urlcolor=blue           % color of external links
}

\usepackage{myterms}
\newcommand{\CMB}{\text{\sc cmb}}
\newcommand{\EBL}{\text{\sc ebl}}

\newcommand{\zgg}{z_{\gamma \gamma}}
\newcommand{\tgg}{t_{\gamma \gamma}}
\newcommand{\ttrans}{\vartheta_{\rm trans}}
\newcommand{\tlat}{\vartheta_{\rm lat}}
\newcommand{\ds}{d_{s}}
\newcommand{\zs}{z_{s}}
\newcommand{\dg}{d_{\gamma_0}}
\newcommand{\Dg}{D_{\gamma_0}}
\newcommand{\Eg}{E_{\gamma_0}}
\newcommand{\vpara}{v_{\parallel}}
\newcommand{\vperp}{v_{\perp}}
\newcommand{\Bhat}{\hat{\bf B}}
\newcommand{\nhat}{\hat{\bf n}}
\newcommand{\Eqns}{Eqs.~(\ref{equation1})-(\ref{equation3}) }

\title{
Morphology of blazar-induced gamma ray halos due to a helical intergalactic magnetic field
}

\author{Andrew J. Long}
\author{and Tanmay Vachaspati}

\affiliation{Physics Department and School of Earth and Space Exploration, Arizona State University, Tempe, Arizona 85287, USA.}

\emailAdd{andrewjlong@asu.edu}
\emailAdd{tvachasp@asu.edu}

\abstract{
We study the characteristic size and shape of idealized blazar-induced cascade halos in the $1-100 \GeV$ energy range assuming various non-helical and helical configurations for the intergalactic magnetic field (IGMF).  
While the magnetic field creates an extended halo, the helicity provides the halo with a twist.  
Under simplifying assumptions, we assess the parameter regimes for which it is possible to measure the size and shape of the halo from a single source and then to deduce properties of the IGMF.  
We find that blazar halo measurements with an experiment similar to Fermi-LAT are best suited to probe a helical magnetic field with strength and coherence length today in the ranges $10^{-17} \lesssim B_{0} / {\rm Gauss} \lesssim 10^{-13}$ and $10 \, {\rm Mpc} \lesssim \lambda \lesssim 10 \, {\rm Gpc}$ where $\Hcal \sim B_0^2 / \lambda$ is the magnetic helicity density. 
Stronger magnetic fields or smaller coherence scales can still potentially be investigated, but the connection between the halo morphology and the magnetic field properties is more involved.  
Weaker magnetic fields or longer coherence scales require high photon statistics or superior angular resolution.
}


\begin{document}
\maketitle

%\setlength{\parindent}{20pt}
%\setlength{\parskip}{2.5ex}

%==================================
% Introduction
%==================================
\section{Introduction}\label{sec:Introduction}

%==========
Several lines of reasoning suggest that the voids between galaxies and galaxy clusters contain a large-scale intergalactic magnetic field (IGMF).  
The presence of even a weak IGMF is sufficient to explain the observed micro-Gauss galactic and cluster magnetic fields \cite{Kronberg:1993vk, Widrow:2002ud, Widrow:2011hs}
through the dynamo amplification.  
Moreover, the IGMF may be a remnant from the early universe as it could have been generated during cosmological phase transitions \cite{Vachaspati:1991nm, Ahonen:1997wh}, in the epoch of matter-genesis \cite{Cornwall:1997ms, Vachaspati:2001nb, Long:2013tha}, or in certain inflationary scenarios \cite{Turner:1987bw}.
An observation of the IGMF and measurement of its energy and helicity spectra, therefore, could serve as a powerful new probe of astrophysics, particle physics, and early universe cosmology.  
(For recent reviews on cosmic magnetic fields, see \rref{Durrer:2013pga, Subramanian:2015lua}.)

%==========
On the observational side, TeV blazars offer one of the best strategies for measurements of the IGMF at redshifts out to $z \sim 1$ \cite{1994ApJ...423L...5A, 1995Natur.374..430P, Neronov:2006hc, Elyiv:2009bx, Dolag:2009iv}.  The blazar initiates an electromagnetic cascade as its TeV gamma rays produce electron and positron pairs (leptons) upon scattering on extragalactic background light (EBL).  
The cascade develops as the leptons inverse-Compton scatter on cosmic microwave background (CMB) photons producing secondary GeV gamma rays.
Since the initial TeV gamma ray has a mean free path of $\gtrsim 10 \Mpc$, the cascade develops outside of the host halo where it probes the IGMF.   
In the presence of an IGMF, the charged leptons are deflected by the magnetic field and the blazar acquires a halo of GeV gamma rays.
In the weak field regime, the leptons experience a gentle deflection and the angular extent of the magnetically broadened cascade (MBC) goes as $\Theta \propto B$.  
In the strong field regime, the leptons are so dramatically deflected that the secondary emission is isotropized into an extended pair-halo (PH).  

%==========
The hunt for cascade halos is ongoing at a number of gamma ray observatories \cite{Aleksic:2010gi, Abramowski:2014uta, Alonso:2014ewa} as well as independent collaborations \cite{Neronov:1900zz, Tavecchio:2010mk, Ando:2010rb, 2011APh....35..135E, Chen:2014rsa}.   
Most recently \rref{Chen:2014rsa} reports evidence for GeV halos around 24 low redshift blazars, which are revealed in a stacked analysis of the Fermi-LAT gamma ray data.  
Assuming a magnetically broadened cascade, {\it i.e.} weak bending approximation, they infer the 
IGMF field strength to be $B_{0} \sim 10^{-17}-10^{-15} \Gauss$ where $B_0$ is the magnetic field
strength at the present cosmological epoch.

%==========
While there has been extensive analytic and numerical work on the relationship between halo size and magnetic field strength, the literature contains little discussion of what information could be extracted from the halo shape.  
In principle the halo size and shape together (morphology) may encode not only the magnetic field strength but also its helicity\footnote{  
A helical magnetic field has a larger amplitude in either left- or right-circularly polarized modes.}.

%==========
There are two compelling reasons for considering a {\it helical} IGMF.  
First, magnetic helicity is a prediction of many models of magnetogenesis from the matter-genesis epoch \cite{Cornwall:1997ms, Vachaspati:2001nb, Long:2013tha}, cosmological inflation \cite{Caprini:2014mja, Atmjeet:2014cxa, Bartolo:2014hwa, Cheng:2014kga, Fujita:2015iga, Campanelli:2015jfa}, and other early universe scenarios \cite{Jackiw:1999bd, Joyce:1997uy, DiazGil:2007dy}.  
Second, the evolution of a magnetic field within the magneto-hydrodynamic (MHD) approximation is 
known to be more robust to dissipation if it is helical \cite{Kahniashvili:2012uj}. 
So, if a magnetic field was generated by causal processes in the early universe, it has a much better chance of surviving if it is helical.  

%==========
The helicity of the IGMF can be observed if charged particles propagate in this field, such as in the case of cosmic rays \cite{Kahniashvili:2005yp} and cascade gamma rays  \cite{Tashiro:2013bxa, Tashiro:2014gfa}. 
The parity violating IGMF helicity leads to certain non-trivial parity-odd correlation functions of the arrival directions of cosmic rays and cascade gamma rays.
An advantage of seeking a parity odd signature is that other sources of noise are expected to be parity even and hence do not contribute to the signature on average.  
The parity-odd correlation, called $Q$, has been evaluated for diffuse gamma ray data obtained by the Fermi Gamma Ray Telescope, and provides evidence for an IGMF of strength $\sim 10^{-14}\Gauss$ on distance scales $\sim 10\Mpc$ with left-handed helicity \cite{Tashiro:2013ita, Chen:2014qva}.  

%==========
In this work we endeavor to develop an understanding of how a helical IGMF leads to parity-violating features in the shape of blazar-induced cascade halos.  
To that end, we study an idealized system:  we focus on a few toy models of the IGMF, we do not model astrophysical sources, we do not include stochasticity in the development of the cascade, and we do not include foreground (noise).  
In this setting we can focus on the relationships between the parameters of the magnetic field model, {\it i.e.} the field strength, coherence length, and helicity, and the resulting size and shape of the cascade halo.  
We find that there are several regimes in the range of interesting physical parameters that each lead to qualitatively different halo morphology.
We identify the region of parameter space where measurements of cascade halo size and shape are best suited to probe the helical IGMF.  

%==========
In \sref{sec:Cascade} we review the physics giving rise to the cascade halo.  
In \sref{sec:Arrival} we study the propagation of particles through the cascade and derive a set of equations that can be solved to find the size and shape (morphology) of the halo.  
In \sref{sec:Examples} we apply the results of \sref{sec:Arrival} to calculate the halo shape for five specific magnetic field configurations, three non-helical and two helical.  
In \sref{sec:Sensitivity} we analyze which regions of the magnetic field parameter space could be probed by measurements of cascade halos.  
We conclude in \sref{sec:Conclusion} with a discussion of potential directions for future work.

%==================================
% The Blazar-Induced Cascade Halo
%==================================
\section{The Blazar-Induced Cascade Halo}\label{sec:Cascade}

\begin{figure}[t]
\hspace{0pt}
\vspace{-0in}
\begin{center}
\includegraphics[width=0.95\textwidth]{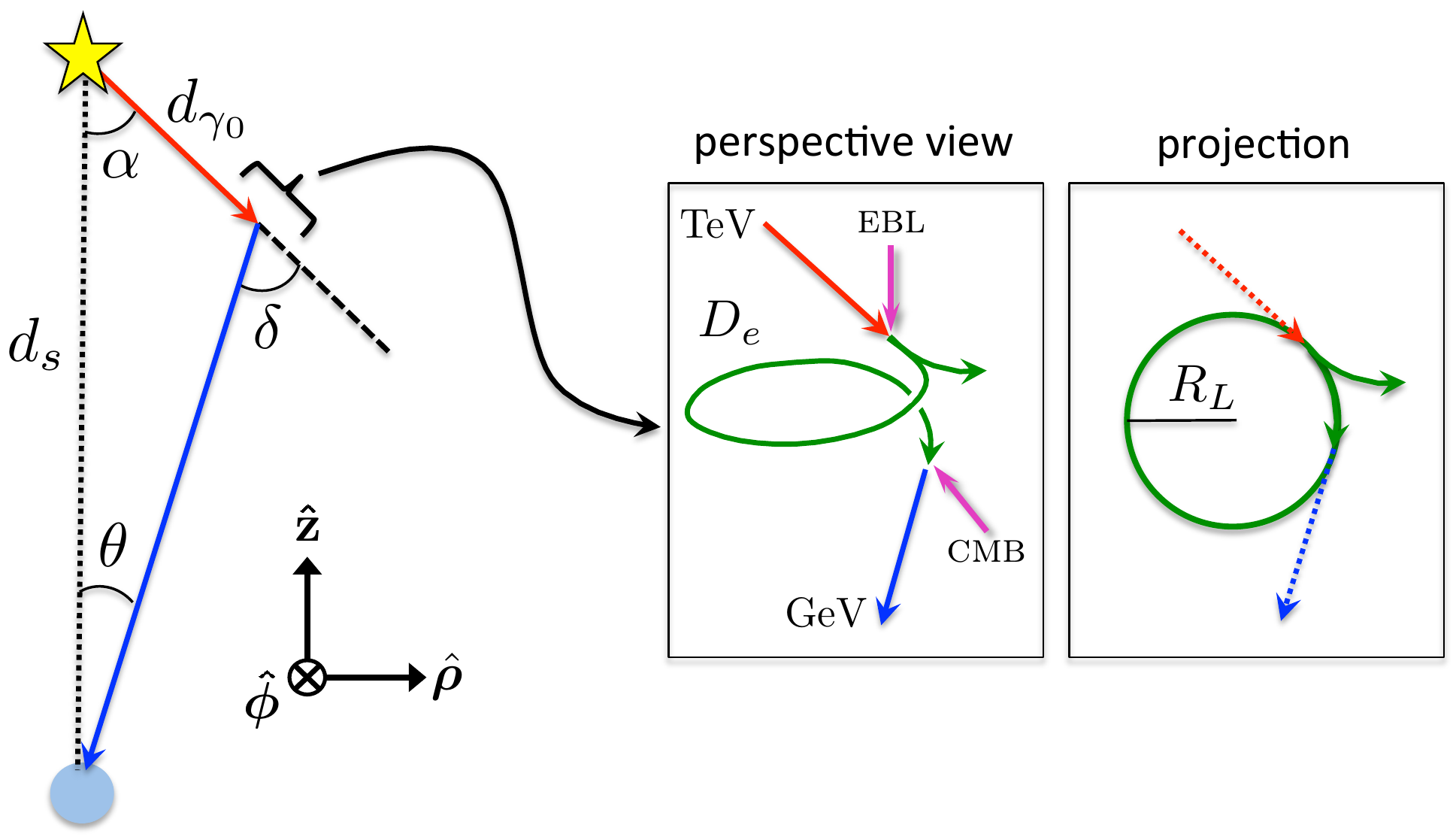}
\caption{
\label{fig:triangle}
The geometry of gamma ray propagation.  A TeV gamma ray (red arrow) travels off of the line of sight (black dashed line) connecting the blazar (yellow star) to Earth (pale blue dot).  Pair production occurs and either the electron or positron (green arrow) is deflected back toward the line of sight.  Inverse Compton scattering creates a GeV photon (blue arrow) that eventually reaches Earth.  The lepton need not make a complete orbit as we have shown here.  
}
\end{center}
\end{figure}

%==========
As depicted in Fig.~\ref{fig:triangle}, a blazar's TeV gamma rays initiate an electromagnetic cascade when they scatter on extragalactic background light.  
If the cascade occurs in the presence of a magnetic field, the blazar acquires a halo of GeV photons.  
Then the shape and angular extent of the halo are related to the magnetic field in the neighborhood of the blazar.  
This section is a review of the physics giving rise to the halo following \rref{Neronov:2009gh}.  

%=========

%==========
The comoving distance from the Earth to a blazar (source) at redshift $\zs$ is
\begin{align}\label{eq:dE_def}
	\ds
	= \frac{c}{a_{0} H_{0}} \int_{0}^{\zs} \frac{dz}{\sqrt{\Omega_{m}(1+z)^3 + \Omega_{\Lambda}}}
	\simeq (1 \Gpc) \frac{\zs}{0.24} 
\end{align}
where we have used the measured values of the cosmological parameters \cite{Planck:2015xua} and assumed $\zs \ll 1$ in the last equality.  
The blazar emits $O(1-10 \TeV)$ gamma rays into a jet or pair of jets \cite{Costamante:2013sva}.  
In the simplest model of the jet, radiation is uniform within a cone of half opening angle $\theta_{\rm jet}$, and typically $\theta_{\rm jet} \simeq 5^{\circ}$ corresponding to a solid angle of $\Omega_{\rm jet} \simeq 0.024 \sr$.   

%==========
A gamma ray with energy $\Eg \sim \TeV$ at redshift $\zgg$ is likely to scatter on optical and infrared extragalactic background light (EBL) and produce an electron-positron pair.  
The mean free path is given by $\Dg = \langle \sigma_{\gamma \gamma} n_{\EBL} \rangle^{-1}$ where $\sigma_{\gamma \gamma}$ is the pair production cross section and $dn_{\EBL}(\epsilon,\zgg)/d\epsilon$ is the spectrum of EBL photons at redshift $\zgg$.  
Assuming $n_{\EBL}(\zgg) = (1+\zgg)^{-2} n_{\EBL}(z=0)$ 
it was argued in \rref{Neronov:2009gh} that the mean free path can be reliably approximated as 
\begin{align}\label{eq:Dg_def}
	\Dg
	\simeq (80 \Mpc) \frac{\kappa}{(1+\zgg)^2} \left( \frac{\Eg}{10 \TeV} \right)^{-1} \per
\end{align}
\rref{Neronov:2009gh} estimates a range of values $0.3 < \kappa < 3$ for the dimensionless coefficient, 
and we will take $\kappa = 1$ hereafter.  
The comoving mean free path is given by $\dg = (1+\zgg) \Dg$.  
As long as $\dg \ll \ds$ we can approximate $\zgg \approx \zs$ and write the comoving mean free path as 
\begin{align}\label{eq:dg_def}
	\dg
	\simeq (80 \Mpc) \frac{1}{(1+\zs)} \left( \frac{\Eg}{10 \TeV} \right)^{-1} \per
\end{align}

%==========
After pair production the charged leptons acquire an energy 
$E_{e} \approx \Eg / 2$, 
and, to accuracy $m_e/E_e \sim 10^{-6}$ (the inverse boost factor), they travel in the same direction as the initial TeV gamma ray.  
The leptons produce secondary $\gamma$-rays through inverse Compton (IC) scattering on cosmic microwave background (CMB) photons.  
If $E_{\CMB}$ is the energy of a CMB photon at redshift $\zgg$ then energy conservation gives the energy of the corresponding IC photon to be 
\begin{align}
	E_{\gamma}^{\prime} 
	= \frac{4}{3} E_{\CMB} \frac{E_e^{2}}{m_e^2} 
\end{align}
where $m_{e} \simeq 0.511 \MeV / c^2$ is the electron mass.  
When this photon reaches Earth ($z=0$) it will have been redshifted to an energy of $E_{\gamma} = (1+\zgg)^{-1} E_{\gamma}^{\prime}$.  
If the spectrum of the TeV blazar is known, one can use the spectra of the EBL and CMB to calculate the spectrum of GeV gamma rays arriving at Earth.  
For our purposes only the average relationships are required.  
At redshift $\zgg$ the average energy of a CMB photon is $\langle E_{\CMB}(\zgg) \rangle = \epsilon_{\CMB} \simeq (6 \times 10^{-4} \eV) (1+z_{\gamma \gamma})$, and the average energy of a GeV photon arriving at Earth is then
\begin{align}\label{eq:Eg_to_Ee}
	E_{\gamma} 
	= \frac{4}{3} (1+z_{\gamma \gamma})^{-1} \epsilon_{\CMB} \frac{E_e^{2}}{m_e^2} 
	\simeq (77 \GeV) \left( \frac{\Eg}{10 \TeV} \right)^2 \per
\end{align}
Evidently an initial spectrum of gamma rays from $\Eg \sim 1-10 \TeV$ are transferred into cascade photons with $E_{\gamma} \sim 1-100 \GeV$ energies.  
The scattering of leptons on CMB photons is a stochastic process that occurs with a typical mean free path $l_{\rm mfp}$.  
On average the leptons lose their energy via IC within the electron cooling distance 
\begin{align}\label{eq:De_def}
	D_{e} 
	= \frac{3 m_{e}^2 c^4}{4 \sigma_{T} U_{\CMB} E_{e}}
	\simeq (31 \kpc) \left( \frac{E_{e}}{5 \TeV} \right)^{-1} \left( \frac{1+\zgg}{1.24} \right)^{-4} \com
\end{align}
where $\sigma_{T} \simeq 6.65 \times 10^{-25} \cm^2$ is the Thomson scattering cross section and $U_{\CMB} \simeq (1+\zgg)^4 (0.26 \eV / {\rm cm}^3)$ is the CMB energy density at redshift $\zgg$.  

%=========
The halo emerges because TeV gamma rays directed off of the line of sight can induce a cascade that is deflected back toward the line of sight by the magnetic field.  
At the point of pair production (redshift $\zgg$) let ${\bf B}$ and ${\bf v}$ be the magnetic field and the velocity of the lepton, respectively.  
Assuming that the coherence length of the magnetic field $\lambda \approx |{\bm B}|/|\nabla {\bm B}|$ is much larger than $D_{e}$, the electron or positron will probe an effectively homogeneous magnetic field.  
In this background the lepton follows a helical trajectory with gyroradius (Larmor radius)
\begin{align}\label{eq:RL_def}
	R_{L} = R \frac{|{\bf v}_{\perp}|}{c} 
	\qquad \text{where} \qquad
	R \equiv \frac{E_{e}}{e |{\bf B}|} \com
\end{align}
and $-e$ is the charge of the electron.
The component of ${\bf v}$ perpendicular to ${\bf B}$ is ${\bf v}_{\perp} = {\bf v} - ({\bf v} \cdot \hat{\bf B}) \hat{\bf B}$ where unit vectors are denoted by a hat.   
Using this expression we can write 
\begin{align}\label{eq:RL_to_R}
	R_{L} = R \, \sqrt{ 1 - ( \hat{\bf v} \cdot \hat{\bf B})^2 } \per
\end{align} 
If the magnetic field is frozen into the plasma, then its energy density redshifts like radiation, and we have $|{\bf B}| = B_{0} (1+\zgg)^2$ with $B_{0}$ the field strength today.  
Then we can we estimate 
\begin{align}\label{eq:R_numerical}
	R 
	\simeq (3.5 \Mpc) \left( \frac{E_{e}}{5 \TeV} \right) \left( \frac{B_{0}}{10^{-15} \Gauss} \right)^{-1} \left( \frac{1+\zs}{1.24} \right)^{-2} 
\end{align} 
where we have also used $\zgg \approx \zs$.  

%=========
By approximating $\zs \approx \zgg$ throughout our analysis we assume that 
$\dg / \ds \ll 1$. Using \erefs{eq:dE_def}{eq:dg_def} this ratio is
\begin{align}\label{eq:dg_over_dE}
	\frac{\dg}{\ds} \simeq 0.18 \left( \frac{E_{\gamma}}{10 \GeV} \right)^{-1/2} \left( \frac{1 + \zs}{1.24} \right)^{-1} \left( \frac{\ds}{1\Gpc} \right)^{-1} 
\end{align}
where $\ds$ and $\zs$ are related by \eref{eq:dE_def}.  
We will primarily be interested in $E_{\gamma} \gtrsim 10 \GeV$ and $d_{s} \gtrsim 0.7 \Gpc$ where the approximation is well-justified.
For lower energy gamma rays or closer sources, the approximation begins to break down as the Earth is located inside of the developing cascade.  

%=========
It is also useful to compare the electron cooling distance $D_{e}$ and gyroradius $R$.  
Using Eqs.~(\ref{eq:De_def}) and (\ref{eq:R_numerical}) and approximating $z_{\gamma\gamma} \approx \zs$ this ratio is 
\begin{align}\label{eq:De_over_R}
	\frac{D_{e}}{R} \simeq 0.067 \left( \frac{B_0}{10^{-15} \Gauss} \right) \left( \frac{E_{\gamma}}{10 \GeV} \right)^{-1} \left( \frac{1 + \zs}{1.24} \right)^{-2} \per
\end{align}
Since $2 \pi R_{L}$ is the circumference of the lepton's orbit, this ratio indicates whether the lepton travels around the orbit many times $D_{e} / R \gg 1$ or whether it makes only a small arc $D_{e} / R \ll 1$.  
In the former case, called the ``pair halo regime,'' the cascade gamma rays are spread out over large angles; in the latter case, called the ``magnetically broadened cascade regime,'' the cascade photons are slightly spread out around the source.

%==================================
% Morphology of the Cascade Halo
%==================================
\section{Morphology of the Cascade Halo}\label{sec:Arrival}

%==========
We are interested in the size and shape of the cascade halo as it appears from Earth.  
In this section we first establish an analytic formalism for calculating halo maps, {\it i.e.} the orientation 
$\hat{\bf n}(E_{\gamma})$ of GeV gamma rays reaching Earth.  
Next we introduce parameters that quantify the halo size and shape, which can be extracted from the halo map.   

%------------------------------------------------------------
% Formalism and Assumptions
%------------------------------------------------------------
\subsection{Formalism and Assumptions}\label{sec:assumptions}

%==========
We move between spherical, cylindrical, and Cartesian coordinates.  
The origin is located at the Earth and $\hat{\bf z}$ is oriented along the line of sight to the blazar.  
The polar and azimuthal angles are denoted by $(\theta,\phi)$, and the spherical and cylindrical radial coordinates are $r$ and $\rho$.  
The sets of unit vectors $\{ \hat{\bf r} , \hat{\bm \theta}, \hat{\bm \phi} \}$, $\{ \hat{\bm \rho} , \hat{\bm \phi}, \hat{\bf z} \}$, and $\{ \hat{\bf x}, \hat{\bf y}, \hat{\bf z} \}$ form right-handed orthonormal coordinate systems.  
It will be useful to note the relationships 
\begin{align}
\begin{array}{l}
	\hat{\bm \rho} = \cos \phi \, \hat{\bf x} + \sin \phi \, \hat{\bf y} \\
	\hat{\bm \phi} = - \sin \phi \, \hat{\bf x} + \cos \phi \, \hat{\bf y}  
\end{array} 
\quad , \quad
\begin{array}{l}
	\hat{\bf x} = \cos \phi \, \hat{\bm \rho} - \sin \phi \, \hat{\bm \phi} \\
	\hat{\bf y} = \sin \phi \, \hat{\bm \rho} + \cos \phi \, \hat{\bm \phi} 
\end{array} \per
\end{align}
For plotting gamma ray arrival directions, it is convenient to introduce the lateral and transverse angular extent as
\begin{align}\label{eq:lat_trans_def}
	\tlat = \theta \, \cos \phi
	\qquad {\rm and} \qquad 
	\ttrans = \theta \, \sin \phi \per
\end{align}
The distinction between the lateral and transverse directions is arbitrary.  
Although other mappings are possible, this one has the convenient feature that the Euclidean norm is equal to the polar angle $\theta = \sqrt{\tlat^2 + \ttrans^2}$.  
A gamma ray arriving at Earth is specified by an energy $E_{\gamma}$ and an orientation, which can be expressed as the pair $(\theta,\phi)$ or equivalently $(\tlat,\ttrans)$ or $\hat{\bf n} = \hat{\bf r}$.  

%=========
We seek to study the development of the cascade {\it semi-analytically} by deriving a set of trigonometric equations that can be solved (analytically if possible, numerically if not) to find the orientation of GeV gamma rays reaching Earth.  
In this sense our approach differs from a purely numerical simulation in which TeV gamma rays are ejected from the blazar in random directions, the paths of the lepton and GeV gamma rays are calculated, and trajectories that do not intersect the Earth are discarded. As we will see, inspection of \fref{fig:triangle} leads to a set of three equations: 
the first arises from the trigonometry of the triangle, the second arises from the geometry of the lepton's orbit, and 
the third ensures that the GeV gamma ray intersects with the line of sight.
We solve these three equations for a given magnetic field configuration and gamma ray energy $E_{\gamma}$ to obtain the orientation of the GeV gamma ray at Earth ($\theta,\phi$) as well as the bending angle $\delta$.  
This semi-analytic approach is applicable thanks to the following two well-justified approximations.  

%=========
First, we assume that the lepton samples a homogeneous magnetic field, and thus its path is a simple helix.  
If $\lambda$ is the coherence length of the magnetic field, this condition is expressed as $D_{e} \ll \lambda$.  
We saw in \eref{eq:De_def} that typically $D_{e} \sim 100 \kpc$, and since we will be interested in $\lambda > 1 \Mpc$ this assumption is well-justified.  
If we were interested in smaller coherence scales, where the lepton motion is diffusive, then our approach would not be applicable.  

%=========
Second, we assume that the displacement of the lepton can be neglected.  
This ensures that, to a good approximation, the two gamma rays lie in the same plane and that they form the legs of a triangle as in \fref{fig:triangle}.  
To verify that the longitudinal displacement along the field line is negligible, we need $|{\bf v}_{\parallel}| D_{e} / |{\bf v}| \ll \ds , \dg$; to ensure that the transverse displacement around the orbit is negligible we need ${\rm Min}[D_{e},R_{L}] \ll \ds , \dg$.  
In Eqs.~(\ref{eq:dE_def}),~(\ref{eq:dg_def}),~and~(\ref{eq:R_numerical}) we saw 
that typically $\ds \sim 1 \Gpc$, $\dg \sim 100 \Mpc$, and $D_{e}, R \sim 100 \kpc$.  
Then over all of the relevant parameters space we have $D_{e} \ll \dg$, and the assumption is very well-justified.  

%=========
Although our approach only requires the above two approximations, we also use the following two assumptions as a matter of convenience.
First, it is important to remark that many aspects of the cascade are stochastic in nature: the spectrum of TeV gamma rays emitted by the blazar, the distance traveled by the TeV gamma rays before pair production, the spectrum of EBL photons, the distance traveled by the leptons before IC, the number of IC photons emitted before electron cooling becomes appreciable, and the spectrum of CMB photons that are up-scattered.  
For simplicity we neglect the stochastic spread in each of these various parameters, and we fix them equal to their average values given in \sref{sec:Cascade}.  
As a result, we obtain a deterministic relationship between the energy of gamma rays reaching Earth and their orientations on the sky; we call this function the halo map $\hat{\bf n}(E_{\gamma})$.  
These halo maps are useful tools for studying the connection between halo morphology and the underlying magnetic field since they can be calculated quickly, without sophisticated numerical simulation, and they capture the characteristic features of the halo size and shape.  
It is expected that properly taking account of the stochasticity will lead to a significantly ``smeared'' version of the halo maps shown here, since the variance in the random parameters is typically $O(1)$.  
We discuss the stochastic smearing further in the conclusions, \sref{sec:Conclusion}, as a direction for future work.

%=========
As a second simplification, we neglect the blazar's jet structure and assume that emission from the blazar is isotropic.  
If we were to properly treat the angular distribution of radiation from the blazar, then only a portion of our halo maps would be visible.  
We return to this point in the conclusions, \sref{sec:Conclusion}, where we also show a few halo maps that have been calculated with the jet restriction in place.  

%------------------------------------------------------------
% Constraint Equations
%------------------------------------------------------------
\subsection{Constraint Equations}\label{sec:constraint}

%=========
We obtain the first in the set of three equations by applying the law of sines to the triangle in \fref{fig:triangle}:  
\begin{align}\label{eq:law_of_sines}
	\sin \theta = \frac{\dg}{\ds} \ \sin \delta \per
\end{align}
Here $\theta$ is the polar angle that a GeV gamma ray arriving at Earth makes with the line of sight to the blazar, and the bending angle $\delta$ is the angle between the orientation of the initial TeV gamma ray and the final GeV gamma ray.  
The second equation is a relationship for the bending angle $\delta$:  
\begin{align}\label{eq:delta}
	1 - \cos \delta 
	= \Bigl( 1 - (\hat{\bf v}_{i} \cdot \hat{\bf B})^2 \Bigr) \Bigl( 1 - \cos (D_{e} /  R) \Bigr) \per
\end{align}
To derive \eref{eq:delta}, we write the magnetic field at the point of pair production as ${\bf B} = B \, \hat{\bf n}_{\parallel}$, and we write the initial lepton velocity as ${\bf v}_{i} = v_{\parallel} \, \hat{\bf n}_{\parallel} + v_{\perp} \, \hat{\bf n}_{\perp}$ where $\hat{\bf n}_{\parallel} \cdot \hat{\bf n}_{\perp} =0$.  
Then $\vpara = {\bf v}_{i} \cdot \hat{\bf B}$ and $\vpara^2 + \vperp^2 = v^2 \approx c^2$.  
The lepton velocity at time $t$ after pair production is 
\begin{align}
\label{v(t)}
	{\bf v}(t) = v_{\parallel} \, \hat{\bf n}_{\parallel} + v_{\perp} \cos (\omega t) \, \hat{\bf n}_{\perp} \mp v_{\perp} \sin (\omega t) \, \hat{\bf n}_{\parallel} \times \hat{\bf n}_{\perp}
\end{align}
where $\omega = v_{\perp} / R_L = c / R$ is the angular frequency of the orbital motion.  
The sign ambiguity in the last term is related to the charge of the lepton; the $(-)$ is for positrons and $(+)$ is for electrons.    
The lepton travels an electron cooling distance\footnote{Throughout our analysis we assume that the lepton always travels an electron cooling distance, and that IC scattering only occurs once, at this point.  In reality the mean free path of the lepton is shorter than $D_{e}$ and multiple IC photons are emitted.  These multiple emissions are an example of the stochastic effects, discussed previously, that we neglect for simplicity.  We perform a preliminary investigation of the stochasticity in \sref{sec:Conclusion}.''}, $D_{e}$, in time $\tau = D_{e} / c$.  
Since the gamma rays are approximately tangential to the lepton trajectory, the orientation of the initial TeV 
gamma ray is $\hat{\bf v}(0)$, the orientation of the final GeV gamma ray is $\hat{\bf v}(D_{e}/c)$, and the bending angle satisfies $\cos \delta = \hat{\bf v}(0) \cdot \hat{\bf v}(D_{e}/c)$, which gives \eref{eq:delta}.

%==========
Before discussing the third equation, it is useful to consider the limits of small and large lepton deflection.  
Recall that $D_{e}/R = \omega \tau$ is the angular deflection of the lepton as it travels a distance $D_{e}$ around the gyro-circle, and we saw in \eref{eq:De_over_R} that $D_{e} / R \propto B_{0} / E_{\gamma}$.  
For sufficiently weak magnetic field or high gamma ray energy we have $D_{e} / R \ll 1$ and \erefs{eq:law_of_sines}{eq:delta} reduce to 
\begin{align}\label{eq:small_deflect}
	\theta \approx \sqrt{ 1 - ( \hat{\bf v}_{i} \cdot \hat{\bf B})^2} \, \Theta_{\rm ext}
	\qquad \text{and} \qquad
	\delta \approx \sqrt{ 1 - (\hat{\bf v}_{i} \cdot \hat{\bf B})^2 } \, \frac{D_e}{R} 
\end{align}
where
\begin{align}\label{eq:Theta_ext_def}
	\Theta_{\rm ext} \equiv 
	\frac{\dg D_{e}}{\ds R} \per
\end{align}
In $\Theta_{\rm ext}$ we find the familiar expression for the angular extent of the halo \cite{Neronov:2006hc}.  
Using the numerical estimates from \sref{sec:Cascade} we have 
\begin{align}\label{eq:Theta_ext}
	\Theta_{\rm ext}
	& \simeq (0.68^{\circ}) \left( \frac{B_{0}}{10^{-15} \Gauss} \right) \left( \frac{E_{\gamma}}{10 \GeV} \right)^{-3/2} \left( \frac{\ds}{1\Gpc} \right)^{-1} \left( \frac{1+\zs}{1.24} \right)^{-3} \per 
\end{align}
Since $\Theta_{\rm ext} \propto B_{0}$ the angular extent of the halo grows larger as the field strength is increased, which is characteristic of the MBC regime.  

%==========
In the opposite regime where the lepton deflection is large we have $D_{e} / R \gg 1$, and the lepton makes multiple orbits before IC occurs.  
If $(\hat{\bf v}_{i} \cdot \Bhat)^2 < 1/2$ then \eref{eq:delta} has a discrete set of solutions at energies $E_{\gamma}^{(n)}$
where the halo reaches a maximum angular extent 
\begin{align}\label{eq:large_deflect}
	\theta \approx \Theta_{\rm max}
	\qquad \text{and} \qquad
	\delta \approx \frac{\pi}{2}
\end{align}
with 
\begin{align}\label{eq:Theta_max}
	\Theta_{\rm max} \equiv {\rm arcsin} \left( \frac{\dg}{\ds} \right) 
	\simeq (10^{\circ}) \left( \frac{E_{\gamma}^{(n)}}{10 \GeV} \right)^{-1/2} \left( \frac{1 + \zs}{1.24} \right)^{-1} \left( \frac{\ds}{1\Gpc} \right)^{-1} \per
\end{align}
In this limit, known as the PH regime, the angular extent of the halo is not proportional to the magnetic field strength, but instead it is restricted only by the geometry of the TeV gamma ray propagation.  

%=========
We finally turn to the third constraint equation, which needs to enforce that the TeV and GeV gamma rays approximately lie in a plane of constant $\phi$, as shown in \fref{fig:triangle}.  
To ensure that the charged lepton is not deflected out of this plane, the Lorentz force ${\bf F} = (e/c) \, {\bf v} \times {\bf B}$ must be normal to $\hat{\bm \phi}$.  
Of course, as the lepton follows its helical path, the direction of its velocity changes and so too does the direction of the Lorentz force.  
Then we must average the Lorentz force over the trajectory of the lepton.  

%==========
Let ${\bf x}(t)$ be the helical trajectory of the lepton, ${\bf v}(t) = d{\bf x}/dt$ be its velocity, $\tgg$ be the time of pair production, and $\tau = D_{e} / c$ be the time elapsed before IC up-scattering.  
Neglecting the cosmological expansion, which is not relevant on such short time scales, the impulse imparted on the charged lepton at redshift $\zgg$ is given by 
\begin{align}
	{\bf J} = \pm \frac{e}{c} \int_{\tgg}^{\tgg+\tau} \ud t \, {\bf v}(t) \times {\bf B}({\bf x}(t),t) 
\end{align}
where the $\pm$ is related to the charge on the lepton.  
If the magnetic field is static and homogeneous over the path of the lepton, we can pull it out of the integral, and the integrand contains only ${\bf v}(t)$.  
Defining the time averaged electron velocity as 
\begin{align}\label{eq:vavg_to_Dx}
	{\bf v}_{\rm avg} 
	\equiv \frac{1}{\tau} \int_{\tgg}^{\tgg+\tau} \ud t \, {\bf v}(t) 
	\com
\end{align}
we can write the impulse as 
\begin{align}
	{\bf J} = \pm \frac{e \tau}{c} \, {\bf v}_{\rm avg} \times {\bf B} \per
\end{align}
Since the magnetic field does no work, the magnitude $|{\bf v}(t)| \approx c$ is fixed. 
Then using the geometry shown in \fref{fig:triangle}, we see that ${\bf v}_{\rm avg}$ must bisect the angle $\delta$, and it can be written as 
\begin{align}\label{eq:v_avg}
	\hat{\bf v}_{\rm avg} 
	= \sin \left( \frac{\delta}{2} - \theta \right) \, \hat{\bm \rho} - \cos \left( \frac{\delta}{2} - \theta \right) \, \hat{\bf z} \per
\end{align}
Writing also 
\begin{align}\label{eq:Bhat}
	\Bhat = b_{\rho} \hat{\bm \rho} + b_{\phi} \hat{\bm \phi} + b_{z} \hat{\bf z} 
\end{align}
we have 
\begin{align}\label{eq:vBphi}
	\hat{\bf v}_{\rm avg} \times \hat{\bf B} \cdot \hat{\bm \phi} = - b_{\rho} \cos \left( \frac{\delta}{2} - \theta \right) - b_{z} \sin \left( \frac{\delta}{2} - \theta \right) = 0 \per
\end{align}
Typically $b_{\rho}$ or $b_{z}$ will depend on the azimuthal angle $\phi$, and then this equation fixes the plane (normal to $\hat{\bm \phi}$) in which lie the two gamma rays and the line of sight.  

%==========
Now we summarize the three constraint equations.
To ensure that the motion remains in the plane, we impose \eref{eq:vBphi}:
\begin{align}\label{equation1}
	b_{\rho} \cos \left( \frac{\delta}{2} - \theta \right) + b_{z} \sin \left( \frac{\delta}{2} - \theta \right) = 0 
\end{align}
where \eref{eq:Bhat} gives the decomposition of the magnetic field into cylindrical coordinates.  
From the geometry of the gamma ray trajectories, we have the law of sines in \eref{eq:law_of_sines}, 
\begin{align}\label{equation2}
	\sin \theta = \frac{\dg}{\ds} \ \sin \delta \com
\end{align}
and finally the bending angle is given by \eref{eq:delta}, 
\begin{align}\label{equation3}
	1 - \cos \delta 
	= \Bigl( 1 - \bigl( b_{\rho} \sin(\delta - \theta) - b_{z} \cos(\delta - \theta) \bigr)^2 \Bigr) \Bigl( 1 - \cos (D_{e} /  R) \Bigr) 
	\com
\end{align}
where we have written the initial lepton velocity as 
\begin{align}\label{eq:vi_def}
	\hat{\bf v}_{i} = \sin (\delta - \theta) \hat{\bm \rho} - \cos (\delta - \theta) \hat{\bf z} \per
\end{align}
The constraints in \Eqns can also be derived from the single vector equation
\begin{equation}\label{constraint}
	d_{\gamma 0} {\hat {\bm v}}_i + \Delta {\bm x} + L {\hat {\bm v}}_f + d_s {\hat {\bm z}} = 0 \com
\end{equation}
which ensures that the cascade photon reaches Earth.
Here $L$ is the distance from the IC scattering to the observation point and $\Delta {\bm x}$ is the displacement 
of the lepton between pair production and IC scattering. To find $\Delta {\bm x}$, we can integrate \eref{v(t)}.  
However, in the limit that $D_e \ll d_\gamma, d_s$, the $\Delta {\bm x}$ term can be dropped from the equation.  
Trigonometric manipulation of \eref{constraint} then once again leads to the above constraint equations.

%------------------------------------------------------------
% Shape Parameter
%------------------------------------------------------------
\subsection{Shape Parameter}\label{sec:shape}

%==========
In the next section we consider various static magnetic field configurations $\hat{\bf B}({\bf x})$, and we 
solve \Eqns for $\theta, \phi,$ and $\delta$ to determine the halo map $\hat{\bf n}(E_{\gamma})$.  
Having solved \Eqns we construct the halo map as the radial unit vector
\begin{align}
	\hat{\bf n}(E_{\gamma}) & = \sin \theta \, \cos \phi \, \hat{\bf x} + \sin \theta \, \sin \phi \, \hat{\bf y} + \cos \theta \, \hat{\bf z} 
\end{align}
that points from the Earth toward the arriving gamma ray of energy $E_{\gamma}$.  
Note that $\nhat(E_{\gamma})$ may be multi-valued meaning that gamma rays of a particular energy may appear from multiple directions in the sky.
All of the information about the halo size and shape is contained in the function $\hat{\bf n}(E_{\gamma})$.  
For instance, the size of the halo is given by 
$\Theta(E_{\gamma}) = \arccos [ \hat{\bf n}(E_{\gamma}) \cdot \hat{\bf z} ]$.  

%==========
There are many ways to quantify the halo shape and orientation.  
Since we are interested in probing magnetic helicity, we are motivated to consider a parity-odd $Q$-statistic \cite{Tashiro:2013bxa, Tashiro:2014gfa}, which is sensitive to the sign of the magnetic helicity.  
The statistic is defined as the triple product of vectors sampled from the halo map $\hat{\bf n}(E_{\gamma})$ at three different energies.  
For illustrative purposes, we consider a few energy combinations:  
\begin{subequations}\label{eq:Q_def} 
\begin{align}
	Q_{10,30,50} & = \hat{\bf n}_{10} \times \hat{\bf n}_{30} \cdot \hat{\bf n}_{50} \\
	Q_{25,30,35} & = \hat{\bf n}_{25} \times \hat{\bf n}_{30} \cdot \hat{\bf n}_{35} \\
	Q_{8,10,12} & = \hat{\bf n}_{8} \times \hat{\bf n}_{10} \cdot \hat{\bf n}_{12} \\
	Q_{38,40,42} & = \hat{\bf n}_{38} \times \hat{\bf n}_{40} \cdot \hat{\bf n}_{42} 
\end{align}
\end{subequations}
where $\hat{\bf n}_{\#}$ is shorthand notation for $\hat{\bf n}(E_{\gamma} = \# \GeV)$.  
We will see that different energy combinations are sensitive to different ranges of parameters, namely magnetic field strength and coherence length. Since $\nhat(E_{\gamma})$ may be multi-valued in general, one can calculate 
$Q$ by first averaging over the multiple arrival directions for a given energy.

%==========
It is also useful to write the triple product $Q_{abc} = \nhat_{a} \times \nhat_{b} \cdot \nhat_{c}$ as 
\begin{align}\label{eq:Q_alt}
	Q_{abc} = \sin \vartheta_{ab} \sin \varphi_{abc}
\end{align}
where $0 \leq \vartheta_{ab} \leq \pi$ is the angle between $\hat{\bf n}_{a}$ and $\hat{\bf n}_{b}$, and $-\pi/2 \leq \varphi_{abc} \leq \pi/2$ is the angle between $\hat{\bf n}_{c}$ and its projection onto the plane spanned by $\hat{\bf n}_{a}$ and $\hat{\bf n}_{b}$.  
If $\nhat_{a}$ and $\nhat_{b}$ are collinear then $\vartheta_{ab} = 0$ and $Q_{abc}$ vanishes; if $\nhat_{c}$ lies in the same plane as $\nhat_{a}$ and $\nhat_{b}$ then $\varphi_{abc}=0$ and $Q_{abc}$ also vanishes.  
The sign of $Q_{abc}$ is controlled by the sign of $\varphi_{abc}$, which depends on whether $\nhat_{c}$ is ``in front of'' or ``behind'' the plane normal to $\nhat_{a} \times \nhat_{b}$.  
Since it is odd under reflections one sees that $Q_{abc}$ is a measure of parity violation in the halo map.

%==================================
% Halo Morphology for Specific Magnetic Field Configurations
%==================================
\section{Halo Morphology for Specific Magnetic Field Configurations}\label{sec:Examples}

\begin{figure}[h]
\hspace{0pt}
\vspace{-0in}
\begin{center}
\includegraphics[width=0.45\textwidth]{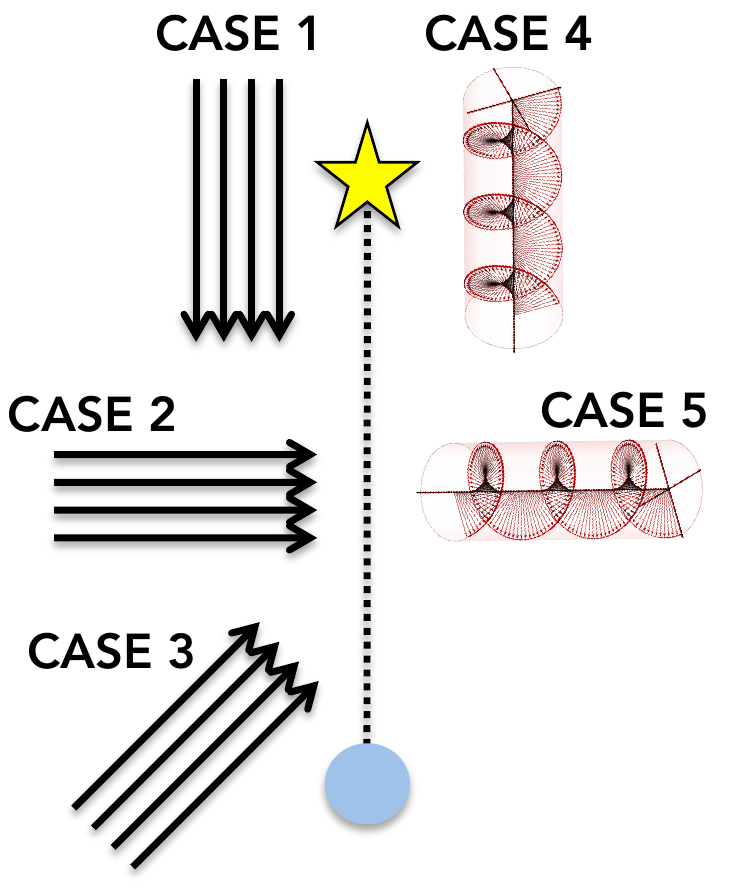} 
\caption{
\label{fig:5cases}
The five magnetic field configurations that we consider.  Cases 1-3 are homogeneous field configurations that have different orientations with respect to the line of sight with the blazar.  Cases 4-5 are helical field configurations with their wavevectors oriented either along or normal to the line of sight.
}
\end{center}
\end{figure}

%==========
In this section we study the size and shape of the GeV halo for various specific magnetic field configurations.  
The configurations we consider are simplified and do not realistically model the intergalactic magnetic field.  
However, these examples serve to illustrate the parametric relationships between the field configuration and the halo morphology.  
Previous studies have focused on the size information alone, and we will see that the shape information provides insight into the magnetic field's orientation and helicity.  
We consider five different non-helical and helical magnetic field configurations, as shown in \fref{fig:5cases}.  

%==========

%------------------------------------------------------------
% Uniform Magnetic Field Parallel to Line of Sight
%------------------------------------------------------------
\subsection{Case 1:  Uniform Magnetic Field Parallel to Line of Sight}\label{sec:case1}

%==========
The simplest configuration is a homogeneous magnetic field oriented along the line of sight with the blazar,
\begin{align}\label{case1:B_config}
	\hat{\bf B} 
	= - \hat{\bf z} 
	\per
\end{align}
The three constraint equations, \Eqns, reduce to 
\begin{subequations}\label{case1:equations}
\begin{align}
	& \sin(\delta/2 - \theta) =0 \\
	& \sin \theta = \frac{\dg}{\ds} \sin \delta \\
	& 1 - \cos \delta 
	= \sin^2 (\delta-\theta) \Bigl( 1 - \cos (D_{e} /  R) \Bigr)
	\per
\end{align}
\end{subequations}
There is a trivial solution with $\delta = \theta = 0$ corresponding to gamma rays oriented along the line of sight, $\hat{\bf v} = - \hat{\bf z}$, that are not deflected by the magnetic field.  
There is also a nontrivial solution, 
\begin{equation}
	\delta = 2\theta = 2 \cos^{-1}\left (\frac{\ds}{2d_\gamma} \right ) 
	\quad , \quad
	\frac{D_e}{R} = (2n+1) \pi
\end{equation}
where $n$ is an integer.  
Since the azimuthal angle $\phi$ does not appear in these equations, the solution will be rotationally symmetric about the line of sight to the blazar.  
Also, since $D_e / R$ and $\dg / \ds$ depend on energy, there will only be a discrete set of energies for which a solution exists. 
The solution can be understood in physical terms:  the velocity component along the magnetic field 
remains constant, and the velocity component perpendicular to the magnetic field gets reflected, and so the 
triangle in \fref{fig:triangle} is an isosceles triangle.
Apart from these solutions, the magnetic field deflects other gamma rays away from the line of sight, and they do not reach Earth.

%------------------------------------------------------------
% Uniform Magnetic Field Normal to Line of Sight
%------------------------------------------------------------
\subsection{Case 2:  Uniform Magnetic Field Normal to Line of Sight}\label{sec:case2}

%==========
Next we consider a homogeneous magnetic field that is oriented normal to the line of sight with the blazar
(see Fig.~\ref{fig:5cases}).
Without loss of generality we can align the Cartesian coordinate system with the magnetic field such that 
\begin{align}\label{case2:B_config}
	\hat{\bf B} 
	= \hat{\bf y} 
	= \sin \phi \, \hat{\bm \rho} + \cos \phi \, \hat{\bm \phi}
	\per
\end{align}
and \Eqns reduce to 
\begin{subequations}\label{case2:equations}
\begin{align}
	& \sin \phi \, \cos(\delta/2 - \theta) = 0 \\
	& \sin \theta = \frac{\dg}{\ds} \sin \delta \\
	& 1 - \cos \delta 
	= \Bigl( 1 - \sin^2 (\delta - \theta) \, \sin^2 \phi \Bigr) \Bigl( 1 - \cos (D_{e} /  R) \Bigr)
	\per
\end{align} 
\end{subequations}
For a given gamma ray energy $E_{\gamma}$ there is a solution 
\begin{align}\label{case2:solution}
	\phi = 0 , \pi 
	\quad , \qquad
	\sin \theta = \frac{\dg}{\ds} \sin \delta
	\quad , \quad {\rm and} \quad 
	\cos \delta = \cos \frac{D_{e}}{R}
\end{align} 
where $E_{\gamma}$ enters through $\dg / \ds$ and $D_{e} / R$, see \erefs{eq:dg_over_dE}{eq:De_over_R}.  
Recall that $D_{e}/R > 0$ is unbounded from above but $0 \leq \delta \leq \pi$ and $0 \leq \theta \leq \pi/2$.  
For this magnetic field configuration, the trajectories of all the gamma rays lie in the $y=0$ plane where $\phi = 0,\pi$.  
In the limit of small lepton deflection, $D_{e} / R \ll 1$, the solution further simplifies to 
\begin{align}\label{case2:Theta_ext}
	\theta \approx \Theta_{\rm ext} = \frac{\dg D_{e}}{\ds R} 
\end{align}
as in \erefs{eq:small_deflect}{eq:Theta_ext_def}.  

%=========
\begin{figure}[p]
\hspace{0pt}
\vspace{-0in}
\begin{center}
\includegraphics[width=0.45\textwidth]{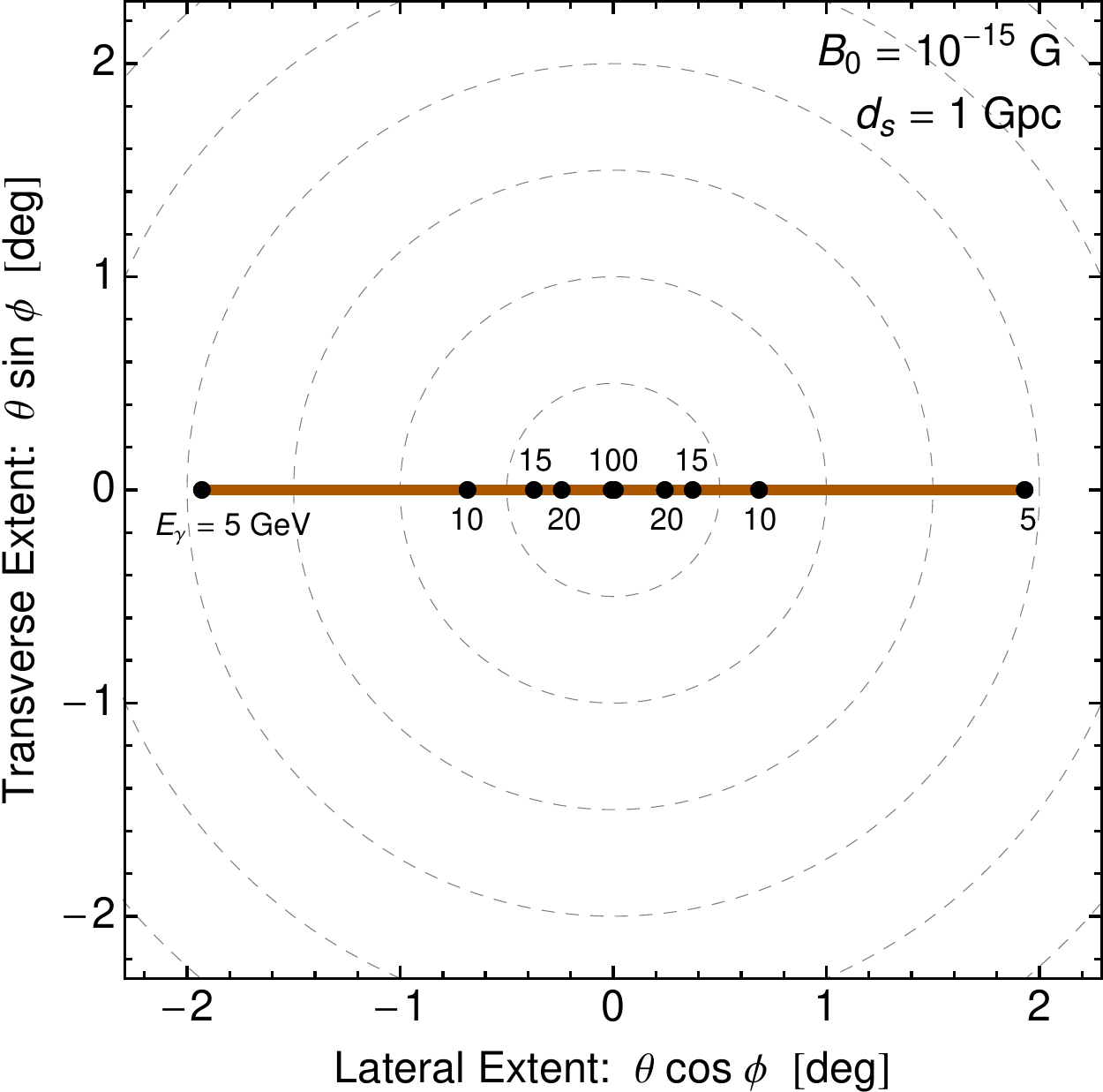} 
\caption{
\label{fig:case2_arrival}
The halo map for Case 2.  Gamma rays arrive at Earth collimated into a line with zero transverse extent.  The black dots indicate $E_{\gamma} = 100,20,15,10,$ and $5 \GeV$ with the higher energy gamma rays arriving near the origin.  The gray dashed lines are curves of constant polar angle $\theta$.  
}
\end{center}
\end{figure}

%=========
\begin{figure}[p]
\hspace{0pt}
\vspace{-0in}
\begin{center}
\includegraphics[width=0.69\textwidth]{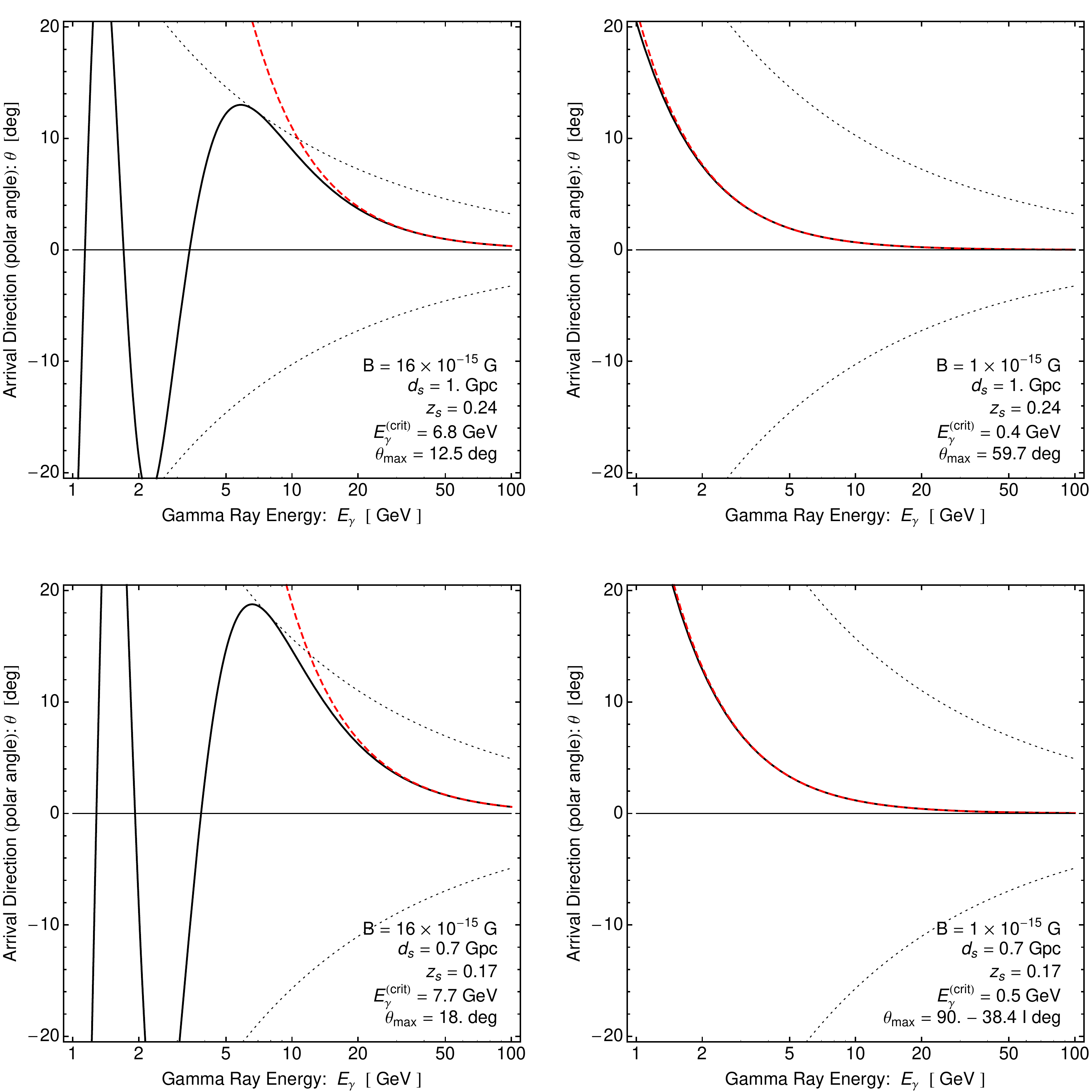} 
\caption{
\label{fig:case2_grid}
Angular extent of the halo as a function of gamma ray energy for Case 2.  
Each panel shows the exact solution (black line) from \eref{case2:solution}, the small bending approximation (red dashed) from \eref{case2:Theta_ext}, and the geometric limit (black dashed) from \eref{eq:Theta_max}.  
The top-right panel shows the same parameters as in \fref{fig:case2_arrival}.  
}
\end{center}
\end{figure}

%==========
The halo map corresponding to the solution in \eref{case2:solution} is shown in \fref{fig:case2_arrival}.  
We have mapped the polar and azimuthal angles $(\theta,\phi)$ into the lateral and transverse angular extent $(\tlat, \ttrans)$ using \eref{eq:lat_trans_def}.  
The $y$-component of velocity is conserved for $\Bhat = {\bf y}$, and only gamma rays lying in the plane $y=0$ normal to the magnetic field are deflected back toward the line of sight.  
The halo map $\nhat(E_{\gamma})$ is double-valued as there are precisely two gamma ray arrival directions for each energy.  
Along one branch the electron was deflected back toward Earth and generated the secondary IC gamma ray, and on the other branch it was the positron.

%==========
The energy dependence of the polar angle is shown in \fref{fig:case2_grid} for various parameter combinations.  
The highest energy gamma rays experience the smallest deflection, and they arrive closest to the line of sight ($\theta = 0$).  
In this regime, the small angle approximation is valid, and we have the scaling $\theta \approx \Theta_{\rm ext} \propto E_{\gamma}^{-3/2}$ from \eref{eq:Theta_ext_def}.  
Lower energy gamma rays are found farther from the line of sight, and there is an energy gradient.  
As the energy decreases further, the bending angle $\delta = D_{e} / R \propto 1 / E_{\gamma}$ continues to grow, as per \eref{eq:De_over_R}, and eventually $\delta = \pi/2$ where the electron experiences a deflection of $90^{\circ}$.  
This corresponds to an energy 
\begin{align}\label{eq:Eg_crit}
	E_{\gamma}^{\rm (crit)} \simeq (0.43 \GeV) \left( \frac{B_{0}}{10^{-15} \Gauss} \right) \left( \frac{1+\zs}{1.24} \right)^{-2} \com
\end{align}
which also serves to indicate where the small bending approximation breaks down.  
Since $\sin \delta = 1$ is maximized at this point, the halo achieves a maximum angular extent ({\it cf.} \eref{eq:Theta_max})
\begin{align}\label{eq:theta_max}
	\Theta_{\rm max} = {\rm arcsin} \left( \frac{\dg}{\ds} \right) 
	\simeq (59^{\circ}) \left( \frac{\ds}{1\Gpc} \right)^{-1} \left( \frac{B_{0}}{10^{-15} \Gauss} \right)^{-1/2} \per
\end{align}
Still lower energy gamma rays arrive closer to the line of sight, because the lepton is bent more than $90^{\circ}$.  

%==========
For this case the halo map is wide in lateral extent and narrow in transverse extent as seen in \fref{fig:case2_arrival}.
In principle one could measure the orientation of the magnetic field by measuring the orientation of the halo map.  
This is an example of how shape information can probe additional aspects of the IGMF beyond just its field strength.  
Of course, we have assumed that the magnetic field is uniform over the scale probed by the TeV gamma rays, {\it i.e.} if the magnetic field coherence length is $\lambda$, then we have implicitly assumed 
$\lambda \gg \dg \sim 100 \Mpc$.  
In a realistic setting it is more likely that the magnetic field forms domains smaller than $\dg$.  
If the magnetic field is statistically isotropic across domains, then different leptons probe random orientations of the field, and the halo will resemble a more familiar, rotationally symmetric halo map.  
Apart from this isotropization, the discussion of this section is largely unchanged, and specifically \eref{case2:solution} still gives the relationship between the gamma ray energy and polar angle.

%------------------------------------------------------------
% Uniform Magnetic Field with Arbitrary Orientation
%------------------------------------------------------------
\subsection{Case 3:  Uniform Magnetic Field with Arbitrary Orientation}\label{sec:case3}

%==========
Next we consider a homogeneous magnetic field that has a component along the line of sight to the blazar.  
This is a generalization of the previous two cases.  
Without loss of generality we can write the magnetic field configuration as 
\begin{align}\label{case3:B_config}
	\hat{\bf B} 
	& = \left( \cos \beta \, \hat{\bf y} - \sin \beta \, \hat{\bf z} \right) 
	= \cos \beta \sin \phi \, \hat{\bm \rho} + \cos \beta \cos \phi \, \hat{\bm \phi} - \sin \beta \, \hat{\bf z}
\end{align}
where the skew angle $\beta$ controls the component of $\Bhat$ along the line of sight.  
\Eqns reduce to 
\begin{subequations}\label{case3:equations}
\begin{align}
	& \sin \phi = \tan \beta \, \tan(\delta/2 - \theta) \\
	& \sin \theta = \frac{\dg}{\ds} \sin \delta \\
	& 1 - \cos \delta = \Bigl( 1 - \sin^2 \beta \frac{\cos^2(\delta/2)}{\cos^2(\delta/2-\theta)} \Bigr) \Bigl( 1 - \cos (D_{e} /  R) \Bigr) 
\end{align}
\end{subequations}
where we have used the equation for $\sin \phi$ to simplify the third equation.  

%=========
\begin{figure}[t]
\hspace{0pt}
\vspace{-0in}
\begin{center}
\includegraphics[width=0.45\textwidth]{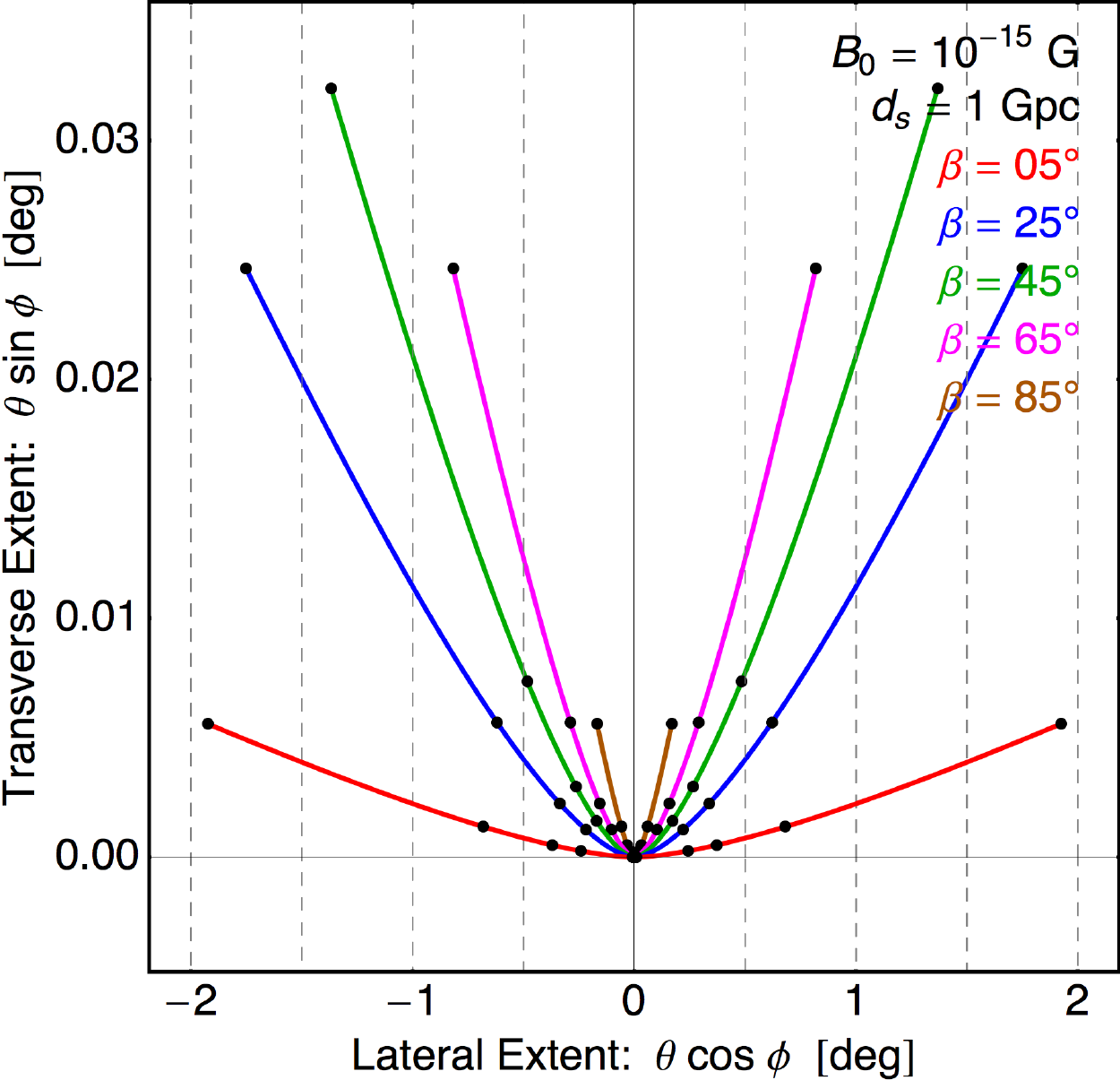} \hfill
\includegraphics[width=0.45\textwidth]{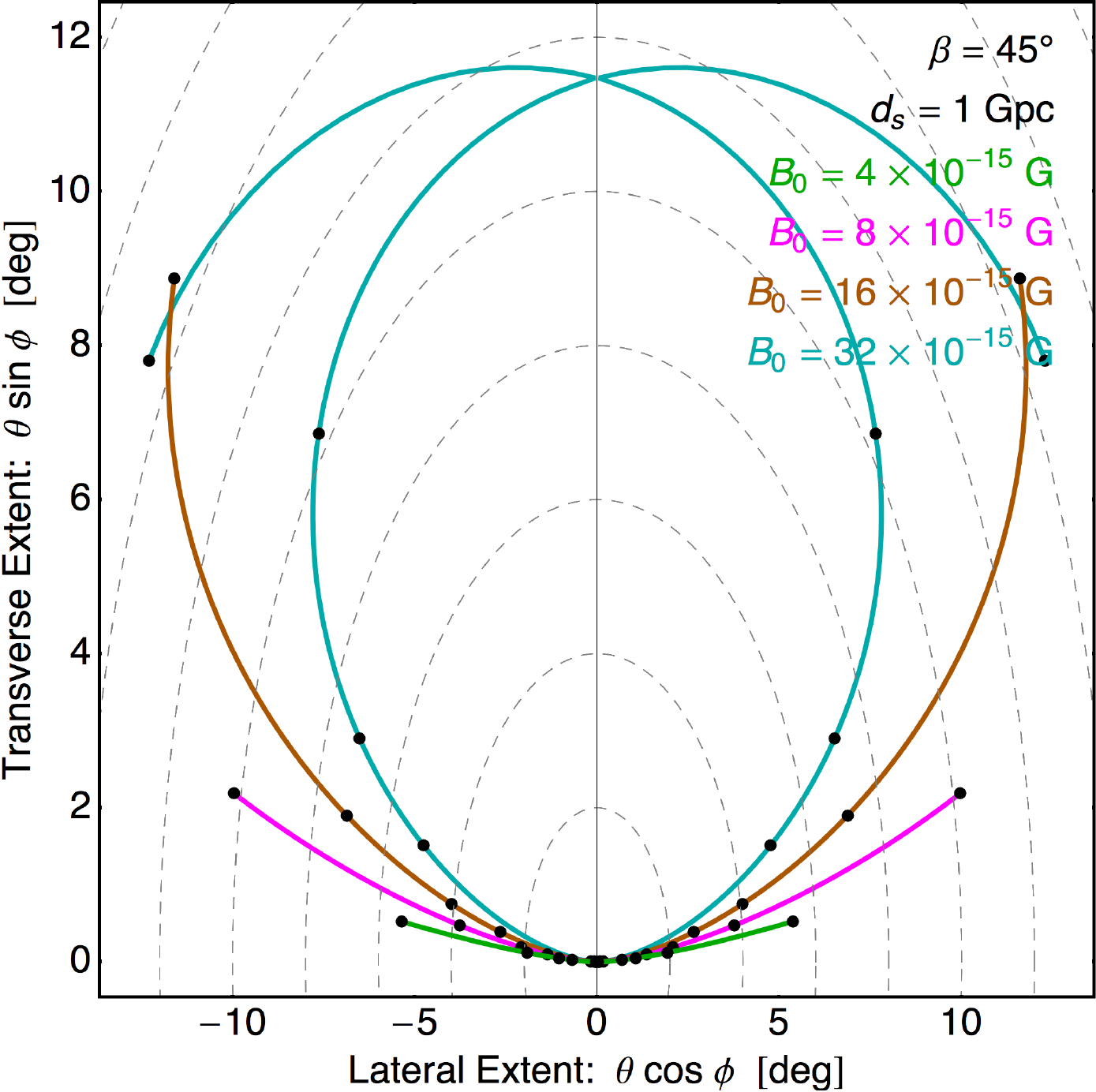}
\caption{
\label{fig:case3_arrival}
The halo map for Case 3.  
If we had taken $\beta < 0$ then the halo maps would be reflected in the vertical direction.  
The black dots and gray dashes have the same meaning as in \fref{fig:case2_grid}.  
}
\end{center}
\end{figure}

%==========
Numerically solving \eref{case3:equations} leads to the halo maps shown in \fref{fig:case3_arrival}.  
In comparing with Case 2 from \sref{sec:case2} we see that the halo map is no longer restricted to a line, but instead the halo acquires a transverse extent.  
We vary the skew angle $\beta$ in the left panel of \fref{fig:case3_arrival}.  
In the limit that $\beta$ goes to zero, we regain the line-like halo map of Case 2, and in the limit that $\beta$ goes to $90^{\circ}$, we regain point-like halo map of Case 1.  
We vary the field strength in the right panel of \fref{fig:case3_arrival}.  
In the PH regime where $B_{0} \lesssim 10 \times 10^{-15} \Gauss$, the halo size is proportional to the field strength (green and magenta curves), while in the MBC regime where $B_{0} \gtrsim 10 \times 10^{-15} \Gauss$, the halo size is limited by the geometry (brown and teal curves).  

%==========
Our analytic approach to calculating the halo map has an advantage over numerical shooting techniques insofar as we can solve for the halo map analytically in certain limiting regimes.  
To demonstrate this point, first consider the limit of small skew angle $\beta \ll 1$ in which Case 3 reduces to Case 2.  
In this limit, \eref{case3:equations} becomes
\begin{align}\label{case3:solutionA}
	\phi \approx \beta \, \tan(\delta/2 - \theta)
	\quad , \qquad
	\theta \approx {\rm arcsin} \left( \frac{\dg}{\ds} \sin \frac{D_{e}}{R} \right)
	\quad , \quad {\rm and} \quad 
	\delta \approx \frac{D_{e}}{R} \com
\end{align} 
and it follows that 
\begin{align}
	& \tlat \approx {\rm arcsin} \left( \frac{\dg}{\ds} \sin \frac{D_{e}}{R} \right) \\
	& \ttrans \approx \beta \, \tan \left( \frac{D_{e}}{2R} - {\rm arcsin} \left( \frac{\dg}{\ds} \sin \frac{D_{e}}{R} \right) \right) \, {\rm arcsin} \left( \frac{\dg}{\ds} \sin \frac{D_{e}}{R} \right) \per
\end{align}
This behavior is seen in the left panel of \fref{fig:case3_arrival}.  
Second, consider the small bending angle regime $D_{e} / R \ll 1$ (weak magnetic field) where \eref{case3:equations} becomes 
\begin{align}\label{case3:solutionB}
	\phi \approx \frac{\sin \beta}{2 \cos^2 \beta} \frac{D_{e}}{R}
	\quad , \qquad
	\theta \approx \frac{ \Theta_{\rm ext} }{\cos \beta} 
	\quad , \quad {\rm and} \quad 
	\delta \approx \frac{D_{e}}{R} \cos \beta \com
\end{align}
and $\Theta_{\rm ext} \propto B_{0}$ was given by \eref{eq:Theta_ext_def}.  
Then the lateral and transverse extents are 
\begin{align}
	& \tlat \approx \frac{ \Theta_{\rm ext} }{\cos \beta} \\
	& \ttrans \approx \frac{\sin \beta}{2 \cos^2 \beta} \frac{D_{e}}{R} \frac{ \Theta_{\rm ext} }{\cos \beta} \per
\end{align}
Observe that $\ttrans$ is suppressed with respect to $\tlat$ by an additional factor of $(D_{e} / R) \propto B_{0}$.  

%=========
We quantify the halo shape using the triple product $Q$-statistic, given by \eref{eq:Q_def}.  
Since the magnetic field configuration under consideration is not helical, we expect $Q = 0$.  
In fact this is immediately evident from the symmetry of \fref{fig:case3_arrival}: the gamma rays on the branch in the first quadrant contribute $Q < 0$ while those in the second quadrant contribute $Q > 0$, and upon summing the two branches, they cancel.  
However, this cancellation is possible in part because we have assumed isotropic emission from the blazar.  
In practice, the jet may only illuminate a small patch of the halo map.  
Thus, to demonstrate the parametric dependence and typical scale of $Q$ it is illustrative to calculate the statistic using only the gamma rays in one of the two branches.

\begin{figure}[t]
\hspace{0pt}
\vspace{-0in}
\begin{center}
\includegraphics[width=0.45\textwidth]{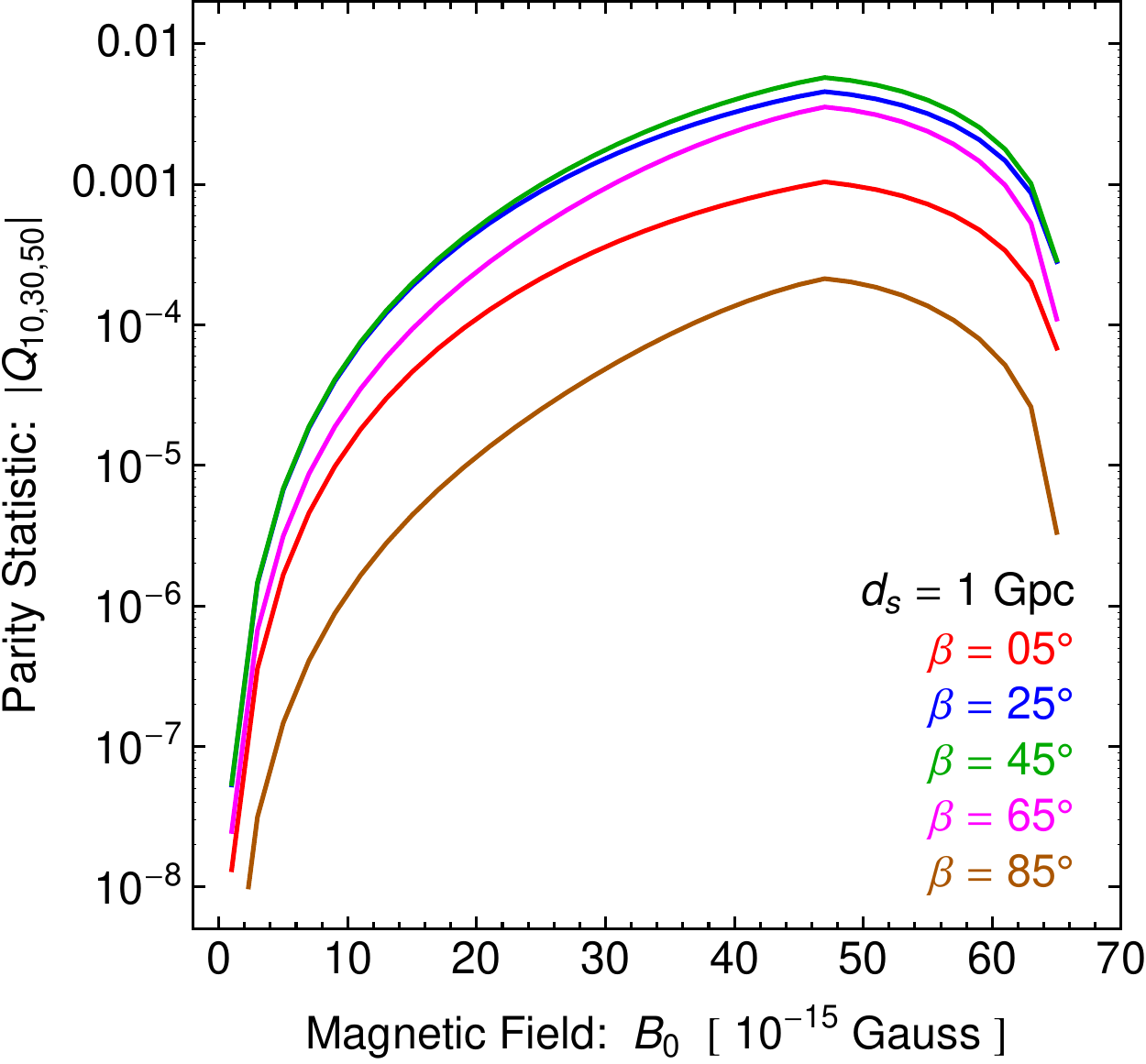} \hfill
\includegraphics[width=0.45\textwidth]{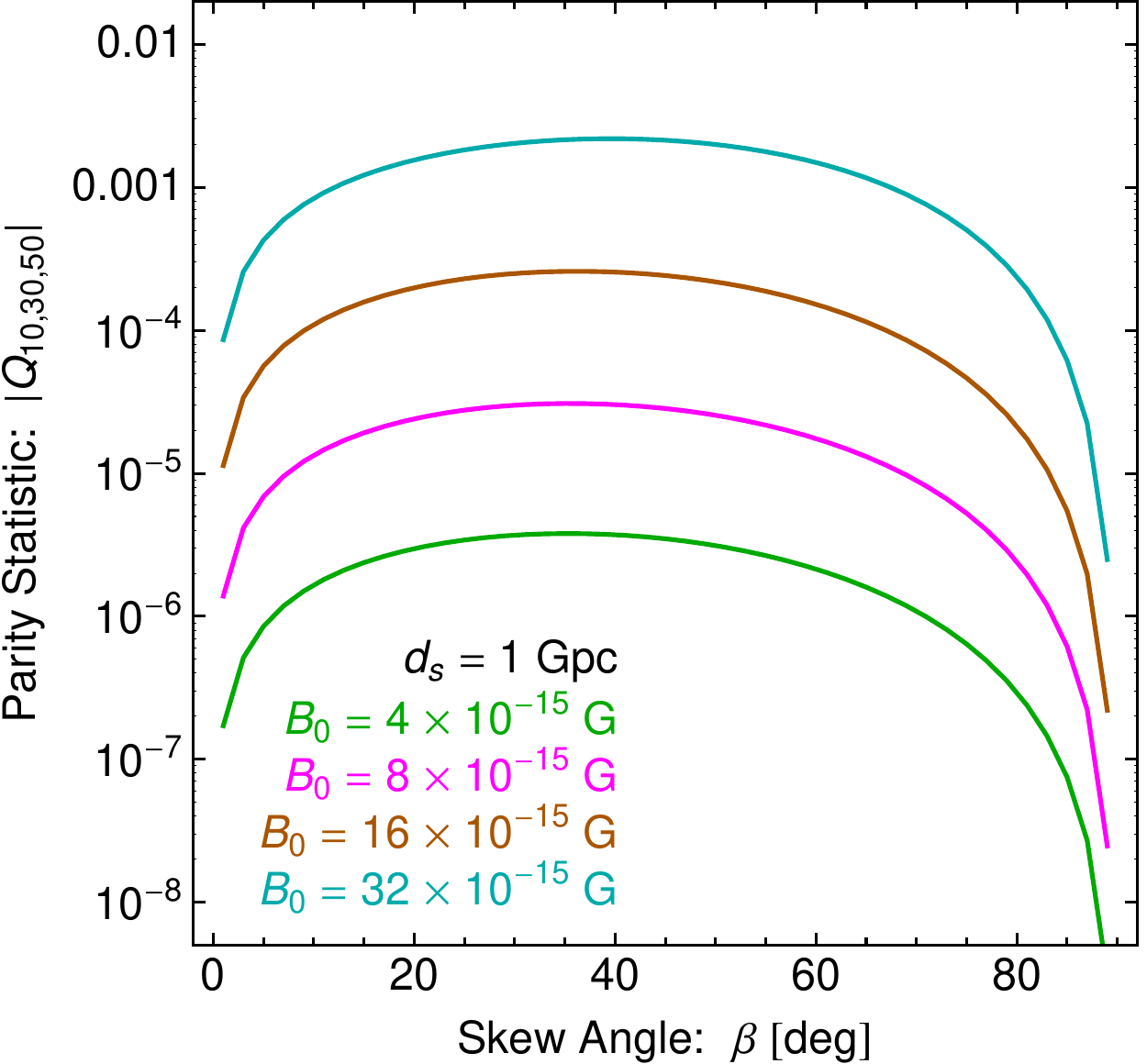} 
\caption{
\label{fig:case3_plotQ}
The magnitude of the parity statistic, given by \eref{eq:Q_def}, for Case 3.  If we had averaged over all gamma rays, we would find $Q_{10,30,50} = 0$ since the gamma rays in the first and second quadrants of \fref{fig:case3_arrival} cancel.  Here we just show $Q_{10,30,50}$ for the gamma rays in the second quadrant.  
}
\end{center}
\end{figure}

%=========
We evaluate $Q_{10,30,50}$ from \eref{eq:Q_def} and show the results in \fref{fig:case3_plotQ}.  
Similar results are obtained for the other energy combinations, and we do not show them here.  
The parameters are chosen to correspond with the halo maps in \fref{fig:case3_arrival}.  
The statistic $Q_{10,30,50}$ becomes small (i) in the limit $\beta \to 0$ where the halo map approaches a straight line, (ii) in the limit $\beta \to 90^{\circ}$ where the halo map approaches a point, and (iii) in the limit $B_{0} \to 0$ where the angular extent of the halo decreases.  
Increasing the field strength grows $Q_{10,30,50}$ until the crossover from the MBC to the PH regime at $B_{0} \sim 50 \times 10^{-15} \Gauss$.  
For larger $B_{0}$ the statistic first decreases and then begins to oscillate, similar to the behavior of the halo size, seen in \fref{fig:case2_grid}.  

%=========

%------------------------------------------------------------
% CASE 4
%------------------------------------------------------------
\subsection{Case 4:  Helical Magnetic Field with Wave Vector Parallel to Line of Sight}\label{sec:case4}

%=========
We now turn our attention to helical magnetic field configurations.  
The simplest configuration consists of a single circular polarization mode with wavelength $\lambda$ and wavevector ${\bf k} = (2\pi / \lambda) \hat{\bf z}$ oriented along the line of sight with the blazar: 
\begin{align}\label{case4:B_config}
	\hat{\bf B} = 
	\cos (\psi + 2 \pi z / \lambda) \, \hat{\bf y} + \sigma \, \sin(\psi + 2 \pi z / \lambda) \, \hat{\bf x} \per
\end{align}
The spatial coordinate $z$ should not be confused with the redshift $\zs$.
The three parameters are the coherence length $\lambda$, the handedness index $\sigma = \pm 1$, and the phase shift $\psi$.  
The handedness index controls the sign of the magnetic helicity density, 
\begin{align}\label{eq:H_def}
	\Hcal \equiv {\bf B} \cdot {\bm \nabla} \times {\bf B} = \sigma \frac{2 \pi}{\lambda} |{\bf B}|^{2} \per
\end{align}
The case $\sigma = 0$ corresponds to a non-helical, linearly polarized plane wave, while $\sigma= +1$ corresponds to left-circular polarization and $-1$ to right.  
In the subsequent analysis, one should bear in mind that varying $\lambda$ at fixed $|{\bf B}|$ implies that the magnetic helicity is being varied.  

\begin{figure}[t]
\hspace{0pt}
\vspace{-0in}
\begin{center}
\includegraphics[width=0.45\textwidth]{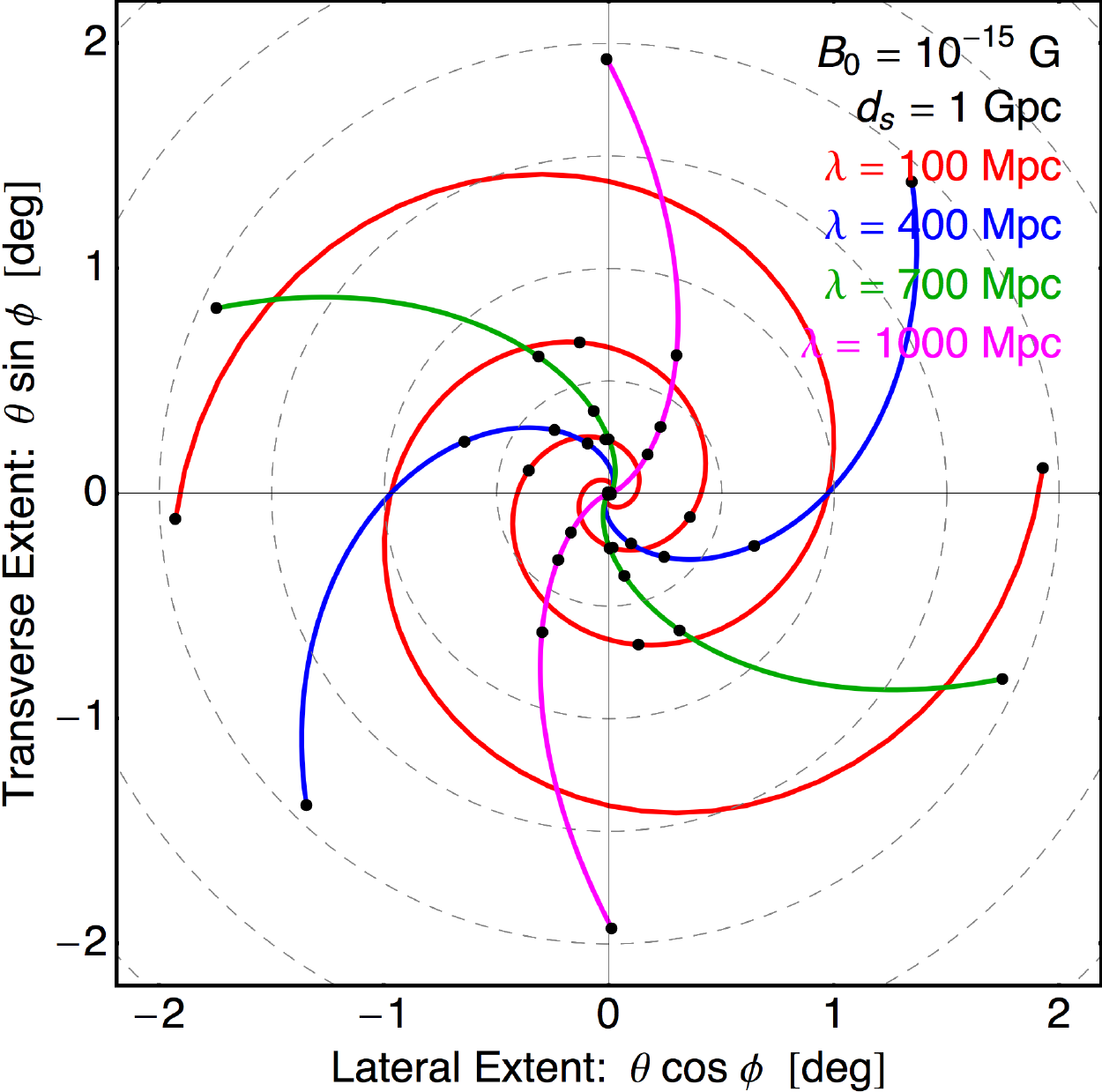} \hfill
\includegraphics[width=0.45\textwidth]{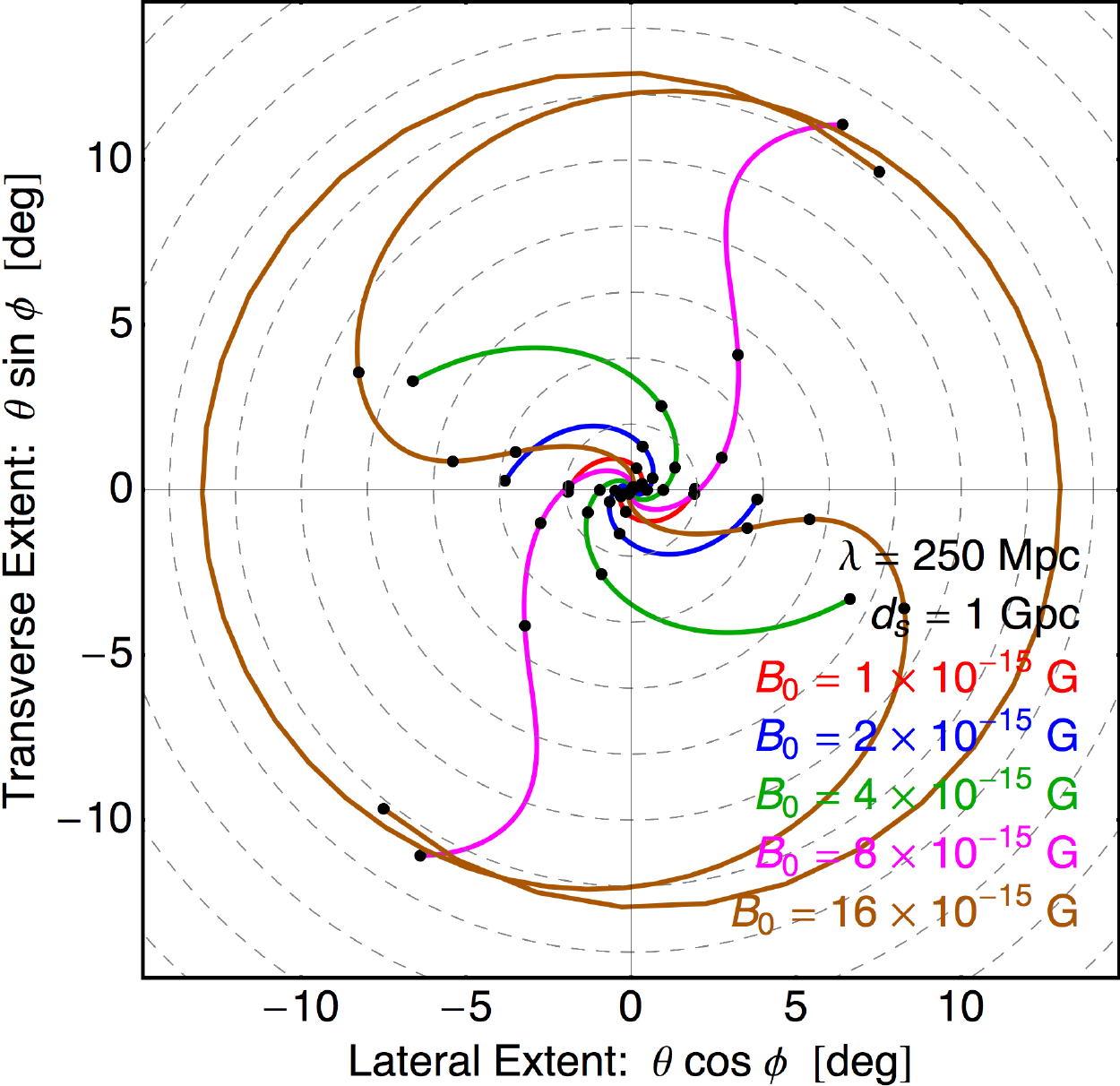} 
\caption{
\label{fig:case4_arrival}
The halo map for Case 4.  
We have taken $\sigma = +1$ and $\psi = 0$.  
For $\sigma = -1$ the handedness of the spiral is reversed, and for a phase shift $\psi \neq 0$ the halo map is uniformly rotated, but otherwise the structure remains unchanged.  
In the left panel, if $\lambda$ is further decreased below $100 \Mpc$ the spiral becomes tighter, and if $\lambda$ is further increased above $1 \Gpc$ the halo map asymptotes to a straight line.  
}
\end{center}
\end{figure}

%=========
It is convenient to define the angle $\beta_{\rm eff} = \psi + 2 \pi z / \lambda$.  
Then using the geometry of \fref{fig:triangle} the longitudinal coordinate at the point of pair production is 
$z = \ds - \dg \cos(\delta-\theta)$, and 
\begin{align}\label{case4:beta_eff}
	\beta_{\rm eff}(\delta,\theta) = \psi + \frac{2 \pi}{\lambda} \left( \ds - \dg \cos(\delta - \theta) \right)
\end{align}
is the effective skew angle.  
For this case, \Eqns reduce to the set of equations
\begin{subequations}\label{case4:equations}
\begin{align}
	& \tan \phi = -\sigma \, \tan \beta_{\rm eff}(\delta,\theta) \\
	& \sin \theta = \frac{\dg}{\ds} \sin \delta \\
	& \cos \delta = \cos \frac{D_{e}}{R} \com
\end{align}
\end{subequations}
which can be solved analytically.  
The solutions are shown in \fref{fig:case4_arrival}.  
The most striking feature in these figures is that the halo map forms a spiral pattern.  
The handedness of the spiral is controlled by the helicity of the magnetic field, parametrized here by $\sigma= \pm 1$.  
As the phase shift $\psi$ is varied, the halo map is uniformly rotated clockwise or counterclockwise.  

%=========

%=========
As we vary the coherence length $\lambda$ the spiral becomes flatter or tighter.  
In the limit of large coherence length $\lambda \gg \dg \sim 100 \Mpc$, the cascade takes place in an effectively homogeneous magnetic field, $\Bhat \approx \hat{\bf y}$.  
Then we regain the behavior of Case 2 from \sref{sec:case2} in which the gamma rays propagate in a plane and arrive at Earth collimated into a line with small transverse extent.  
In the opposite limit of small coherence length, $\lambda \ll \dg \sim 100 \Mpc$, the TeV gamma rays sample the magnetic field at a random phase.  
In terms of the halo map, this translates into a tightly wound spiral with multiple cycles.  
Varying the magnetic field strength has the same effect as in the previous cases.  
In the PH regime where the field is weak, the angular size of the halo grows with increasing field strength, while in the MBC regime where the field is strong, the angular size of the halo is limited by the geometry (brown curve in right panel).  

%=========
\begin{figure}[t]
\hspace{0pt}
\vspace{-0in}
\begin{center}
\includegraphics[width=0.45\textwidth]{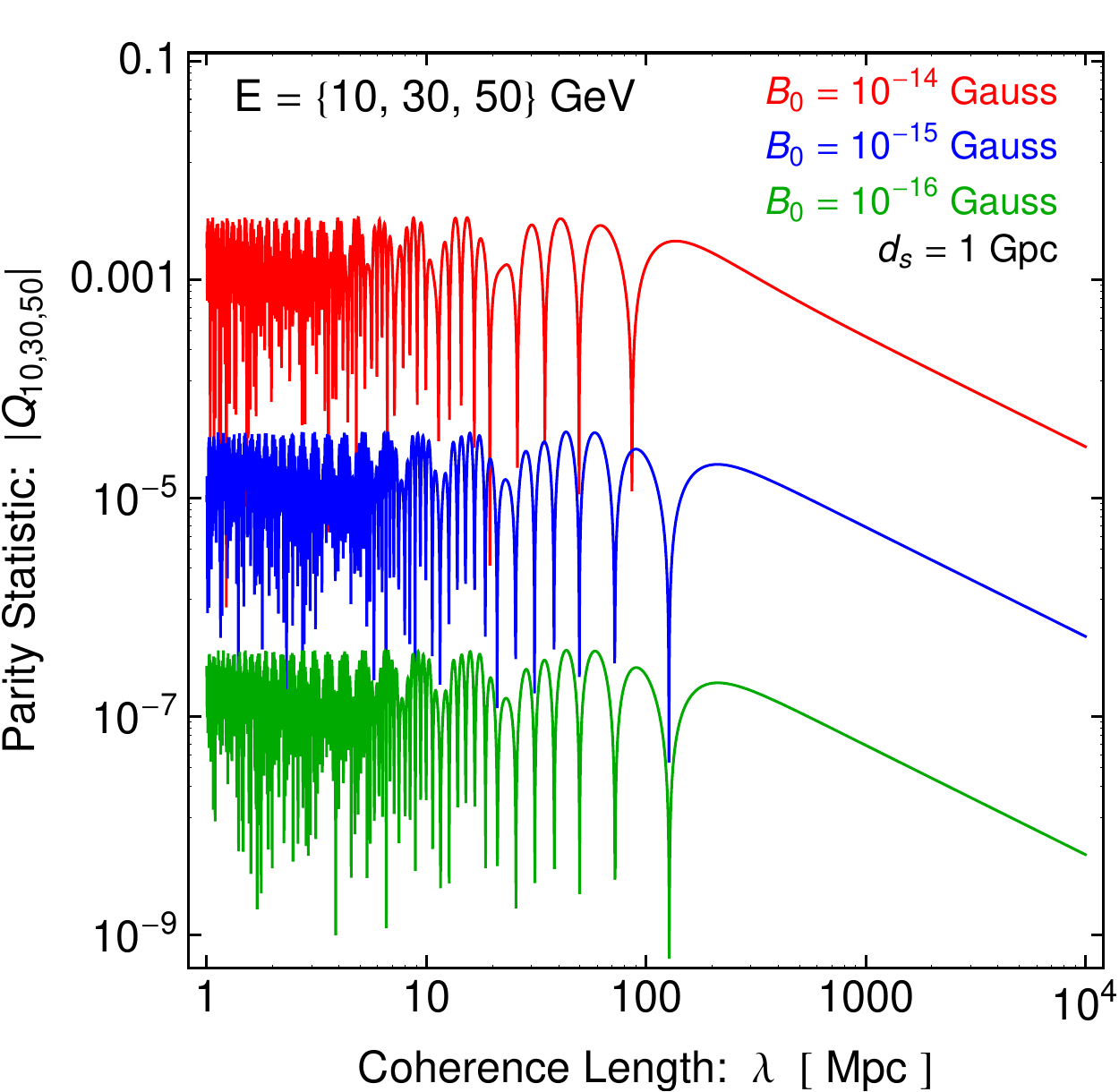} \hfill
\includegraphics[width=0.45\textwidth]{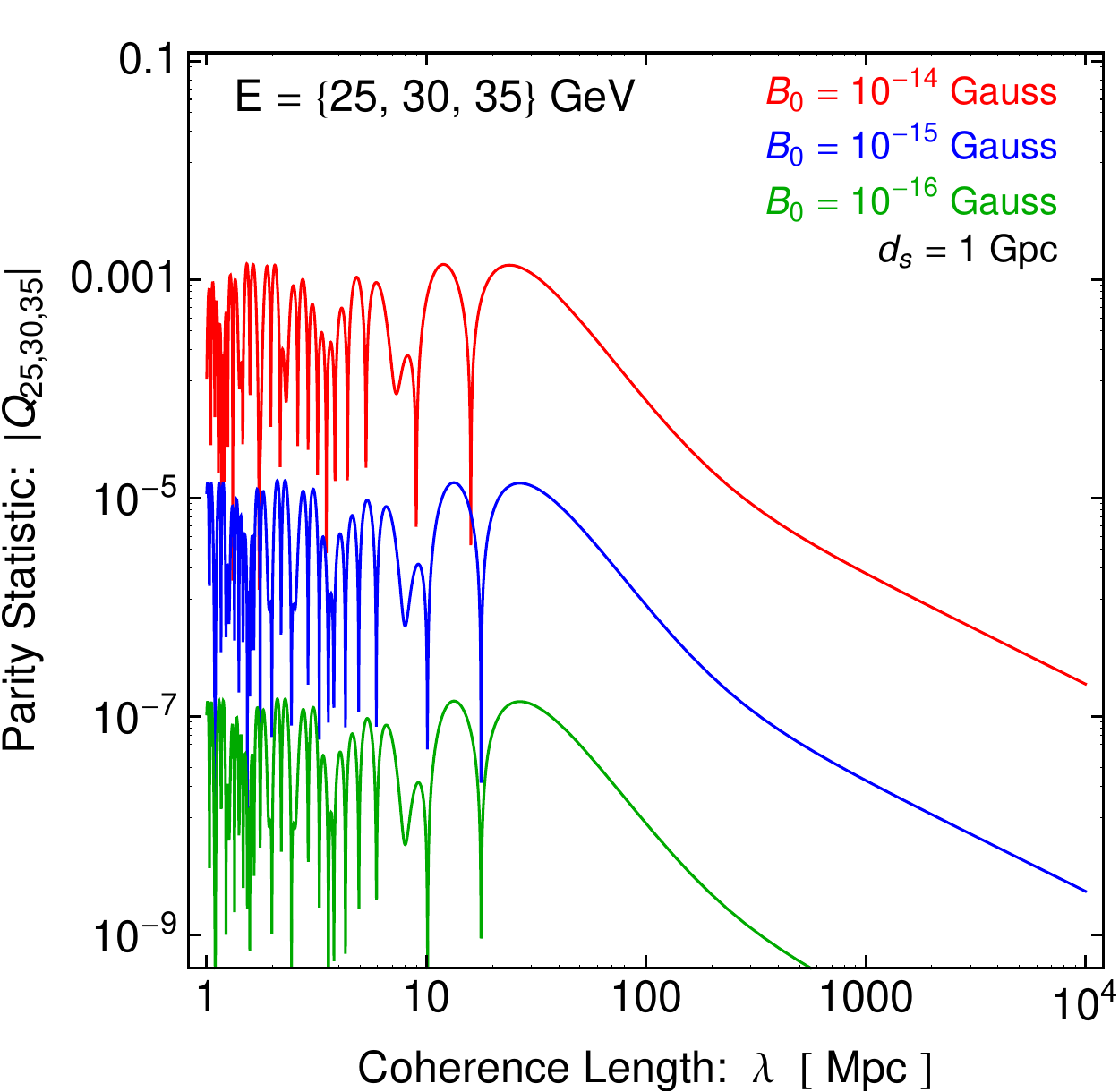} \\
\includegraphics[width=0.45\textwidth]{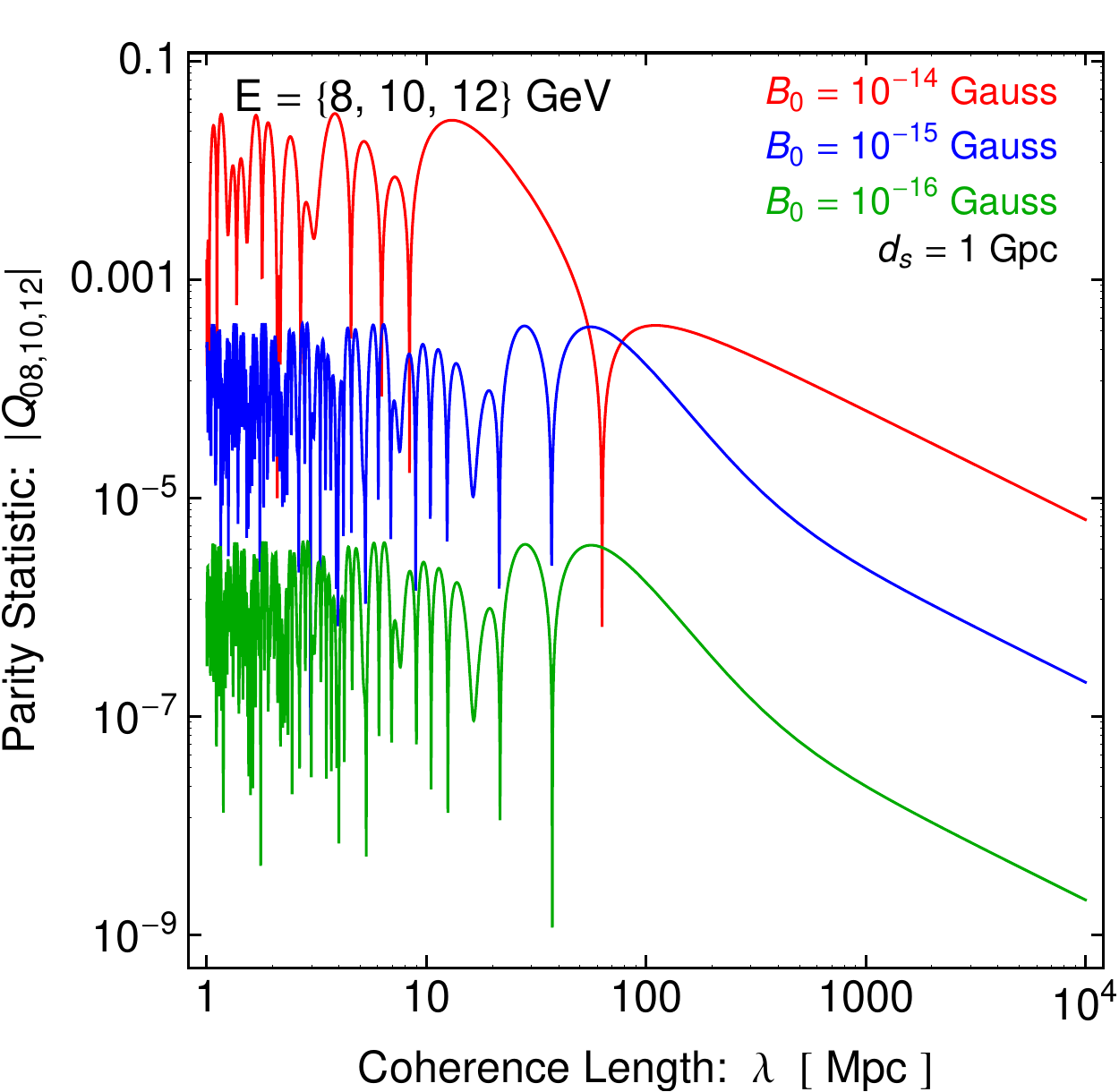} \hfill
\includegraphics[width=0.45\textwidth]{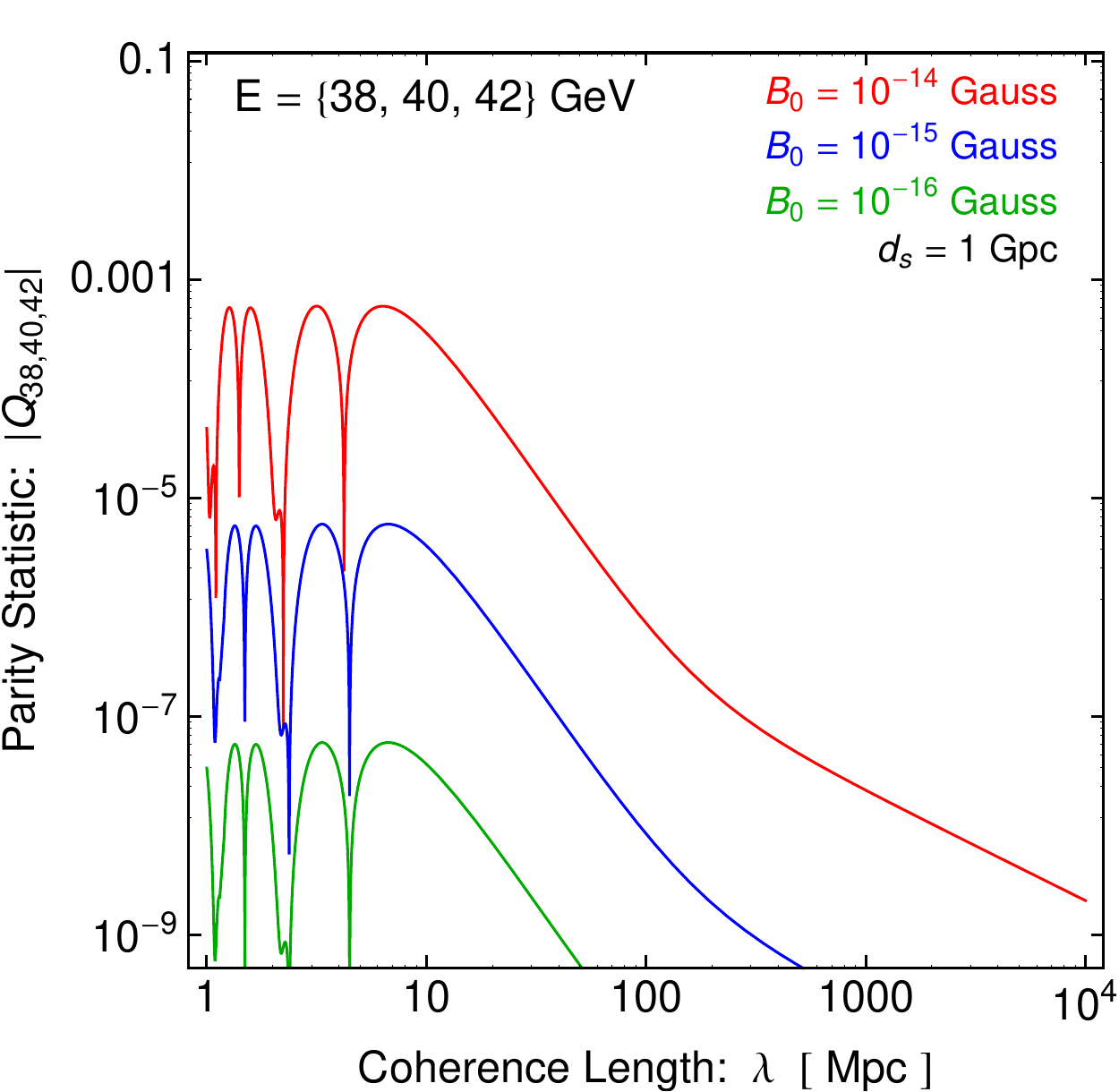} 
\caption{
\label{fig:case4_plotQ}
The parity statistics, given by \eref{eq:Q_def}, for Case 4.  
For large $\lambda$ the sign of the $Q$-statistic is given by ${\rm sign}[Q] = -\sigma = -{\rm sign}[\Hcal]$.  
As the coherence length is lowered, the halo map begins to spiral around the line of sight, as seen in \fref{fig:case4_arrival}, and the sign of $Q$ oscillates.  
Statistics calculated from different energy combinations have the same qualitative behavior, but they differ in the magnitude of $Q$ and the value of $\lambda$ at which the oscillations begin.  
}
\end{center}
\end{figure}

%=========
The halo maps in \fref{fig:case4_arrival} display a clear parity-violation that should be captured by the $Q$-statistics.  
However, the halo map $\nhat(E_{\gamma})$ is double-valued, {\it i.e.} there are two gamma rays at each energy, which correspond to the two branches of the spiral in \fref{fig:case4_arrival}.  
If we first average $\nhat(E_{\gamma})$ over its multiple solutions, we would obtain ${\rm Avg}[\nhat(E_{\gamma})] = 0$ and therefore $Q=0$.  
This result is a consequence of the high symmetry of the system under consideration:  we have assumed that the wavevector of the magnetic field is oriented along the line of sight, and we have allowed for isotropic emission from the blazar.  
As we discussed for Case 3 of \sref{sec:case3}, under more realistic conditions the blazar's jet will only illuminate a part of the full halo map, and then the cancellation is disrupted.  
To model this effect, we calculate $Q$ along a single branch of the spiral.  

%=========
We calculate the $Q$-statistics using \eref{eq:Q_def} and show the parametric dependence on coherence length $\lambda$ and field strength $B_{0}$ in \fref{fig:case4_plotQ}.  
The statistic has the same qualitative behavior for each energy combination: (i) $|Q|$ decreases for small $B_{0}$, (ii) $|Q|$ decreases for large $\lambda$, and (iii) $Q$ oscillates rapidly from positive to negative values for small $\lambda$.  
When the magnetic field is weak, the halo has a small angular extent, as shown in the right panel of \fref{fig:case4_arrival}, and $Q_{abc} = \nhat_a \times \nhat_b \cdot \nhat_c$ decreases as $\nhat_a, \nhat_b,$ and $\nhat_c$ become approximately collinear (see also \eref{eq:Q_alt}).   
When the coherence length is large the halo map resembles a straight line, as seen in the left panel of \fref{fig:case4_arrival}, and $Q_{abc}$ decreases as $\nhat_a, \nhat_b,$ and $\nhat_c$ become approximately coplanar.  
In this regime, the magnetic field appears uniform, and therefore non-helical, on the scale probed by the TeV gamma rays.  
When the coherence length is small the halo map resembles a tight spiral, as seen in the left panel of \fref{fig:case4_arrival}, and the three gamma rays used to construct $Q$ may not lie on the same cycle of the spiral.  
In this case, $Q$ may take positive or negative values depending on the energies at which the spiral is sampled, and as $\lambda$ decreases further and the spiral becomes more tightly wound, the sign of $Q$ oscillates.  
We will use $\lambda_{\rm osc}$ to denote the coherence length at which $Q$ begins to oscillate.  

%=========
The four cases in \fref{fig:case4_plotQ} display important quantitative differences between $Q$-statistics constructed from different energy combinations.  
Comparing the bottom two panels, $|Q_{8,10,12}| > |Q_{38,40,42}|$, we see that lower energy gamma rays leads to a larger value for $|Q|$.  
This is simply because lower energy gamma rays are more easily deflected and lead to a larger halo in the MBC regime, see \eref{eq:Theta_ext}.  
Comparing the top two panels, $\lambda_{\rm osc}(\{10,30,50\}\GeV) > \lambda_{\rm osc}(\{25,30,35\}\GeV)$, we see how the energy spacing affects the coherence length scale below which $Q$ begins to oscillate.  
As $\lambda$ decreases and the halo map becomes a more tightly wound spiral, and the gamma rays from more closely spaced energies, in this case $\{25,30,35\}\GeV$, remain on the same cycle of the spiral longer than more widely spaced energies.  
In the two lower panels the energy spacings are identical but nevertheless $\lambda_{\rm osc}(\{38,40,42\}\GeV) < \lambda_{\rm osc}(\{8,10,12\}\GeV)$.  
This is because the spiraling behavior becomes more pronounced as the gamma ray energy is lowered; see \fref{fig:case4_arrival} and note that the separation between the $5$ and $10 \GeV$ gamma rays is much larger than the separation between the $10$ and $15 \GeV$ gamma rays.

%------------------------------------------------------------
% CASE 5
%------------------------------------------------------------
\subsection{Case 5:  Helical Magnetic Field with Wave Vector Normal to Line of Sight}\label{sec:case5}

%=========
Finally we consider a more generic helical magnetic field configuration.  
We do not require that the wave vector is oriented along the line of sight, but for simplicity we don't allow it to have a general orientation either, and 
instead we require that ${\bf k}$ is normal to the line of sight with the blazar.  
Without further loss of generality we can write ${\bf k} = (2\pi / \lambda) \hat{\bf x}$.  
The magnetic field configuration is
\begin{align}\label{case5:B_config}
	\hat{\bf B} & = 
	\cos (\psi + 2 \pi x / \lambda) \, \hat{\bf y} - \sigma \, \sin (\psi + 2 \pi x / \lambda) \, \hat{\bf z} \nn
	& = \cos (\psi + 2 \pi x / \lambda) \, \sin \phi \, \hat{\bm \rho} + \cos (\psi + 2 \pi x / \lambda) \, \cos \phi \, \hat{\bm \phi} - \sigma \, \sin (\psi + 2 \pi x / \lambda) \, \hat{\bf z} \per
\end{align}
In this case the phase shift $\psi$ plays a nontrivial role.  
For specific values $\psi = 0, \pi/2, \pi, \cdots$ the magnetic field is either symmetric or antisymmetric when reflected across the line of sight ($x \to - x$ and $\phi \to - \phi$), but for general $\psi$ there is no such symmetry.  
Once again the handedness index $\sigma$ controls the magnetic helicity via \eref{eq:H_def}.  

%=========
\begin{figure}[t]
\hspace{0pt}
\vspace{-0in}
\begin{center}
\includegraphics[width=0.45\textwidth]{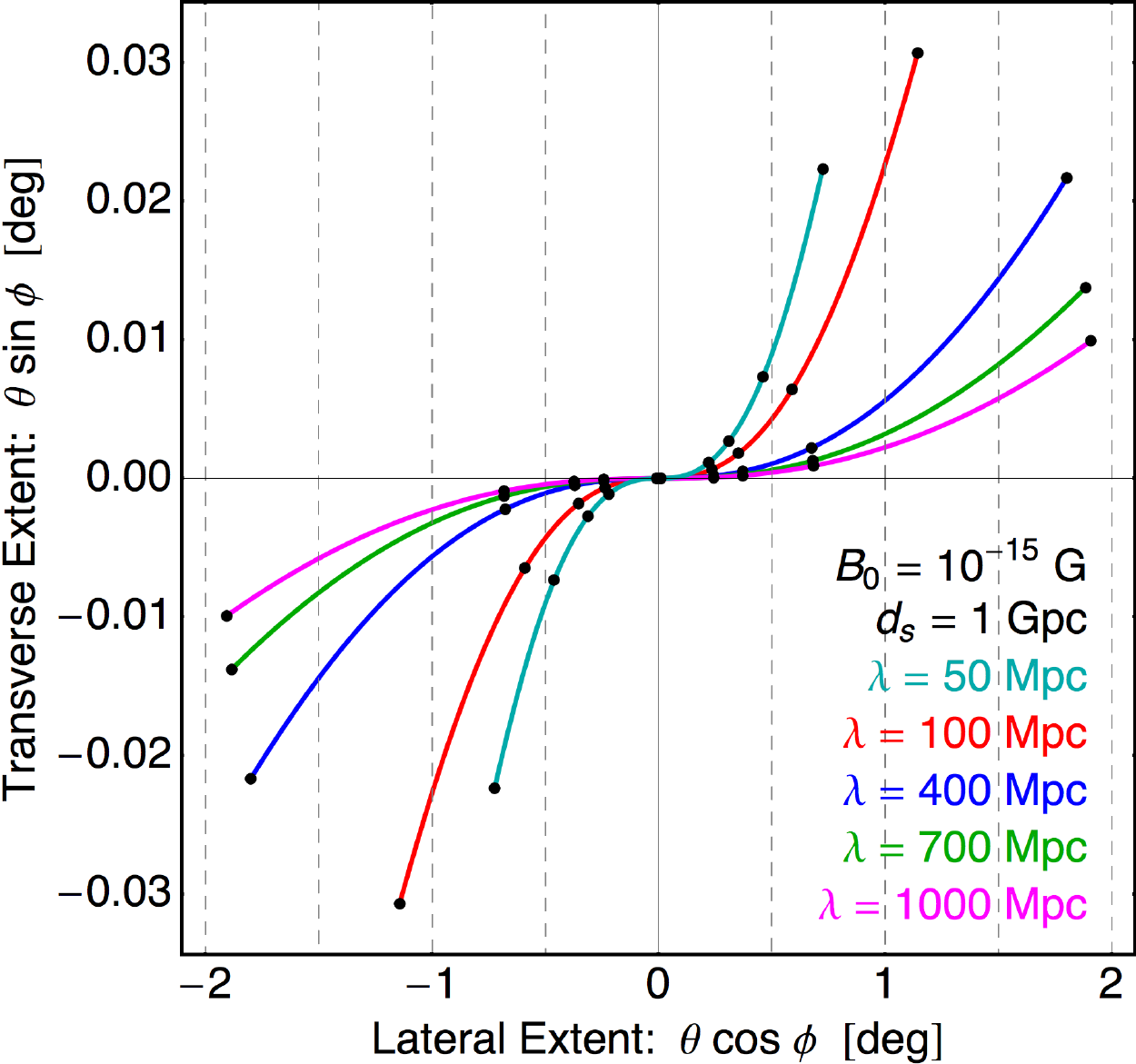} \hfill
\includegraphics[width=0.45\textwidth]{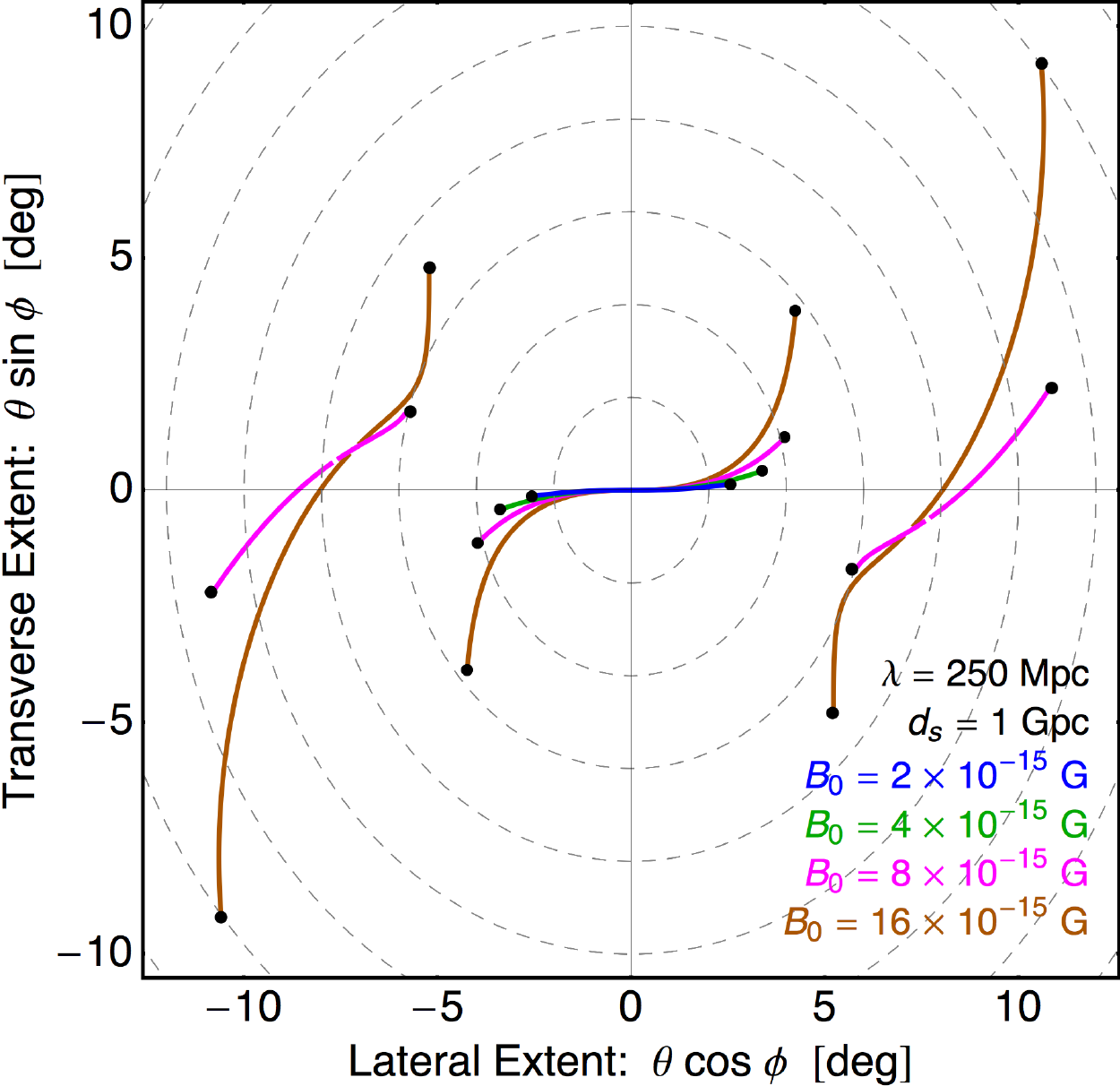} \\
\vspace{0.5cm}
\includegraphics[width=0.45\textwidth]{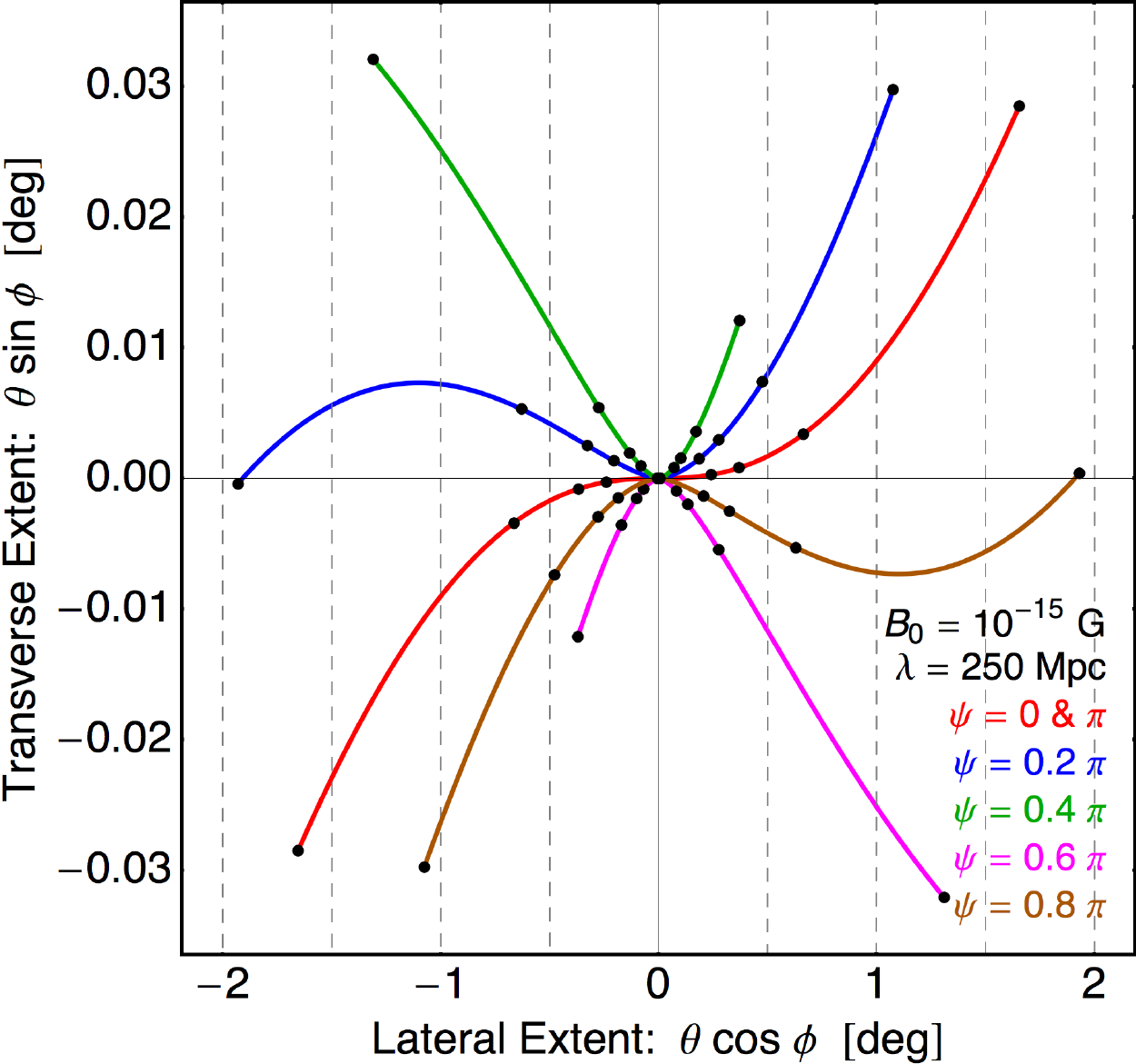} 
\caption{
\label{fig:case5_arrival}
The halo map in Case 5.  We have taken $\sigma = +1$ in all panels, and for $\sigma=-1$ the halo maps are reflected across the vertical axis.  In the first and second panels we fix $\psi = 0$.  In the first and third panels, the black dots have the same meaning as in the previous halo maps:  they indicate $E_{\gamma} = 100,20,15,10,$ and $5 \GeV$.  In the second panel, the black dots show only $E_{\gamma} = 5 \GeV$, the black square indicates $E_{\gamma} = 7 \GeV$ for $B_{0} = 8 \times 10^{-15} \Gauss$, and the black triangle indicates $E_{\gamma} = 11 \GeV$ for $B_{0} = 16 \times 10^{-15} \Gauss$.  
}
\end{center}
\end{figure}

%=========
It is convenient to introduce the effective skew angle
\begin{align}\label{case5:beta_eff}
	\beta_{\rm eff}(\delta,\theta,\phi) = \psi + \frac{2 \pi \dg}{\lambda} \sin(\delta - \theta) \cos \phi
\end{align}
where we have used $x = \dg \sin(\delta-\theta) \, \cos \phi$.  
Then \Eqns reduce to 
\begin{subequations}\label{case5:equations}
\begin{align}
	& \sin \phi = \sigma \, \tan \beta_{\rm eff}(\delta,\theta,\phi) \, \tan \left( \delta / 2 - \theta \right) \\
	& \sin \theta = \frac{\dg}{\ds} \sin \delta \\
	& 1 - \cos \delta = \Bigl( 1 - \sin^2 \beta_{\rm eff}(\delta,\theta,\phi) \frac{\cos^2(\delta/2)}{\cos^2(\delta/2-\theta)} \Bigr) \Bigl( 1 - \cos (D_{e} /  R) \Bigr) 
	\per
\end{align}
\end{subequations}
These equations have a rich and interesting family of solutions, but as a result they cannot be solved analytically as in Case 4.  
Instead we solve \eref{case5:equations} numerically.  

%=========
The halo maps are shown in \fref{fig:case5_arrival}, and we have chosen the same parameters as in \fref{fig:case4_arrival} to facilitate comparison with Case 4.  
In each of the three panels we have taken the handedness index $\sigma = +1$, and the associated parity-violation is evident in the ``S''-like shape of the halo maps.  
Choosing $\sigma = -1$ reflects the halo maps across the vertical axis, which flips the handedness.  

%=========
In the first panel, the coherence length is reduced from $\lambda = 1000 \Mpc$ to $50 \Mpc$.  
The behavior in the large $\lambda$ regime is the same as in Case 4: when the coherence length is much larger than the scale of the cascade, $\dg \sim 100 \Mpc$, the gamma rays probe an effectively homogeneous magnetic field, and we regain the line-like halo map that was originally seen in Case 2 of \sref{sec:case2}.  
The small $\lambda$ behavior is distinctly different than in Case 4 where we encountered a spiral-shaped halo map.  
Now the wavevector crosses the line of sight to the blazar.  
As $\lambda$ is reduced, the surfaces of constant phase become compressed in the lateral direction, and the ``S''-like halo map becomes ``squeezed.''  

%=========
In the second panel, the magnetic field strength is increased from $B_{0} = 2$ to $16 \times 10^{-15} \Gauss$.  
In the small $B_{0}$ regime, we once again regain the behavior of the previous cases:  a smaller field strength translates into a smaller halo.  
In the large $B_{0}$ regime, on the other hand, a new phenomenon emerges: the halo map acquires multiple disconnected branches.  
In this panel only, all of the black dots denote $E_{\gamma} = 5 \GeV$, and we see that for $B_{0} = 8$ and $16 \times 10^{-15} \Gauss$ the halo map $\nhat(E_{\gamma})$ has six distinct values.  
In the previous cases, the halo map was only double-valued.  
The new branches contain only low energy gamma rays, which are more easily deflected, and the bifurcation points are denoted by a black square and black triangle for $B_{0} = 8$ and $16 \times 10^{-15} \Gauss$, respectively.  
As the field strength is further increased or the coherence length lowered, additional branches will emerge.  

%=========
\begin{figure}[t]
\hspace{0pt}
\vspace{-0in}
\begin{center}
\includegraphics[width=0.45\textwidth]{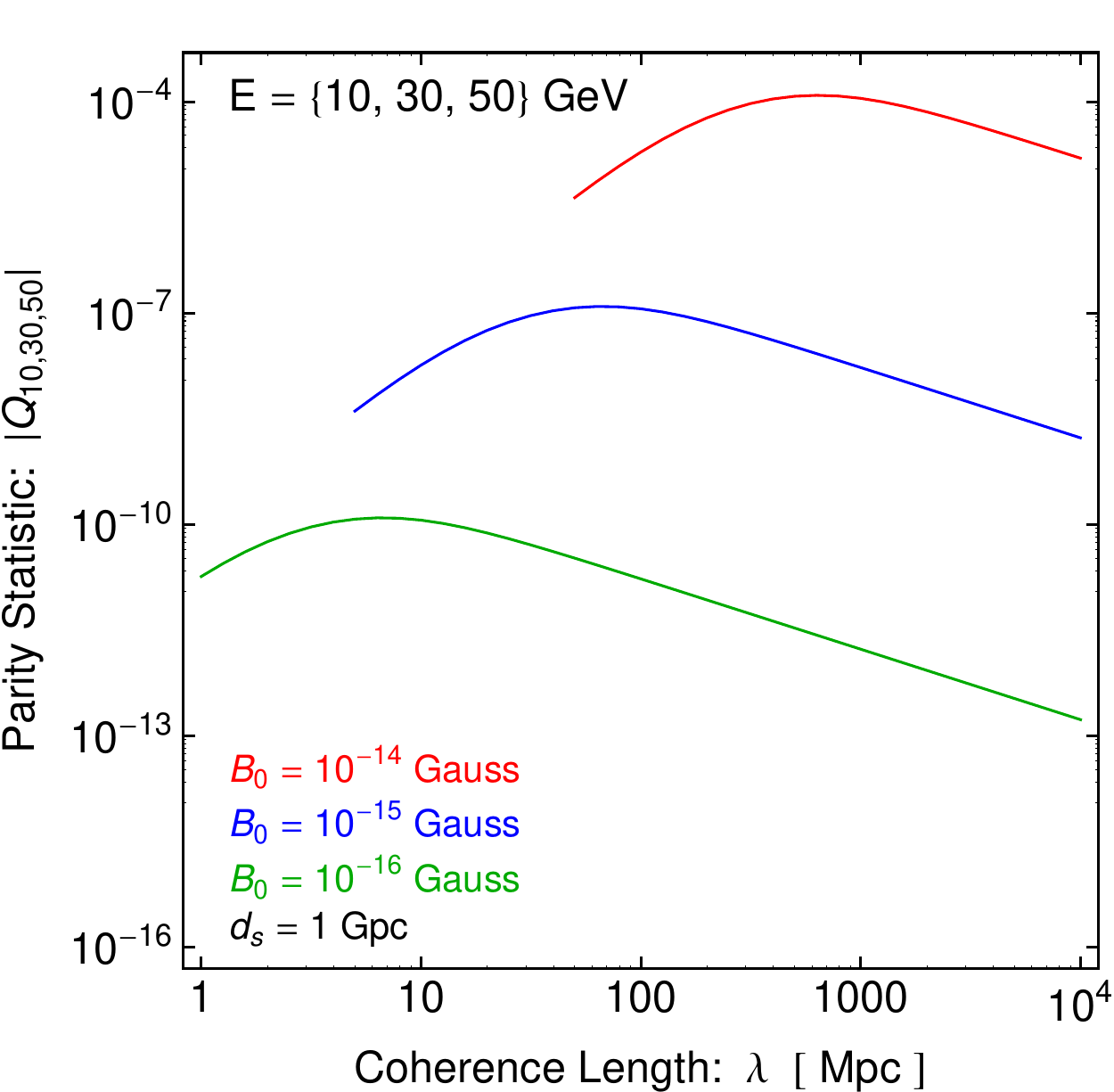} \hfill
\includegraphics[width=0.45\textwidth]{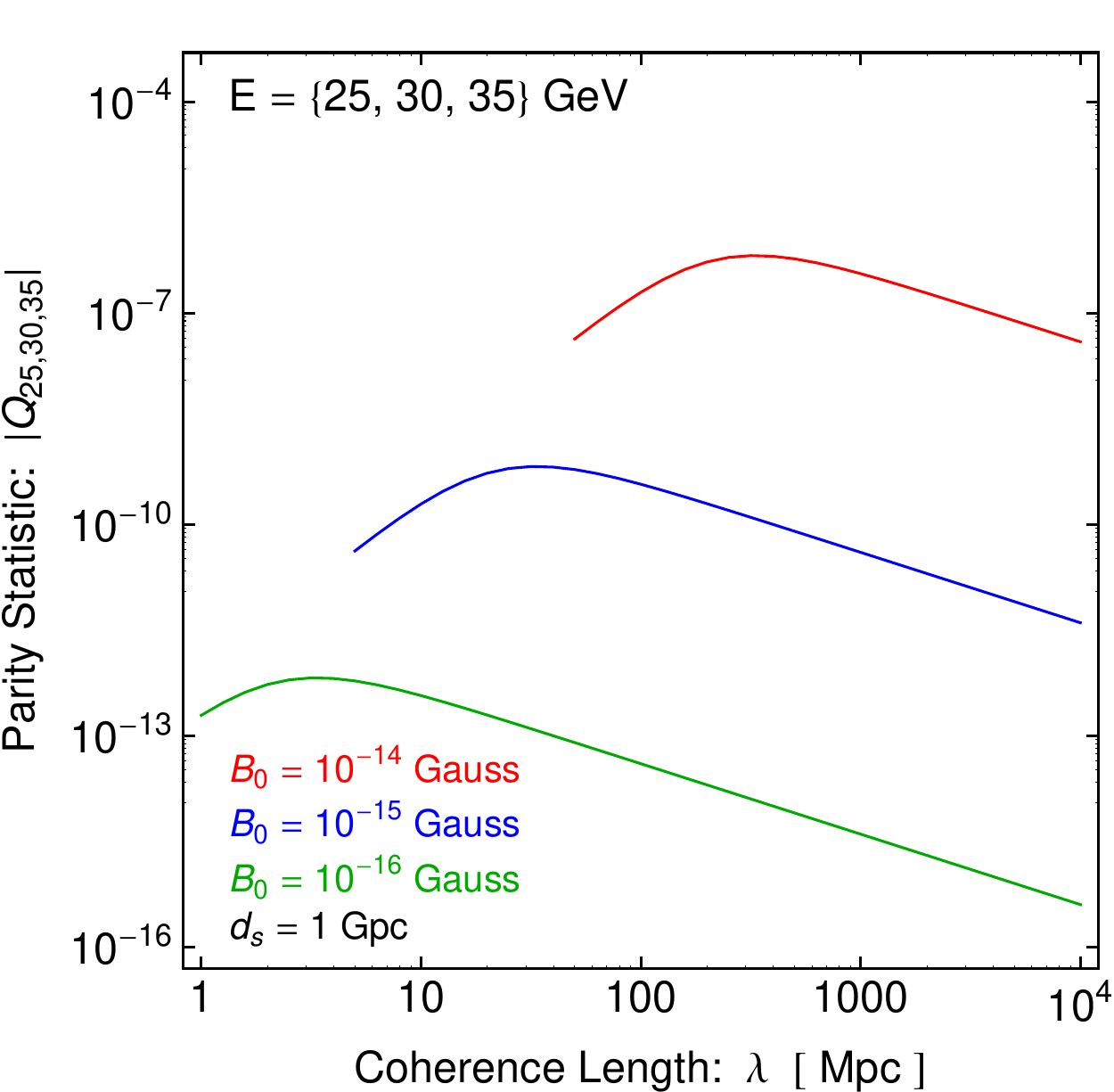} \\
\includegraphics[width=0.45\textwidth]{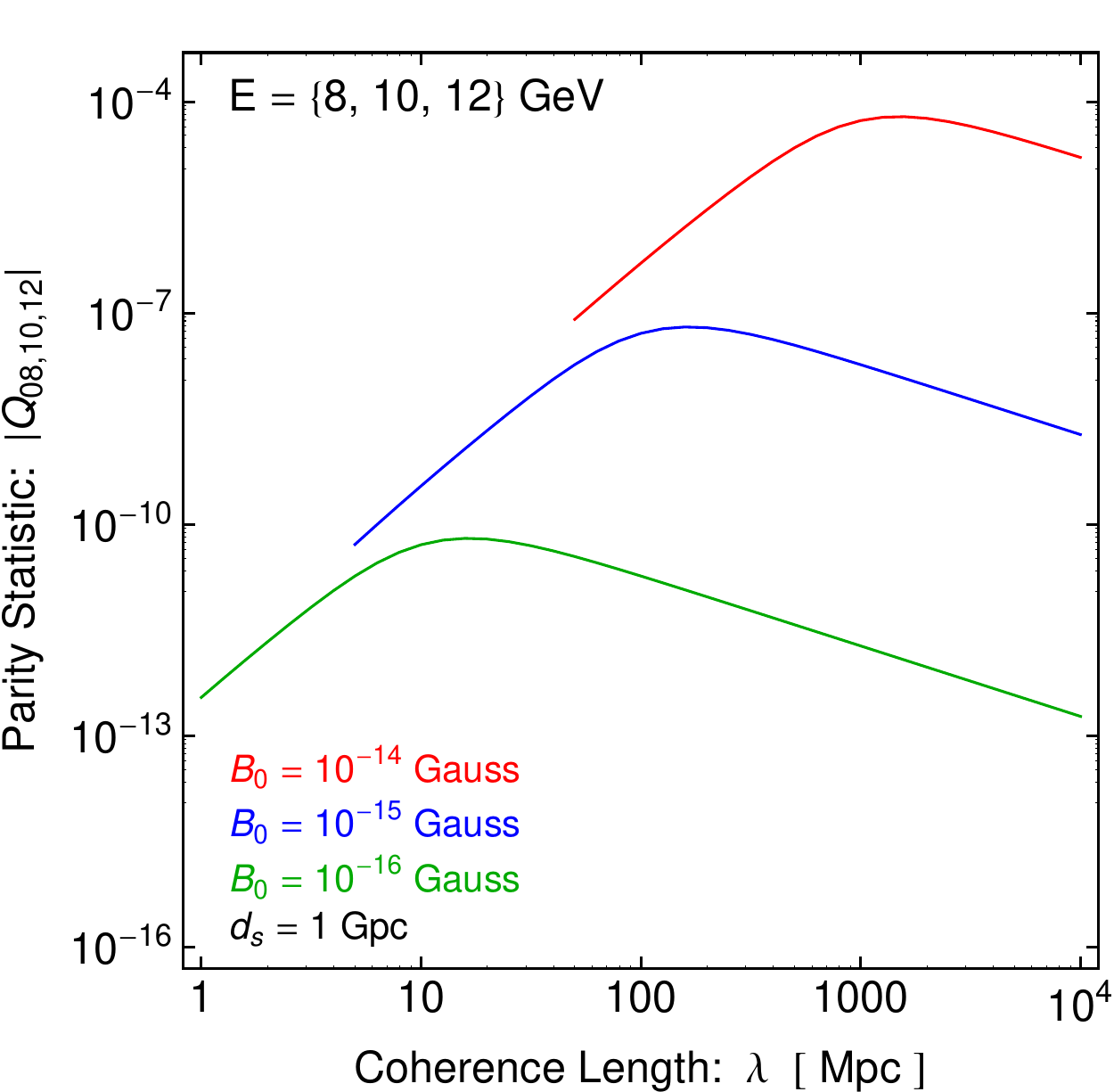} \hfill
\includegraphics[width=0.45\textwidth]{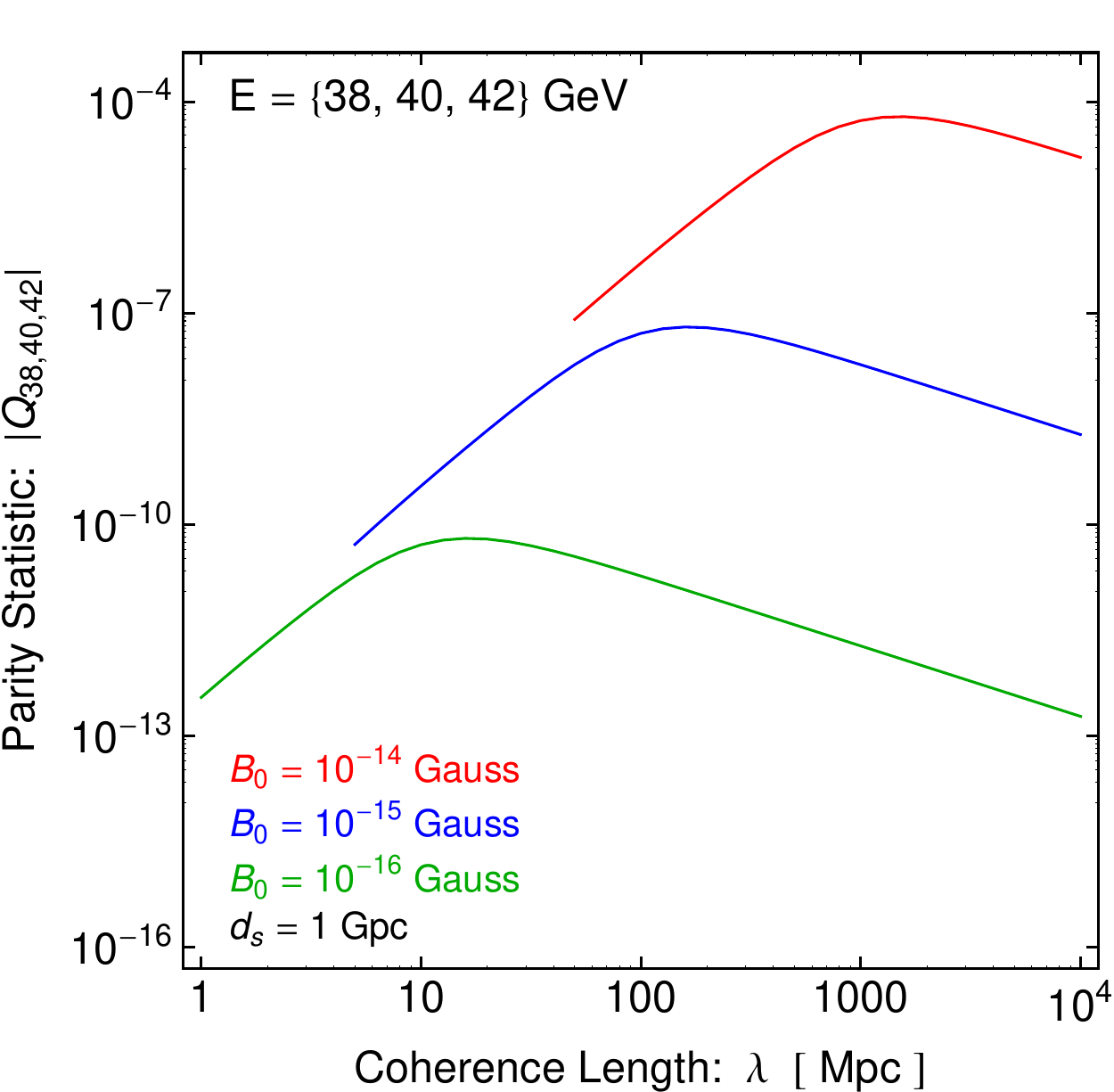} 
\caption{
\label{fig:case5_plotQ}
The parity statistics for Case 5. 
We calculate $Q$ using only the main solution branch. 
The plots are truncated at small $\lambda$ and large $B_{0}$ where numerical issues arise.  
For small $\lambda$ and large $B_{0}$ where there can be multiple branches, this approach probably 
underestimates $Q$. 
}
\end{center}
\end{figure}

%=========
In the third panel, we vary the phase parameter $\psi$ from $0$ to $\pi$.  
If $\psi \neq (0 \mod \pi)$ then the reflection symmetry of the halo map is disrupted.  
This was not the situation in Case 4 where varying $\psi$ simply lead to a uniform rotation of the spiral-like halo map.  

%=========
To quantify the parity-violating features in the halo maps, we calculate the $Q$-statistics using \eref{eq:Q_def} and show the results in \fref{fig:case5_plotQ}.  
As we have seen, the halo map becomes multi-valued for large $B_{0}$ and small $\lambda$, and it displays multiple branches.  
To generate \fref{fig:case5_plotQ} we calculate $Q$ using only the main branch, which is continuously connected to the line of sight.  
By neglecting the branches with larger angular separation from the line of sight, we would presumably underestimate $Q$ in this regime.  

%=========
Over the parameter space shown \fref{fig:case5_plotQ} we have ${\rm sign}[Q] = -\sigma = -{\rm sign}[\Hcal]$, where $\Hcal > 0$ for the left-circular polarization mode.  
In the limit of large $\lambda$ we have the same behavior as in Case 4, see \fref{fig:case4_plotQ}.  
In the limit of small $\lambda$ the $Q$-statistics decrease because the main branch is squeezed, and the angular extent of the halo is smaller.  
Also in comparing with Case 4, we see that the scaling with $B_{0}$ is different:  in \fref{fig:case4_plotQ} we found $|Q| \sim B_{0}^2$ but \fref{fig:case5_plotQ} implies that $|Q| \sim B_{0}^4$ for Case 5.  
As such, the magnitude of $Q$ very quickly decreases with decreasing field strength.

%==================================
% SENSITIVITY ANALYSIS
%==================================
\section{Implications for Helicity Measurement}
\label{sec:Sensitivity}

%=========
By considering particular realizations of the IGMF, our analysis reveals the parametric behavior of the halo size and shape on the various magnetic field variables, specifically the field strength $B_0$, 
the coherence length $\lambda$, and the magnetic helicity density $\Hcal = \pm (2\pi/\lambda) B_0^2$.  
In the ``corners'' of the parameter space, where either $B_0$ or $\lambda$ is very large or very small, the halo map behaves in a way that may make a measurement of the magnetic helicity more challenging than in the ``central'' parameter regime.  
For instance, small $B_{0}$ implies a small halo, which may not be distinguishable from a point source with a telescope's finite angular resolution.  
In this section we will demarcate the various parametric regimes and discuss the challenges posed by each.  

%=========
The quantitative analysis in this section uses the simplified field configurations considered in \sref{sec:Examples}, but we expect qualitatively similar results (parameter space boundaries) even if the intervening magnetic field configuration is not of one of the forms we have discussed, {\it e.g.} if the field is stochastically homogeneous and isotropic.  
The inclusion of stochastic variables in the development of the cascade, {\it e.g.} spectrum of the EBL or scattering probabilities, will smear out the halo patterns we have seen, and non-cascade photons introduce noise in our signal and dilute the halo.  
A rigorous evaluation of experimental sensitivities, even for a given set of experimental parameters, will require more information on blazar sources (gamma ray flux and spectrum, jet orientation and structure) and the background noise.

%=========
\ul{\it Strong Field Regime} \\
For a strong magnetic field, at energies such that the gyroradius $R$ is smaller than the typical distance 
traveled by the charged lepton $D_{e}$, see \eref{eq:De_over_R}, we have $D_{e}/R > 1$.  
This has two consequences.  
First, the maximum angular extent of the halo is no longer tied to the magnetic field strength, but instead it is fixed by the geometry as in \eref{eq:Theta_max}.  
Then if a halo is seen, one can infer the presence of an IGMF, but one cannot measure the field strength from the halo size alone.  
Second, since the charged leptons can make a complete orbit around the gyrocircle, the GeV gamma rays will be emitted isotropically.  
This reduces the flux by roughly $\Omega_{\rm jet} / 4\pi$, where $\Omega_{\rm jet}$ is the solid angle of the blazar's jet, and makes it more difficult to see the halo.  

\begin{figure}[t]
\hspace{0pt}
\vspace{-0in}
\begin{center}
\includegraphics[width=0.45\textwidth]{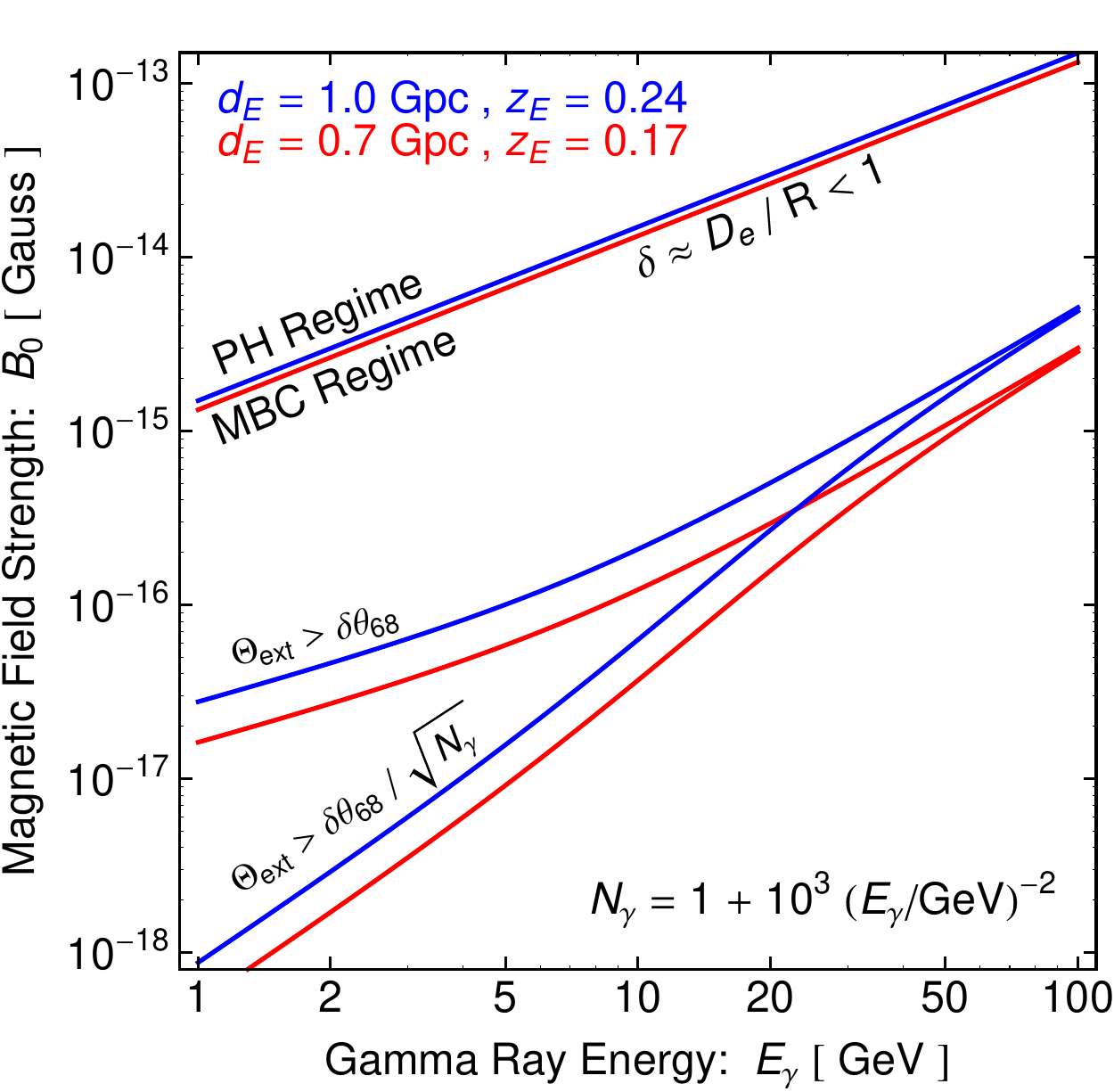} 
\caption{
\label{fig:sensitivity_BE}
The reach of a gamma ray telescope like Fermi-LAT to probing an IGMF of strength $B_{0}$ using cascade gamma rays of energy $E_{\gamma}$.  
We consider blazars located at both $\ds = 1 \Gpc$ (blue) and $0.7 \Gpc$ (red) from Earth.  
The two upper lines labeled $\delta \approx D_{e}/R < 1$ demarcate the boundary between the MBC regime, where a measurement of the halo size furnishes a measurement of the field strength, and the PH regime, where the halo size is insensitive to the field strength.  
The two middle curves labeled $\Theta_{\rm ext} > \delta \theta_{68}$ indicate the field strength below which the halo size is smaller than the Fermi-LAT $68\%$ confinement radius.  
Finally, we suppose that many cascade photons are observed $N_{\gamma} = 1 + 10^{3} (E_{\gamma}/{\rm GeV})^{-2}$, and the two lower curves, $\Theta_{\rm ext} > \delta \theta_{68} / \sqrt{N_{\gamma}}$, indicate the approximate field strength below which the halo cannot be resolved.  
}
\end{center}
\end{figure}

%=========
Therefore in the strong field regime, measurements of the cascade halo have a reduced capacity to probe the parameters of the IGMF.  
To avoid this regime, we require 
\begin{align}
	\frac{D_{e}}{R} \lesssim 1 \com
\end{align}
which leads to an upper bound on the magnetic field strength, 
\begin{align}
	B_{0} \lesssim (15 \times 10^{-15} \Gauss) \left( \frac{E_{\gamma}}{10 \GeV} \right) \left( \frac{1 + \zs}{1.24} \right)^{2}
\end{align}
upon using the expression for $D_{e}/R$ in \eref{eq:De_over_R}.  
We plot this boundary in \fref{fig:sensitivity_BE}.  
It divides the PH regime $D_{e}/R>1$ from the MBC regime $D_{e}/R<1$.  

%=========
\ul{\it Weak Field Regime} \\
A gamma ray telescope cannot pinpoint the arrival direction of a gamma ray with arbitrary precision.  
If the angular resolution is too poor, then the halo cannot be distinguished from a point source, and one cannot use halo size and shape measurements to probe the IGMF.  
This issue becomes especially relevant for weak magnetic fields, which do not induce much bending and lead to a smaller halo ({\it cf.} \eref{eq:Theta_ext}).  

%=========
Angular resolution is quantified by the point-spread function (PSF), which is the probability distribution function for the angle between the true and reconstructed arrival direction of a gamma ray of given energy.  
The $68\%$ confinement radius is the angular radius containing $68\%$ of the probability. 
The confinement radius of the Fermi-LAT (Large Area Telescope) \cite{Fermi-LAT:2013yma} is well-approximated by the following empirical formula:\footnote{This formula matches Fig.~2 of \cite{Fermi-LAT:2013yma} for the case of on-orbit data (${\rm P6\_V11}$) and front-converting events.  }
\begin{align}\label{eq:confinement_radii}
	& \delta \theta_{68}(E_{\gamma}) \simeq (0.11^{\circ}) \sqrt{ 1 + \left( \frac{E_{\gamma}}{7.9 \GeV} \right)^{-1.62} } \per
\end{align}
Although our analysis is not specific to the Fermi-LAT instrument, we use this confinement radius as fiducial point of reference.

%=========
In the weak field regime, the angular extent of the halo is given approximately by $\theta \sim \Theta_{\rm ext}(E_{\gamma})$ from \eref{eq:Theta_ext}, regardless of the specific field configuration under consideration.  
If $\Theta_{\rm ext}$ is sufficiently large compared to $\delta \theta_{68}$, then the halo can be distinguished from a point source and its angular extent measured.  
In fact, if the detector response is known very well (negligible systematic error), then halos as small as $\delta \theta_{68} / \sqrt{N_{\gamma}}$ can be probed when a large number $N_{\gamma}$ of halo gamma rays are visible.  
Thus we assess when the telescope will be able to distinguish the halo from a point source using
\begin{align}\label{eq:ang_res_limit}
	\Theta_{\rm ext}(E_{\gamma}) \gtrsim \frac{\delta \theta_{68}(E_{\gamma})}{\sqrt{N_{\gamma}(E_{\gamma})}}
\end{align}
where $N_{\gamma}(E_{\gamma})$ is the number of gamma rays collected at energy $E_{\gamma}$.  

%=========
The requirement of sufficient angular resolution in \eref{eq:ang_res_limit} leads to a lower bound on the magnetic field strength,
\begin{align}
	B_{0} \gtrsim \frac{(0.16 \times 10^{-15} \Gauss)}{\sqrt{N_{\gamma}(E_{\gamma})}} \left[1 + \left( \frac{E_{\gamma}}{7.9 \GeV} \right)^{-1.62} \right]^{1/2} \left( \frac{E_{\gamma}}{10 \GeV} \right)^{3/2} \left( \frac{\ds}{1\Gpc} \right) \left( \frac{1+\zs}{1.24} \right)^{3}
\end{align}
which is shown in \fref{fig:sensitivity_BE}.  
For smaller $B_{0}$ the cascade halo is too small to distinguish from a point source given an angular resolution comparable to the Fermi-LAT.  
We show a scenario with small photon counts, $N_{\gamma} = 1$, and large photon counts, $N_{\gamma} = 1 + 10^{3} (E_{\gamma}/{\rm GeV})^{-2}$, which could potentially be achieved in a stacked halo analysis \cite{Ando:2010rb, Chen:2014rsa}.  
The $E_{\gamma}^{-2}$ dependence is included as a crude model of the gamma ray flux:  lower energy gamma rays are more abundant.  

%=========

\begin{figure}[p]
\hspace{0pt}
\vspace{-0in}
\begin{center}
\includegraphics[width=0.45\textwidth]{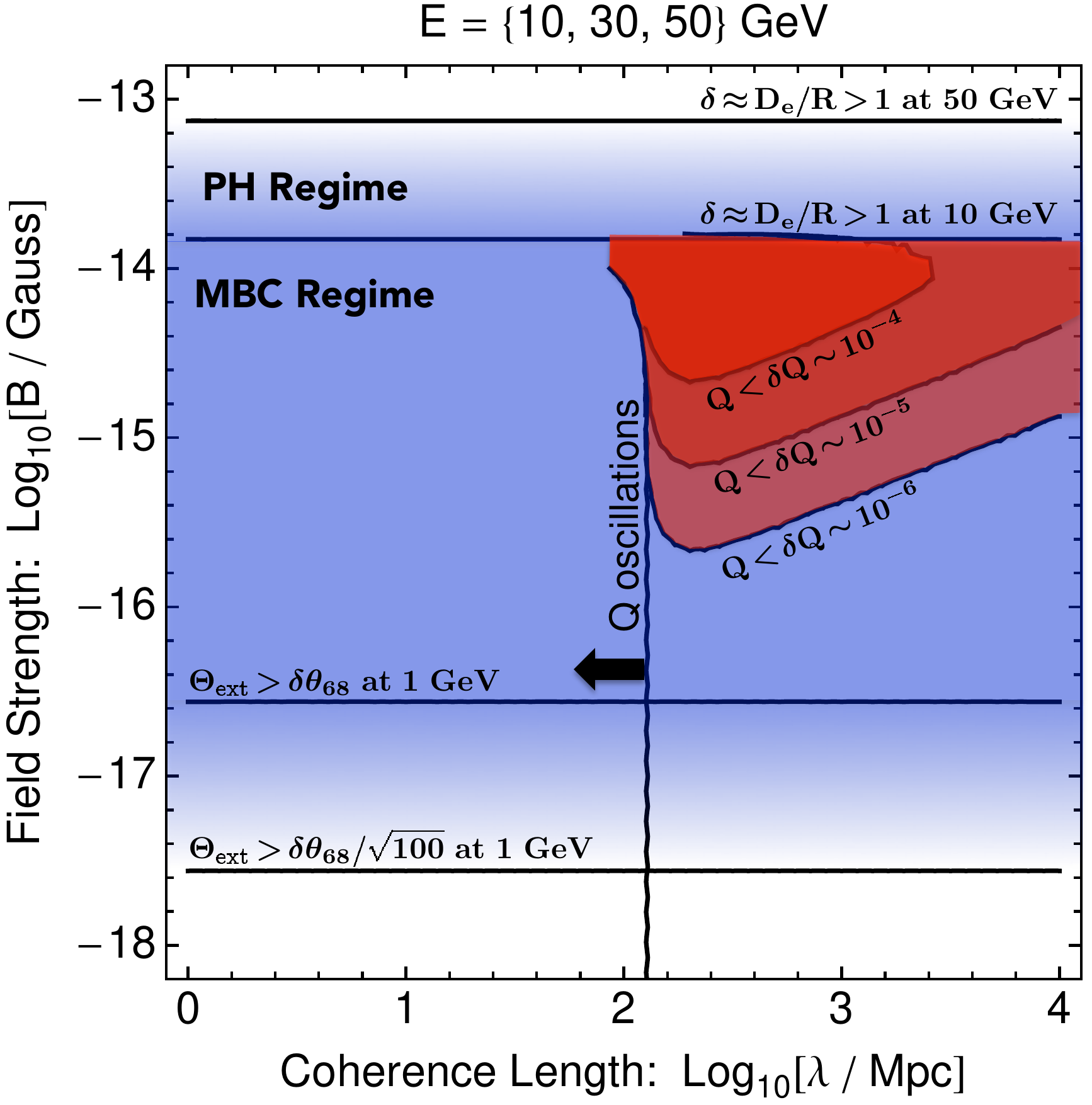} \hfill
\includegraphics[width=0.45\textwidth]{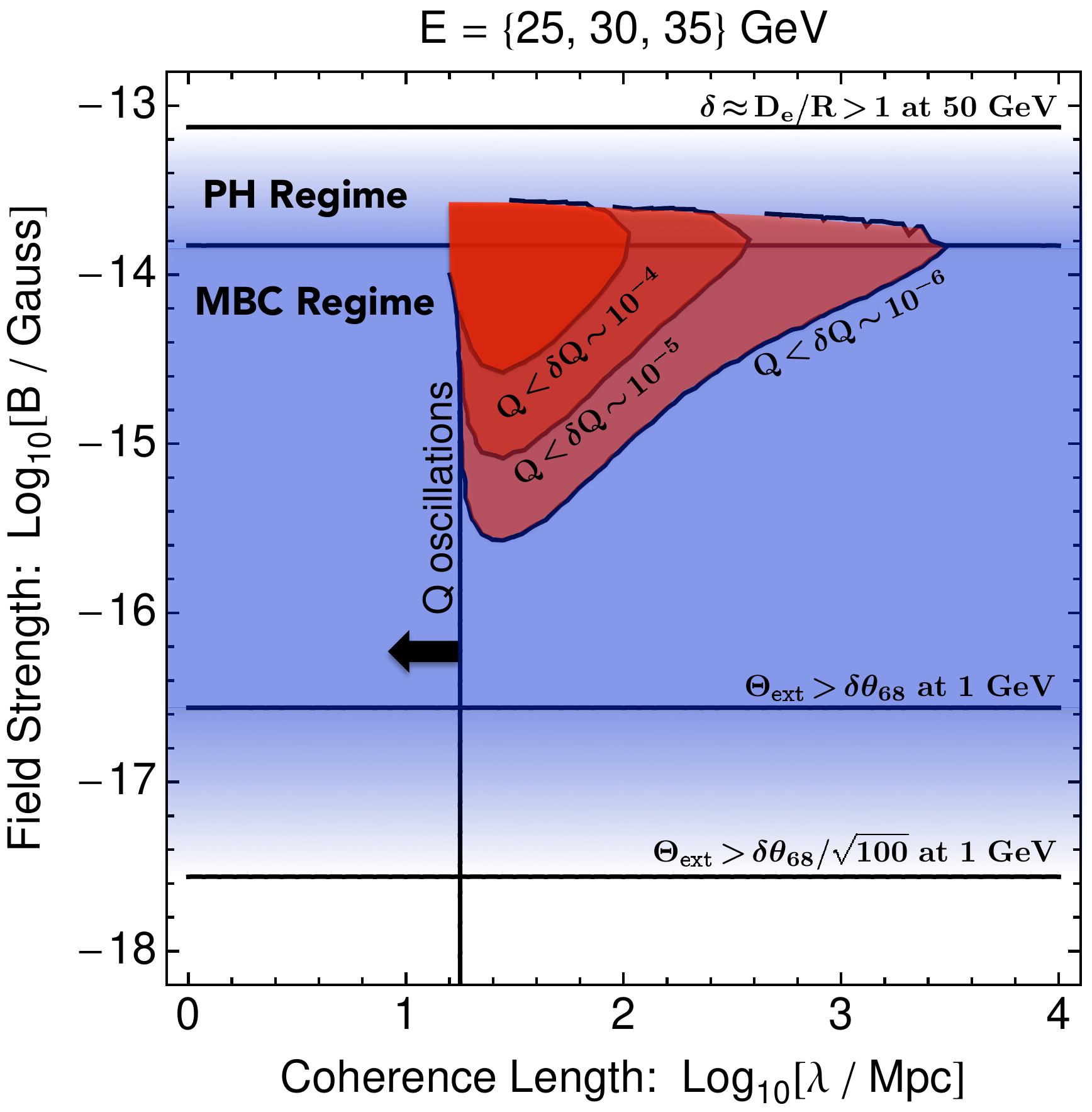} \\
\vspace{0.5cm}
\includegraphics[width=0.45\textwidth]{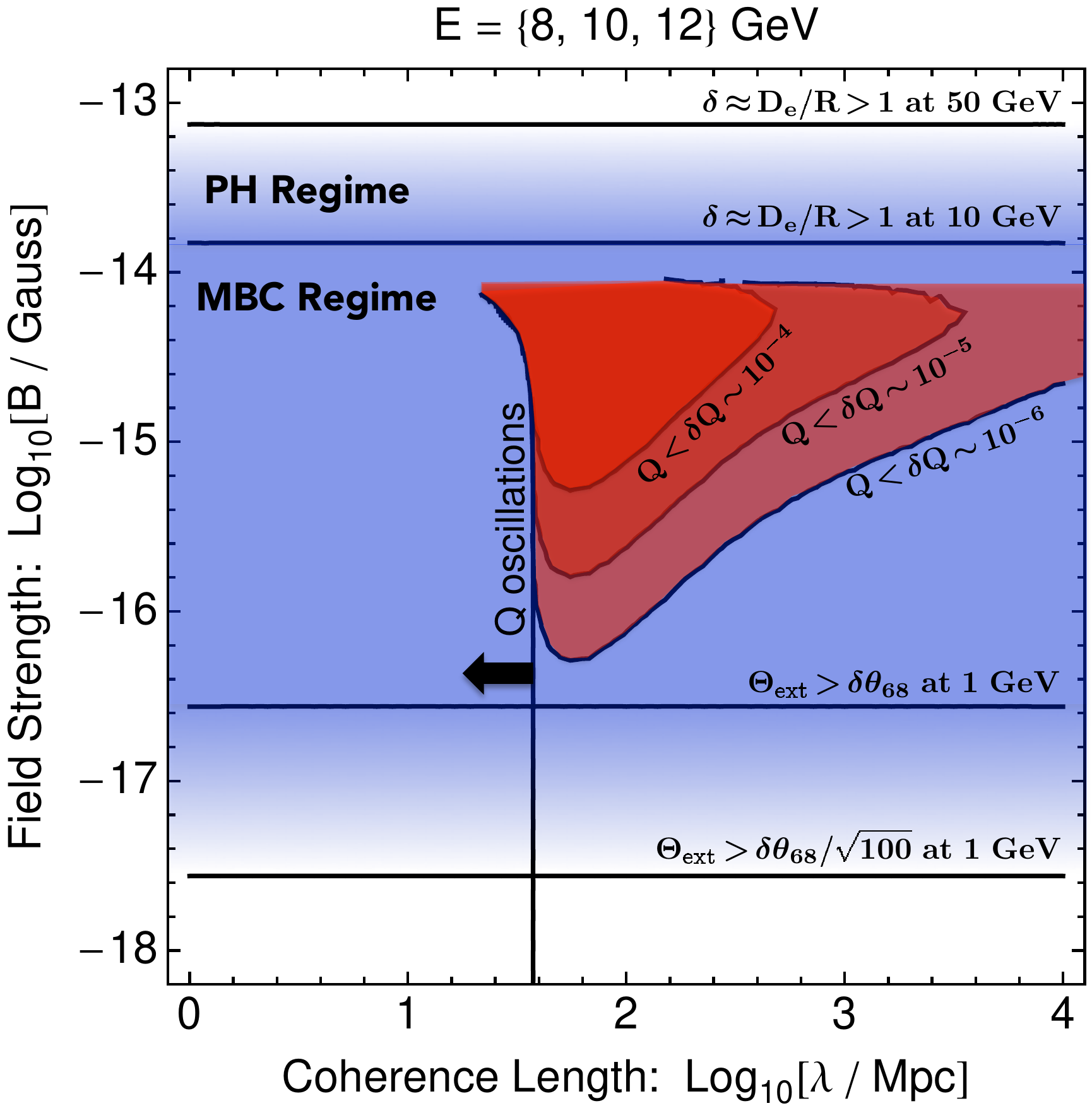} \hfill\includegraphics[width=0.45\textwidth]{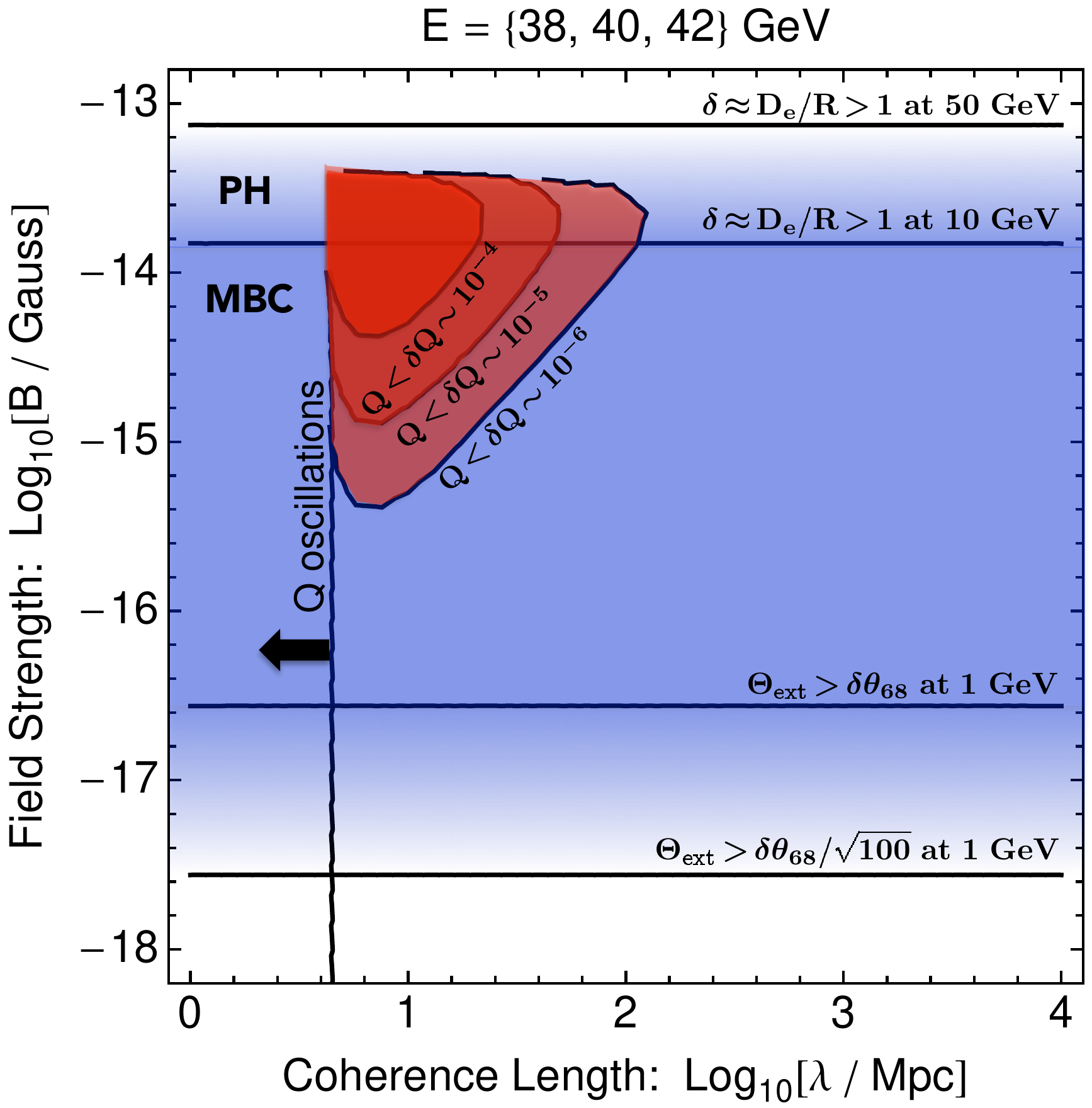} 
\caption{
\label{fig:case4_sensitivity}
This sensitivity plot gives a rough indication of the region of magnetic field parameter space that can be probed with a gamma ray telescope like the Fermi-LAT using four different $Q$-statistics, \eref{eq:Q_def}.  
In the blue shaded regions, the magnetic field strength can be inferred from a measurement of the halo {\it size}.  
For smaller field strength the halo cannot be distinguished from a point source, and for larger field strength (PH regime), one may still be able to measure the halo size, but its connection to the field strength is obscured, see \fref{fig:case2_grid}.  
In the red shaded region, the magnetic helicity can be inferred from a measurement of the halo {\it shape} via the $Q$-statistic.  
(These curves are defined with respect to the helical field configuration in Case 4 of \sref{sec:case4}, and we expect similar boundaries for more general helical configurations.)  
For larger $\lambda$ and smaller $B$ the parity-odd signal is too small to measure given the assumed angular resolution, 
and for smaller $\lambda$ one may still be able to measure the halo shape via $Q$, but its connection to the coherence length is obscured, see \fref{fig:case4_plotQ}.  
}
\end{center}
\end{figure}

%=========
The preceding discussion yields a range of field strengths where measurements of cascade halo sizes are best suited to probing the IGMF.  
This range depends on the energy of the observed gamma rays, as illustrated in \fref{fig:sensitivity_BE}, and taking the extremal values $1 \GeV < E_{\gamma} < 100 \GeV$ leads to $10^{-18} \Gauss < B_{0} < 10^{-13} \Gauss$.  
However, the range also depends on photon fluxes.  
For instance, probing $B_{0} \gtrsim 10^{-14} \Gauss$ requires a sufficient abundance of high energy photons with $E_{\gamma} \gtrsim 10 \GeV$, which remain in the MBC regime, and probing $B_{0} \lesssim 10^{-17} \Gauss$ requires a large number of low energy photons $E_{\gamma} \lesssim 10 \GeV$ so as to beat down the effective angular resolution.  
More conservatively speaking, the range of magnetic field strengths that can be probed using cascade halo size measurements lies between $10^{-18}-10^{-17} \Gauss$ and $10^{-14}-10^{-13} \Gauss$.  
This range is indicated on the parameter space plots of \fref{fig:case4_sensitivity} as the blue shaded region.  
The gradient in the blue color indicates where the boundaries depend on the details of the analysis, just discussed.  
This range is independent of the magnetic field coherence length $\lambda$ as long as $\lambda$ is sufficiently large that the charged lepton probes an effectively homogeneous magnetic field, {\it i.e.}, 
$\lambda \gtrsim D_{e} \sim 100 \kpc$.  
For smaller coherence length, the charged leptons do not follow helical arcs; instead their motion is that of a random walk, and our analysis breaks down.  

%=========
\ul{\it Short Coherence Length Regime} \\
When the coherence length is small, information about spatial inhomogeneities is encoded in features of the halo at small angular scales.  
It is challenging to measure this halo substructure given limitations on angular resolution and photon flux.  
If one is only able to measure the large scale halo morphology, then gamma ray observations have a diminished capacity to probe magnetic helicity on small length scales.  

%=========
As a specific example, recall the study of helical magnetic fields in Case 4 of \sref{sec:case4}.  
In the left panel of \fref{fig:case4_arrival}, the halo of GeV gamma rays form a spiral around the blazar.  
For small $\lambda$ the angular scale of the spiral (separation between subsequent cycles) is smaller than the angular scale of the full halo.  
If we coarse grain on the scale of the halo, {\it e.g.} to model the finite detector resolution, the map would appear rotationally symmetric, but when the small scale behavior is resolved, the spiral can be seen.  

%=========
To study the small scale structure of the halo, we require not only good angular resolution, but also closely spaced gamma ray energies.  
Recall from \erefs{eq:dg_def}{eq:Eg_to_Ee} that the mean free path of the TeV gamma ray is
\begin{align}
	\dg \simeq (180 \Mpc) \left( \frac{1+\zs}{1.24} \right)^{-1} \left( \frac{E_{\gamma}}{10 \GeV} \right)^{-1/2} 
\end{align}
where $E_{\gamma}$ is the energy of the GeV gamma ray reaching Earth.  
Each TeV gamma ray emitted by the blazar can be viewed as a local probe of the magnetic field at the point where pair production occurs.  
If we want to probe the magnetic field on the length scale $\lambda$,
then we should employ gamma rays with energies $E_{1} = E_{\gamma}$ and $E_{2} = E_{\gamma} + \Delta E_{\gamma}$ such that $\dg(E_{1}) - \dg(E_{2}) \approx \lambda$.  
The optimal energy separation is therefore 
\begin{align}\label{eq:DE_to_lambda}
	\Delta E_{\gamma} \approx 1 \GeV \left( \frac{E_{\gamma}}{10 \GeV} \right)^{3/2} \left( \frac{\lambda}{10 \Mpc} \right) \left( \frac{1+\zs}{1.24} \right)
\end{align}
assuming $\Delta E_{\gamma} \ll E_{\gamma}$ for which $\lambda \approx ({\rm d}\ds/{\rm d}E_{\gamma}) \Delta E_{\gamma}$.  
Considerations of gamma ray flux prohibit one from taking $\Delta E_{\gamma}$ arbitrarily small to probe arbitrarily small $\lambda$.  

%=========
Since the $Q$-statistics are constructed from three different gamma ray energies, as in \eref{eq:Q_def}, the associated $\Delta E_{\gamma}$ and $E_{\gamma}$ for a given statistic lead to a corresponding optimal coherence length $\lambda_{\rm opt}$ via \eref{eq:DE_to_lambda}.  
The effects of inhomogeneities on smaller length scales, $\lambda < \lambda_{\rm opt} \approx \dg(E_{\gamma}) - \dg(E_{\gamma}+\Delta E_{\gamma})$, are washed out.  
We have seen this behavior explicitly for Case 4 in \fref{fig:case4_plotQ} and Case 5 in \fref{fig:case5_plotQ}:  as the coherence length decreases below $\lambda \sim \lambda_{\rm opt}$, the $Q$-statistic stops growing and turns over or flattens out.  
To ensure that the optimal coherence length of a particular $Q$-statistic is small enough to probe magnetic field inhomogeneities on the scale $\lambda$, we impose 
\begin{align}\label{eq:lambda_optimal}
	\lambda \gtrsim 10 \Mpc \, \left( \frac{\Delta E_{\gamma}}{1 \GeV} \right) \left( \frac{E_{\gamma}}{10 \GeV} \right)^{-3/2} \left( \frac{1+\zs}{1.24} \right)^{-1} \com
\end{align}
While a non-zero measurement of $Q$ is of interest even for smaller $\lambda$, it becomes difficult to draw a connection between the value of $Q$ and the parameters of the underlying magnetic field.  
This is particularly evident in \fref{fig:case4_plotQ} where $Q$ oscillates with varying $\lambda$, and even ${\rm sign}[Q]$ does not necessarily correspond to the sign of the magnetic helicity.  
Presumably detailed modeling of the cascade development would be required to infer properties of the IGMF at such small $\lambda$.

%=========

%=========
In the parameter space plots of \fref{fig:case4_sensitivity} we indicate the coherence length below which oscillations set in by writing ``Q oscillations.''  
This boundary is drawn at the point where each of the $Q$-statistics has its first zero-crossing, determined numerically, but it is well-approximated by \eref{eq:lambda_optimal}.  
For the energies and energy spacings that we have shown, the $Q$-statistics are best suited to probe coherence lengths in the range $10-100 \Mpc$.  
In order to probe shorter coherence lengths, \eref{eq:lambda_optimal} indicates that we need to consider $Q$-statistics built from larger $E_{\gamma}$ and smaller $\Delta E_{\gamma}$.  
However, this limit is technically challenging:  higher energy gamma rays are less abundant and this problem is exacerbated by small energy bins.  

%=========
\ul{\it Long Coherence Length Regime} \\
We have already seen in the analyses of Secs.~\ref{sec:case4}~and~\ref{sec:case5} that the $Q$-statistic becomes small as the coherence length is made large.  
In this limit, the magnetic field is effectively homogeneous on the scale probed by the cascade, and information about the magnetic helicity, which is encoded in gradients, becomes inaccessible.  
In terms of the halo map, the characteristic ``S''-like curve becomes flattened out (see the left panel of Fig.~\ref{fig:case4_arrival}~or~\ref{fig:case5_arrival}), and the transverse angular extent of the halo becomes small.  
As a consequence angular resolution is a challenge for a measurement of small $Q$ since one needs not only to distinguish the halo from a point, but also distinguish the flattened S from a line.  

%=========
Recall that the triple product $Q_{abc} = \nhat_{a} \times \nhat_{b} \cdot \nhat_{c}$ can be written as in \eref{eq:Q_alt}
\begin{align}
	Q_{abc} = \sin \vartheta_{ab} \, \sin \varphi_{abc}
\end{align}
where $\vartheta_{ab}$ is the angle between $\hat{\bf n}_{a}$ and $\hat{\bf n}_{b}$, and $\varphi_{abc}$ is the angle between $\hat{\bf n}_{c}$ and the plane spanned by the other two vectors.  
Because of the detector's finite angular resolution, the unit vectors can only be localized on the sphere with a precision of $\delta \theta_{68}$.  
This translates into a comparable uncertainty on $\vartheta_{ab}$ and $\varphi_{abc}$, and we estimate the precision with which the $Q$-statistic can be measured as
\begin{align}
	\delta Q_{abc} \approx \sqrt{ \cos^2 \vartheta_{ab} \, \sin^2 \varphi_{abc} + \sin^2 \vartheta_{ab} \, \cos^2 \varphi_{abc} } \ \delta \theta_{68} \per
\end{align}
The factor on the right side depends on the choice of gamma ray energies used to form the triplet as well as the specific magnetic field configuration being considered.  
As a rough estimate, we assume the small angle approximation and take $\vartheta_{ab} \sim \varphi_{abc} \sim \Theta_{\rm ext}(10 \GeV)$ using \eref{eq:Theta_ext}.  
Thus we infer that the Fermi-LAT angular resolution is satisfactory to measure a $Q$-statistic satisfying
\begin{align}\label{eq:Q_res_limit}
	Q > \delta Q \sim \Theta_{\rm ext}(10 \GeV) \, \delta \theta_{68} \sim 3 \times 10^{-5} \frac{B_{0}}{10^{-15} \Gauss} \com
\end{align}
which is analogous to \eref{eq:ang_res_limit} for the halo size.  
As per the discussion surrounding \eref{eq:ang_res_limit}, we expect that large photon counts can weaken the bound.  
Since this is a rough estimate, we consider $\delta Q \simeq 10^{-4}, 10^{-5},$ and $10^{-6}$ in the following analysis.  

%=========
In \fref{fig:case4_sensitivity} we also show the boundaries corresponding to the inequality in \eref{eq:Q_res_limit} for each of the energy combinations considered previously.  
The heavy red shaded region corresponds to $Q > \delta Q \sim 10^{-4}$ where gamma ray observations of cascade halos should be able to measure the strength and helicity of the magnetic field.  
Moving outside of the red shaded region, the helicity measurement becomes more challenging, because either $Q$ is too small to measure given an angular resolution comparable to the Fermi-LAT (large $\lambda$ regime) or $Q$ has a complicated dependence on the magnetic helicity (small $\lambda$ regime).  
Outside of the blue shaded region, even the field strength measurement becomes challenging, because either the halo is too small to distinguish from a point source given the angular resolution (small $B_{0}$ regime) or the halo size is insensitive to the field strength (large $B_{0}$ regime).  
With better statistics one can presumably beat down the angular resolution issues to probe weaker fields.

%==================================
% SUMMARY AND DISCUSSION
%==================================
\section{Summary and Discussion}\label{sec:Conclusion}

%=========
One day gamma ray observations may provide measurements of both the size and the shape of cascade halos.  
Whereas the halo size is tied to the average magnetic field strength, the halo shape can depend on other properties of the magnetic field, particularly its helicity.  
In this work we have attempted to illuminate the connection between magnetic helicity and halo shape by studying the development of the electromagnetic cascade for simplified magnetic field configurations.  
We have taken an analytic approach in which the three equations, (\ref{equation1}),~(\ref{equation2}),~and~(\ref{equation3}), are solved to find the trajectory of gamma rays reaching Earth.  

%=========
By acting as local probes of the magnetic field, blazar-induced cascade halos offer an interesting opportunity to measure properties of the IGMF.  
The physical parameters of the cascade and the detector resolution come together to determine the region of magnetic field parameter space that can be probed with this technique.  
By studying a particular helical magnetic field configuration with its wavevector oriented along the line of sight to the blazar, we have estimated the boundaries of this region in the two dimensional parameter space consisting of field strength $B_{0}$ and coherence length $\lambda$; see \fref{fig:case4_sensitivity}.  
The boundaries are partly determined by the anticipated detector resolution, and an improvement in angular resolution would allow access to smaller $B_{0}$ and larger $\lambda$.  
The boundaries are also depend on the measurement strategy, namely the energy combinations used to form the parity-odd $Q$-statistics via \eref{eq:Q_def}.
For the four statistics considered here, we find that blazar-induced cascade halos are well-suited to probing helical magnetic fields with coherence length $\lambda \gtrsim 10 \Mpc$, see \eref{eq:lambda_optimal}.  
Reaching smaller length scales would require more closely spaced energy combinations, and then one runs into issues with angular resolution and photon flux.  

%=========
There are a number of avenues for future work.  
First, we have considered only the simplest magnetic field configurations in order to draw the connection between halo shape and magnetic helicity.  
These configurations are not likely to be realistic models of the true IGMF. 
It would be straightforward to generalize our analysis and consider field configurations built from multiple modes in random orientations.  
In this case, it may become too difficult to solve Eqs.~(\ref{equation1})-(\ref{equation3}), and instead one would want to simulate the cascade using a shooting algorithm, such as in the analysis of \rref{Elyiv:2009bx}.  

%=========
Second, the parity-odd test statistic $Q$ is not an automatic proxy for the magnetic helicity.  
We have seen in \sref{sec:case4} that even the sign of $Q$ won't necessarily equal the sign of the 
magnetic helicity if the coherence length is small.  
(This was also pointed out in \cite{Tashiro:2014gfa,Chen:2014qva} in the case of statistically homogeneous 
and isotropic magnetic fields.)  
The statistic $Q$ can still be useful in this regime if the three energies at which it is evaluated are finely spaced, though then the photon counts may be small and the resulting error bars will be large.  
Alternately, now that we know qualitatively the effects of magnetic field helicity in the gamma ray pattern, it may be possible to devise improved statistics to detect these patterns. 

\begin{figure}[t]
\hspace{0pt}
\vspace{-0in}
\begin{center}
\includegraphics[width=0.45\textwidth]{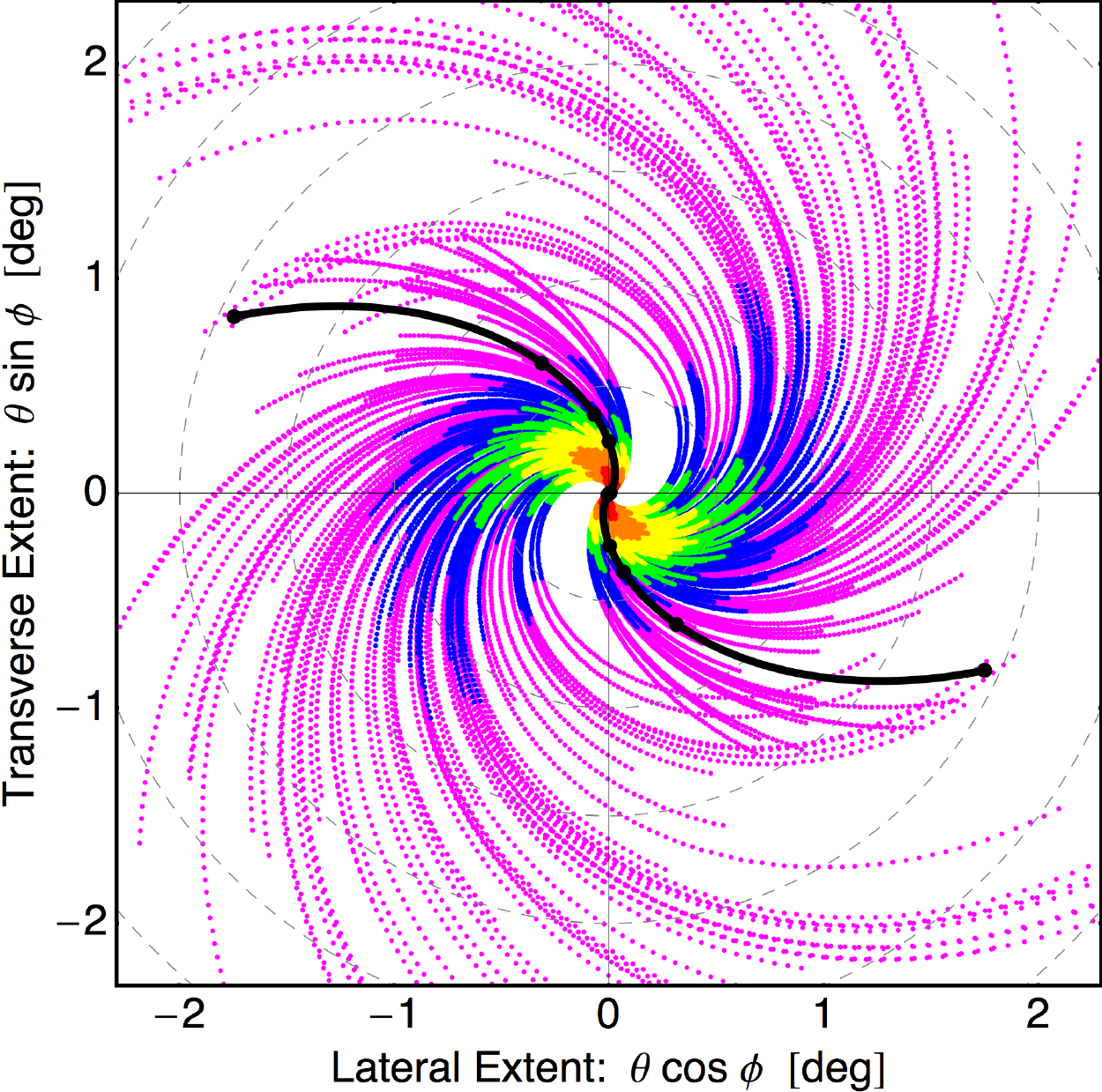} 
\caption{
\label{fig:smearing}
A ``smeared'' halo map for Case 4 with parameters $B_{0} = 10^{-15} \Gauss$, $\lambda = 700 \Mpc$, and $d_{s} = 1 \Gpc$.  The same parameters correspond to the the green curve in \fref{fig:case4_arrival} (left panel), which is also shown on this figure as a black curve.  The colors denote different gamma ray energies: $5 < E_{\gamma} / \GeV < 10$ (magenta), $10 < E_{\gamma} / \GeV < 15$ (blue), $15 < E_{\gamma} / \GeV < 20$ (green), $20 < E_{\gamma} / \GeV < 30$ (yellow), $30 < E_{\gamma} / \GeV < 50$ (orange), and $50 < E_{\gamma} / \GeV < 100$ (red).  
}
\end{center}
\end{figure}

%=========
Third, our analysis neglects the stochastic nature of the cascade's development.  
This simplification allows us to calculate the halo map $\hat{\bf n}(E_{\gamma})$ that gives a deterministic connection between the energy of a gamma ray and its arrival direction on the sky.  
A more realistic model would account for the stochasticity in the gamma ray propagation distances, 
spectra, and so on.  
As a crude model of this effect, we calculate a halo map for Case 4 by replacing $\dg$ and $D_{e}$ in \Eqns with $r_1 \dg$ and $r_2 D_{e}$ where $r_1$ and $r_2$ are randomly selected from a uniform probability distribution over the range $0.5 < r_1, r_2 < 1.5$.  
We calculate $100$ such halo maps in this way for the same parameter set and show the results in \fref{fig:smearing}.  
Although the degree of smearing is significant, the halo map retains the qualitative features that we have seen previously, namely it angular extent and parity-odd spiraling shape.  

%=========

\begin{figure}[p]
\hspace{0pt}
\vspace{-0in}
\begin{center}
\includegraphics[width=0.15\textwidth]{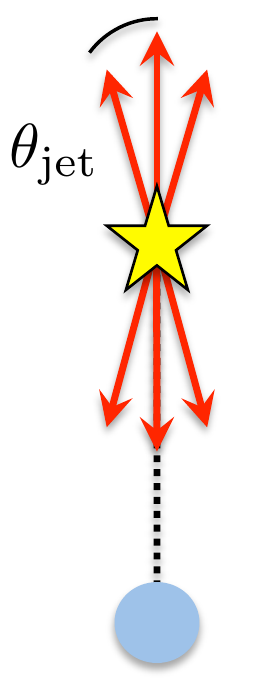} 
\qquad
\includegraphics[width=0.35\textwidth]{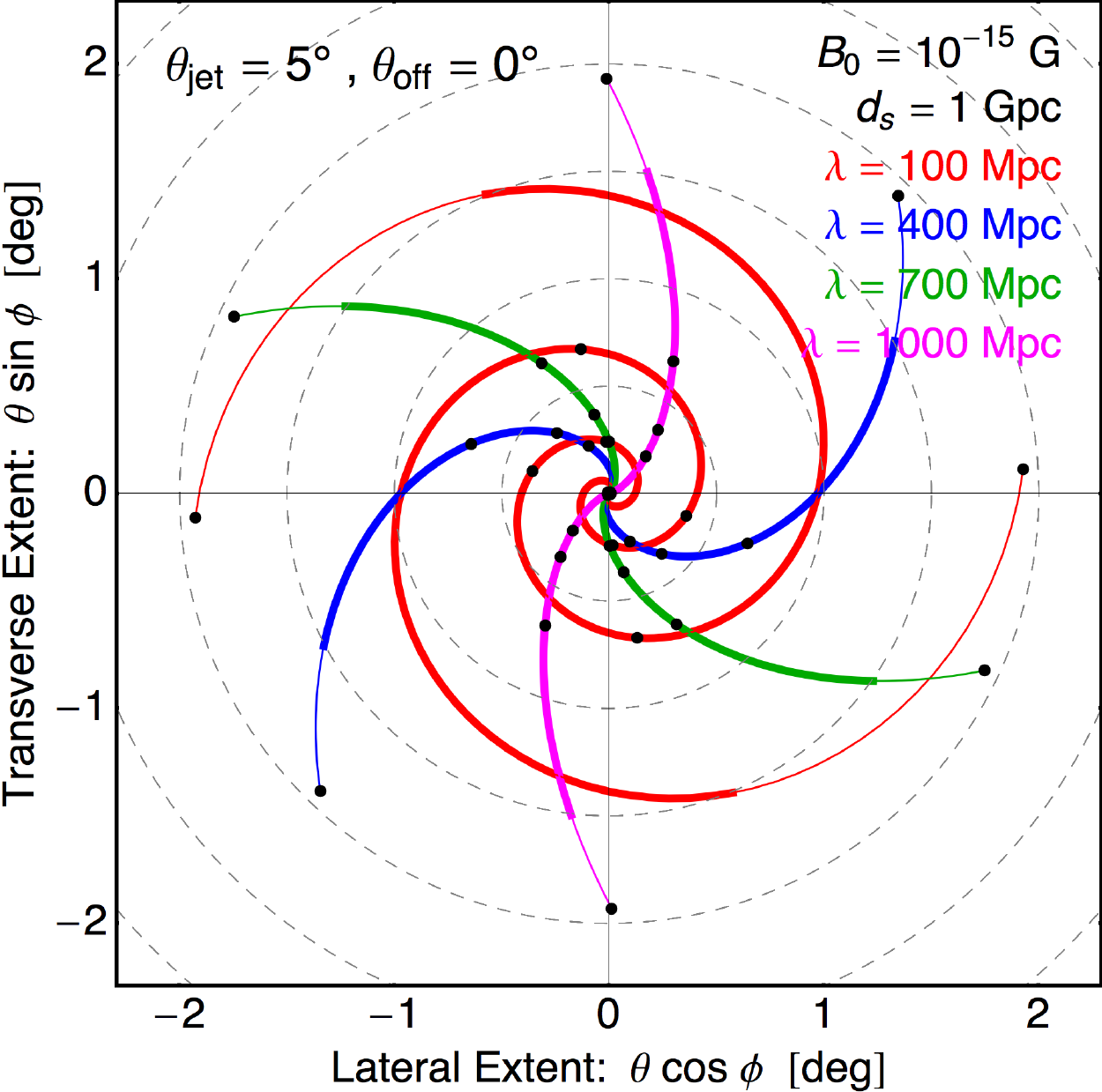} 
\\
\includegraphics[width=0.15\textwidth]{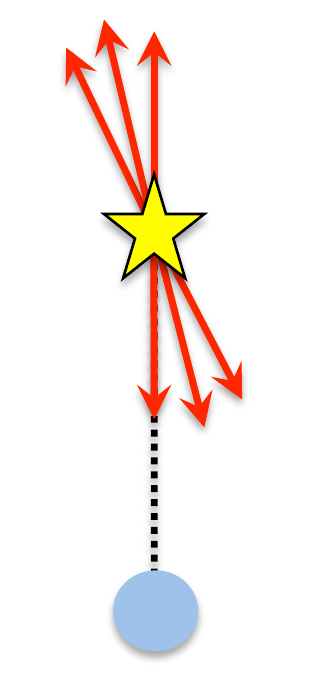} 
\qquad
\includegraphics[width=0.35\textwidth]{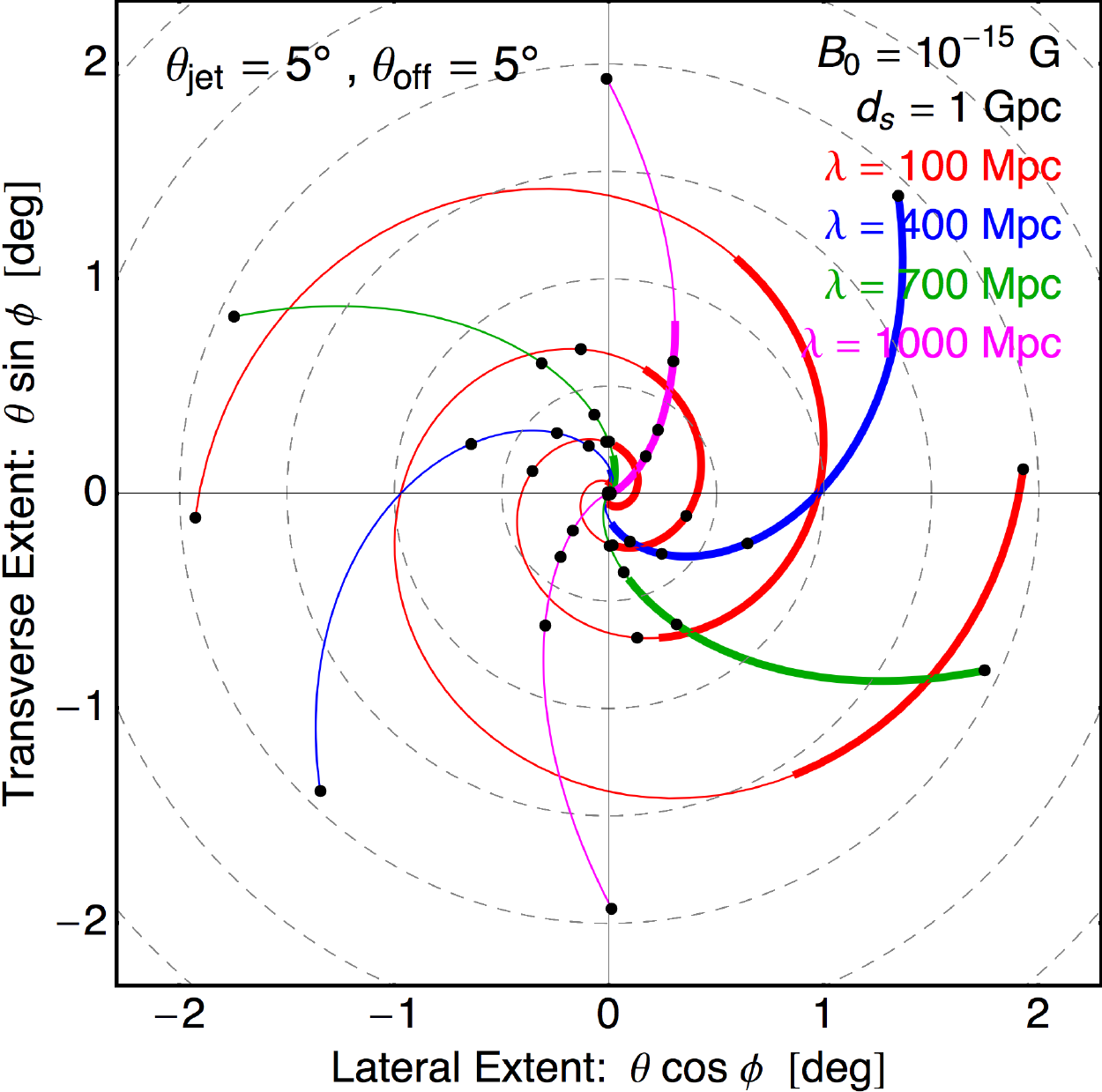} 
\\
\includegraphics[width=0.15\textwidth]{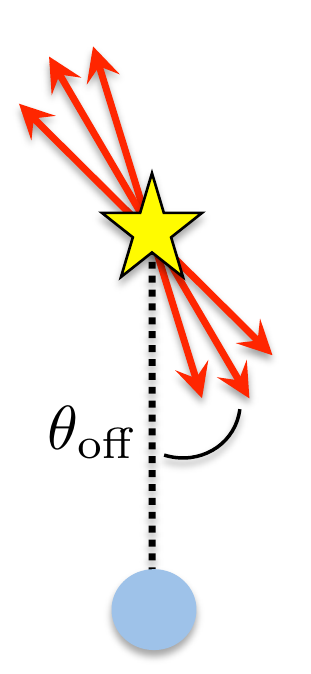} 
\qquad
\includegraphics[width=0.35\textwidth]{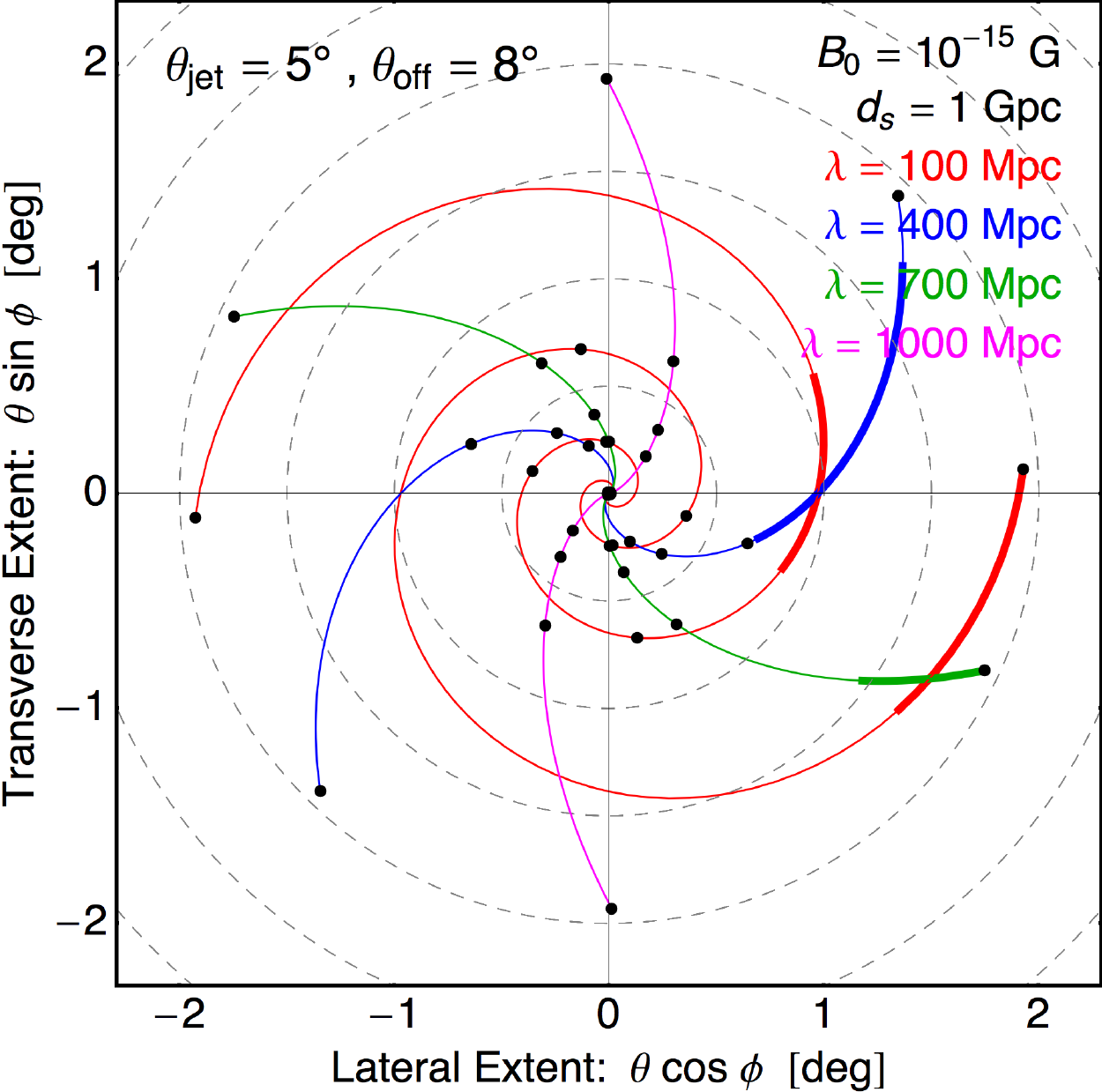} 
\\
\caption{
\label{fig:jet_maps}
The effect of the finite jet angle, $\theta_{\rm jet}$, and offset from the line of sight, $\theta_{\rm off}$, where $\phi_{\rm off} = 0$.  The thin colored curves show the halo map for an isotropically emitting blazar, $\theta_{\rm jet} = 180^{\circ}$, and the thick curves show the halo map for a blazar jet with $\theta_{\rm jet} = 5^{\circ}$ and different offsets $\theta_{\rm off} = 0^{\circ}, 5^{\circ},$ and $8^{\circ}$ in the three rows, respectively.  For $\theta_{\rm off} > 10^{\circ}$ none of the halo is visible with $E_{\gamma} > 5 \GeV$.  
}
\end{center}
\end{figure}

%=========
Fourth, our analysis neglects the jet structure of the blazar, and assumes that the emission is isotropic.  
It is straightforward to include the effect of the jet in our semi-analytic formalism.  
Let
\begin{align}\label{eq:njet_def}
	\nhat_{\rm jet}(\theta_{\rm off}, \phi_{\rm off}) = \sin (\theta_{\rm off}) \hat{\bm \rho}(\theta_{\rm off},\phi_{\rm off}) - \cos (\theta_{\rm off}) \hat{\bf z}(\theta_{\rm off},\phi_{\rm off})
\end{align}
be the orientation of the jet where $\theta_{\rm off}$ and $\phi_{\rm off}$ give the polar and azimuthal offset angles from the line of sight.  
The orientation of the initial TeV gamma ray, $\hat{\bf v}_{i}$, is given by \eref{eq:vi_def}.  
Then the jet criterion,
\begin{align}\label{eq:jet_cut}
	{\rm arccos} \Bigl[ \hat{\bf v}_{i} \cdot \nhat_{\rm jet} \Bigr] < \theta_{\rm jet} \com
\end{align}
ensures that the TeV gamma ray is emitted from within the cone of the jet.  
As a consequence of the jet criterion, only a portion of the full halo map becomes illuminated.  
This behavior is shown in \fref{fig:jet_maps} where we have reevaluated the halo maps from Case 4 using \eref{eq:jet_cut}, and there is a similar effect for the other cases.  
When $\theta_{\rm off} > \theta_{\rm jet}$ the Earth is not contained within the cone of the jet, an the blazar itself may not be seen.  
In this regime, the cascade photons would appear to contribute to the diffuse gamma ray flux.  

%=========

%----------------------------------------------------------------
% Acknowledgements
%----------------------------------------------------------------
\begin{acknowledgments}
We acknowledge illuminating discussions with Sean Bryan, James Buckley, Daniel Chung, Manel Errando, 
Francesc Ferrer, and Angela Olinto. We are especially grateful to Borun Chowdhury for discussion through 
the course of this project.
TV gratefully acknowledges the Clark Way Harrison Professorship at Washington University during the
course of this work.
This work was supported by the Department of Energy at ASU.  
A.J.L. was also supported in part by the National Science Foundation under grant number PHY-1205745.  
\end{acknowledgments}

%----------------------------------------------------------------
% References
%----------------------------------------------------------------
\newpage
\bibliographystyle{JHEP}
\bibliography{refs--Fermi_Sensitivity}

\end{document}